\documentclass[smallextended,final]{svjour3}       
\usepackage[rmargin=2.54cm,lmargin=2.54cm,tmargin=2.54cm,bmargin=2.54cm]{geometry}
\usepackage{amsmath,amsfonts,amssymb}
\usepackage{graphicx}
\usepackage[colorlinks=true,linkcolor=blue,citecolor=blue,urlcolor=blue,breaklinks]{hyperref}
\usepackage{float}
\usepackage{array}
\usepackage{multirow}
\usepackage{tabularx}
\usepackage[title]{appendix}
\usepackage{subfigure}
\DeclareMathOperator{\Tr}{Tr}
\begin{document}
\large
\title{Magneto-electric cooling rate of a multiferroic antiferromagnetic quantum spin system: the cumulative influence of the site-dependent magnetic and electric fields 
}

\titlerunning{Magneto-electric cooling rate of a multiferroic antiferromagnetic quantum spin
system}        


\authorrunning{W. D. WALADI GUEAGNI et al.} 

\author{William Degaulle Waladi Gueagni$ ^{1} $ \and  Lionel Tenemeza Kenfack$ ^{1,*} $ \and Martin Tchoffo$ ^{1,2} $ \and Georges Collince Fouokeng$ ^{1} $\and  Lukong Cornelius Fai$ ^{1} $
}
\institute{ $ ^{1} $Unit\'e de Recherche de Mati\`ere Condens\'ee, d'\'Electronique et de traitement de signal (URMACETS),\\ Department of physics, University of Dschang, PO Box: 67 Dschang, Cameroon. \and \\
$ ^{2} $Centre d'\'Etudes et de Recherches en Agronomie et en Biodiversit\'e, Facult\'e d'Agronomie et des Sciences Agricoles, University of Dschang, P.O. Box: 222 Dschang, Cameroon. \\
$ ^{*} $\email{leonel.kenfack@univ-dschang.org}           
}

\date{Received: \today/ Accepted: xxx}
\maketitle
\begin{abstract}
 The magneto-electrocaloric effect which can be defined as the coupling between magnetocaloric and electrocaloric effects attracts currently considerable attention due to the advantages provided by the caloric effect in designing solid-state refrigeration technologies. The magneto-electrocaloric effect of a multiferroic antiferromagnetic spin system with the Dzyaloshinskii Moriya (DM) interaction is investigated in this paper. The DM interaction is assimilated to a coupling between an external site-dependent electric field and a local electric polarization. The external magnetic is also considered as a site-dependent magnetic field. The spin-wave theory is used as a diagonalization method and through the canonical partition function, some thermodynamic properties such as the Boltzmann entropy and the specific heat capacity are obtained. Then, the adiabatic magnetic, electric, and magnetoelectric cooling rates are also derived. The graphs obtained for the adiabatic magnetic, electric, and magnetoelectric cooling rate show a characteristic behavior of the caloric or multi-caloric effect which is in good agreement with some experimental and theoretical works. Besides, the entropy response due to the variation of the external site-dependent magnetic field exhibits two anomalous entropy peaks indicating the existence of an intermediate phase tuneable by the magnetic and electric site-dependent parameters. Overall, it is demonstrated that the cumulative influence of the site-dependent magnetic and electric fields allows us not only to reveal quantum critical points hidden in a multiferroic quantum spin system but also to control the caloric or multi-caloric effect essential in the construction of solid-state refrigeration devices.
\end{abstract}
\section{Introduction}
 The exposition of a given material to a suitable varying external field instigates its heating or its cooling. This is the caloric effect. Thus, the magnetocaloric effect, the electrocaloric, and elastocaloric effect occurs when a material is exposed to a suitable magnetic field, electric field, and mechanical elastic field respectively. The multiferroic materials which exhibit two or more ferroics orders offer the possibility to obtain multi-caloric effects \cite{1,2,3}. Note that the ferroic material is a material which chooses spontaneously a preferable switchable alignment (anti/ferromagnet is witched by an external magnetic field, ferroelectric material is switched by an external electric field, etc.). The multiferroic materials are then found to be suitable for the technologies made by multi-caloric effects. The magneto-electrocaloric effect is an example of multi-caloric effects resulting from the coupling between magnetocaloric and electrocaloric effects. The caloric phenomena have been already observed by Joule in the mid of 19${}^{th}$ century as the thermo-elastic properties result of solids \cite{4} but the magnetocaloric effect terminology appeared the first times in the literature in the mid of 20${}^{th}$ century from the works of Weiss and Picard on nickel (Ni) \cite{5}. In recent decades and to date, the caloric effect received much effort from both experimental and theoretician physicists due to their promising application in cooling device technologies \cite{6,7,8,9,10}. The magnetocaloric effect which is the most caloric property studied today, allows the creation of magnetic refrigeration recognized to be a viable alternative to conventional vapor-compression refrigeration technology (see the review in ref \cite{11} and references therein for more detail). There are several investigations done concerning the electrocaloric effect in the literature \cite{12,13,14,15,16,17,18}   but so far, there are still few works done for the magneto-electrocaloric effect \cite{19,20,21}. 

\par 
Moreover, the magneto-electrocaloric effect implies the suggestion of the system under the cumulative influence of the magnetic and electric fields. The coexistence of ferroelectricity and magnetism is more explained by the Dzyaloshinskii Moriya (DM) interaction (the combination of the exchange interaction and spin-orbit coupling) which plays a crucial role in the field of study of multiferroic materials. Two theories describe the mechanism of magnetism-driven ferroelectricity in the sense of the DM interaction \cite{22}, the theory of Sergienko and Dagotto ( called Sergienko-Dagotto model) based on the ionic displacement \cite{23} and the theory proposed by Katsura, Nagaosa, and Balatsky (the so-called KNB theory) based on the electronic bias \cite{24}. The main prediction of the KNB theory exploited in this paper is the expression of electric polarization. Based on the definition of the DM interaction and the electric polarization, we have already demonstrated that the DM interaction intrinsically induces ferroelectricity and then ensures the multiferroicity of the system \cite{25}. That idea is better understood with the works of Pradeep Thakur and P. Durganandini \cite{26} and Jozef Sznajd \cite{27}, in which based on the KNB theory they interpreted the Dzyaloshinskii Moriya interaction as the coupling of an external electric field and a local electric polarization. Indeed, the antisymmetric DM interaction, which is the unique spin-spin interaction expressed by using a cross product is defined as $H_{DM} =D\sum _{i}\hat{S}_{i} \times \hat{S}_{i+1} $ where $D$ is the strength of DM interaction and $\hat{S}_{i}$ the spin operator of the atom located on-site $i$. When the strength $D$ is taken in the z-direction, the DM interaction can be assimilated to the coupling of an electric field taken in the y-direction and a local polarization $P\propto \sum _{i}e\times \hat{S}_{i} \times \hat{S}_{i+1} $ by taking the unit vector$e$ along the x-axis and pointing from the site $i$ to the site $i+1$. That is the best way to take into account the interplay between magnetism and electricity in our system with DM interaction.

\par
 Furthermore, it is important to note that there are two kinds of DM interactions such as the uniform and the site-dependent DM interaction also called staggered or modulated DM interaction. A lot of systems have been studied with uniform DM interaction (see \cite{28} and references therein). For instance, Shuling Wang et al. investigated the one-dimensional charge transfer magnets and found that the staggered DM interaction allow to control the transition temperature and the physical properties (electric polarization and magnetic properties) of the such system  \cite{29},  Avalishvili et al. studied the ground state phase diagram of a spin   XXZ Heisenberg chain with modulated DM interaction and theirs results reveal the formation of four states with two types of phase transitions such as Benezinskii-Kosterlitz-Thoules type and Ising type  \cite{30}, the study of the effect of the staggered DM interaction on a quasi-two-dimensional Shastry-Sutherland lattice by Tianqi Chen et al. show that the such interaction significantly modify the magnetization plateaux \cite{31}.  The effect of staggered DM interaction allows us to explore the phase diagram and to tune the phase transitions phenomena in multiferroics.

\par
 Indeed, the influence of both site-dependent magnetic and electric fields on the quantum phase transition occurring in multiferroic antiferromagnets is one of the main interests of this work. The caloric effect which is intimately related to the quantum critical points appears to be the best issue to address that investigation. One of the quantifiers of the caloric effect is the adiabatic cooling rate, a physical parameter very helpful in the study of quantum phase transitions \cite{32}. The adiabatic cooling rate, defined as an adiabatic temperature change of a system in response to a suitable applied external field is closely related to the generalized Gr\"{u}neisen ratio \cite{33}. Thus, the adiabatic magnetic and electric cooling rate interpreted as a particular case of the generalized Gr\"{u}neisen ratio change theirs sign when the magnetic and electric fields (which are the parameters governing the zero-temperature quantum phase transition) cross their critical values.

\par
 In this work, both the site-dependent DM interaction and the site-dependent magnetic field is taking into account. Since the DM interaction is assimilated to the coupling of electric field and electric polarization, the electric field will be staggered. Throughout this work, the terminology ``site-dependent field'' will be used to denote the ``staggered or random field" because the fields considered are closely related to the site parity of the atoms. Notice that there exist in the literature systems dealing with the site-dependent magnetic field \cite{34,35}. So far, to the best of our knowledge, the magneto-electrocaloric effect in a multiferroic antiferromagnet has not yet been investigated with both the site-dependent magnetic field and site-dependent electric field.

\par
 Today, the multiferroic antiferromagnetic system attracts considerable attention because it is susceptible to offer the new opportunity in multi-caloric technologies \cite{36,37,38,39}, as also demonstrated in this work.  Indeed, because of their multifunctional physical properties, robustness against external perturbation and very fast operation, multiferroic antiferromagnetic system tends to replace ferromagnetic materials \cite{40} . 

\par
 The main aim of this work is to investigate in detail the cumulative influence of the site-dependent magnetic field and site-dependent electric field on the magneto-electrocaloric effect in a multiferroic antiferromagnetic quantum spin system. The quantum phase transition phenomena are also addressed. 

\par
 The rest of the paper is structured as follows, the Hamiltonian and spin-wave theory are presented in Sec. II, the entropy and specific heat capacity properties are derived in Sec. III, the adiabatic cooling rates are studied in Sec. IV and the paper is closed in Sec. V with a conclusion.  
   
\section{The Hamiltonian and spin-wave theory }\label{S2}
The model considered for a multiferroic antiferromagnetic is an anisotropic Heisenberg model with Dzyaloshinskii Moriya interaction and on-site easy-axis anisotropic interaction \cite{41,42} as shown in the previous work \cite{25}. More precisely, it is an antiferromagnetic lattice divided into two interpenetrating sublattices a and b, and the full Hamiltonian is given as:
\begin{equation} \label{1}
\begin{split}
&H=-g\mu _{B} \sum _{i}\left[B_{0} \left(1+(-1)^{i} \lambda \right)+B_{A} \right]\hat{S}_{a,i}^{z}  -g\mu _{B} \sum _{j}\left[B_{0} \left(1+(-1)^{j} \lambda \right)-B_{A} \right]\hat{S}_{b,j}^{z}
\\& 
+\sum _{i,\delta }J\left(\hat{S}_{a,i}^{x} \hat{S}_{b,i+\delta }^{x} +\hat{S}_{a,i}^{y} \hat{S}_{b,i+\delta }^{y} +\Delta \hat{S}_{a,i}^{z} \hat{S}_{b,i+\delta }^{z} \right) +\sum _{j,\delta }J\left(\hat{S}_{b,j}^{x} \hat{S}_{a,j+\delta }^{x} +\hat{S}_{b,j}^{y} \hat{S}_{a,j+\delta }^{y} +\Delta \hat{S}_{b,j}^{z} \hat{S}_{a,j+\delta }^{z} \right)
\\&
+\left[\sum _{i,\delta }E_{0} \left(1+(-1)^{i} \tau \right)\left(\hat{S}_{a,i}^{x} \hat{S}_{b,i+\delta }^{y} -\hat{S}_{a,i}^{y} \hat{S}_{b,i+\delta }^{x} \right) -\sum _{j,\delta }E_{0} \left(1+(-1)^{j} \tau \right)\left(\hat{S}_{b,j}^{x} \hat{S}_{a,j+\delta }^{y} -\hat{S}_{b,j}^{y} \hat{S}_{a,j+\delta }^{x} \right) \right]
\\&
-D'\left[\sum _{i}\left(\hat{S}_{a,i}^{z} \right)^{2}  +\sum _{j}\left(\hat{S}_{b,j}^{z} \right)^{2}  \right]
\end{split} 
\end{equation} 
where $g$ is the gyromagnetic factor, $\mu _{B} $ is the Bohr magneton, $J $ is the exchange interaction, $\Delta$ the Heisenberg anisotropy parameter, $\hat{S}_{a,i} $ and $\hat{S}_{b,j} $ represent the spin operators of the $ith\; (jth)$ atom on sublattice  $a$ and $b$, respectively, with each sublattice containing $N$ atoms. The indices $i$ and $j$ label the $N$ atoms in sublattice $a$ and $b$, respectively, whereas the vector $\delta $ connects atom $i(j)$ with its nearest neighbors. $B_{0} \left(1+(-1)^{i} \lambda \right)$ is the applied external site-dependent magnetic field in the z-direction. $B_{A} $ is the anisotropy field, assumed to be positive which approximates the effect of the crystal anisotropic energy with the property of turning for a positive magnetic moment $\mu _{B} $ to align the spin on sublattice $a$ in the positive z-direction and spins on sublattice $b$ in the negative z-direction. $E_{0} \left(1+(-1)^{i} \tau \right)$ is the strength of the z-component of the staggered DM interaction assimilated in this work to the external site-dependent electric field in the y-direction coupling to the local electric polarization by taking the spin chain along the x-direction. The case with $B_{0} =0\left(E_{0} =0\right)$ corresponds to the absence of the external magnetic field (electric field) while the case with $\lambda =0\left(\tau =0\right)$ corresponds to the uniform magnetic field (electric field). So $\lambda $ and $\tau $ are the magnetic site-dependent parameter and the electric site-dependent parameter respectively. $D'$ is the on-site easy-axis anisotropy parameter for anisotropy interaction also looking as potential energy term, assumed to be positive. Also, for sake of simplicity, we have considered only the nearest neighbor interaction.

\par The Holstein-Primakoff transformations allow us to express the spin operators of the system to bosonic operators,
\begin{equation} \label{2} 
\hat{S}_{a,i}^{+} =\sqrt{2S-a_{i}^{+} a_{i} } a_{i}    ,\,\,\,  \hat{S}_{a,i}^{-} =a_{i}^{+} \sqrt{2S-a_{i}^{+} a_{i} }   ,\,\,\,  \hat{S}_{a,i}^{z} =S-a_{i}^{+} a_{i} ,   
\end{equation} 
\begin{equation} \label{3} 
\hat{S}_{b,j}^{+} =b_{j}^{+} \sqrt{2S-b_{i}^{+} b_{i} }   , \,\,\,  \hat{S}_{b,j}^{-} =\sqrt{2S-b_{i}^{+} b_{i} } b_{j}   , \,\,\,    \hat{S}_{b,j}^{z} =b_{j}^{+} b_{j} -S,   
\end{equation} 
then consider the situation that our system is in the low-temperature and low-excitation limit \cite{R11}, such that the thermal averages $\mathrm{<} a_{i}^{+} a_{i}\mathrm{>}$ and $\mathrm{<} b_{j}^{+} b_{j}\mathrm{>}$ are very small compared to $2S$. By considering only the first term of the expansion in power series of the square root part of Eqs.~(\eqref{2} and  \eqref{3}) in terms of $a_{i}^{+} a_{i}/2S$ and $b_{j}^{+} b_{j}/2S$,
\begin{equation*}
\begin{split}
&\left( 1-\frac{1}{2S} a_{i}^{+} a_{i} \right) ^{\frac{1}{2} } =1-\frac{1}{4S} a_{i}^{+} a_{i} +\hdots,
\\&
\left( 1-\frac{1}{2S} b_{j}^{+} b_{j} \right) ^{\frac{1}{2} } =1-\frac{1}{4S} b_{j}^{+} b_{j} +\hdots,
\end{split}
\end{equation*}
the spin ladder operators can be approximated as $S_{a,i}^{+} \approx \sqrt{2S} a_{i}$ and $S_{b,j}^{+} \approx \sqrt{2S} b_{j}^{+}$. Note that in this approximation the products containing more than three bosonic operators are neglected. Thus, we neglect the product of four bosonic operators and denote by $M$  the numbers of nearest neighbor, the Hamiltonian in Eq.~\eqref{1} is rewritten in spin-wave approximation \cite{R11} as, 
\begin{equation} \label{4} 
\begin{split}
&H=H_{0} +\left[g\mu _{B} (B_{+} +B_{A} )+2SD'+J\Delta MS\right]\sum _{i}^{}a_{2i}^{+} a_{2i} 
+\left[-g\mu _{B} (B_{+} -B_{A} )+2SD'+J\Delta MS\right]\sum _{j}^{}b_{2j}^{+} b_{2j} 
\\&
+\left[g\mu _{B} \left(B_{-} +B_{A} \right)+2SD'+J\Delta MS\right]\sum _{i}^{}a_{2i+1}^{+} a_{2i+1} 
+\left[-g\mu _{B} \left(B_{-} -B_{A} \right)+2SD'
+J\Delta MS\right]\sum _{j}^{}b_{2j+1}^{+} b_{2j+1} 
\\&
+J\Delta S\left(\sum _{i,\delta }^{}a_{2i+\delta }^{+} a_{2i+\delta }  +\sum _{j,\delta }^{}b_{2j+\delta }^{+} b_{2j+\delta }  \right)
+J\Delta S\left(\sum _{i,\delta }^{}a_{2i+1+\delta }^{+} a_{2i+1+\delta }  +\sum _{j,\delta }^{}b_{2j+1+\delta }^{+} b_{2j+1+\delta }  \right)
\\&
+SJ\left[\sum _{i,\delta }^{}\left(a_{2i} b_{2i+\delta } +a_{2i}^{+} b_{2i+\delta }^{+} \right) +\sum _{j,\delta }^{}\left(b_{2j}^{+} a_{2j+\delta }^{+} +b_{2j} a_{2j+\delta } \right) \right]  
\\&
+SJ\left[\sum _{i,\delta }^{}\left(a_{2i+1} b_{2i+1+\delta } +a_{2i+1}^{+} b_{2i+1+\delta }^{+} \right) +\sum _{j,\delta }^{}\left(b_{2j+1}^{+} a_{2j+1+\delta }^{+} +b_{2j+1} a_{2j+1+\delta } \right) \right]
\\&
-iSE_{+} \left[\sum _{i,\delta }^{}\left(a_{2i}^{+} b_{2i+\delta }^{+} -a_{2i} b_{2i+\delta } \right) -\sum _{j,\delta }^{}\left(b_{2j} a_{2j+\delta } -b_{2j}^{+} a_{2j+\delta }^{+} \right) \right]
\\&
-iSE_{-} \left[\sum _{i,\delta }^{}\left(a_{2i+1}^{+} b_{2i+1+\delta }^{+} -a_{2i+1} b_{2i+1+\delta } \right) -\sum _{j,\delta }^{}\left(b_{2j+1} a_{2j+1+\delta } -b_{2j+1}^{+} a_{2j+1+\delta }^{+} \right) \right],
\end{split}
\end{equation} 
where $H_{0} =-2g\mu _{B} SNB_{A} -2N(MJ\Delta +D')S^{2} $, $B_{\pm } =B_{0} (1\pm \lambda )$ and $E_{\pm } =E_{0} (1\pm \tau )$ are the site-dependent magnetic field and electric field respectively, in which $B_{+} (E_{+} )$ and $B_{-} (E_{-} )$ denote the site-dependent magnetic field (electric field) for the even sites $2i$ or $2j$  and for the odd sites $2i+1$ or $2j+1$ respectively. It is worth mentioning that the site-dependent field can be modeled by using the geometric configuration. The staggered DM interaction have been introduced in the XY model through the zigzag geometry see Ref.\cite{R12}.   

Using the inverse of the Fourier transformation \cite{43} of the bosonic operators for even and odd atom sites of each sublattice given by
\begin{equation}
\begin{split}
&a_{l:even} =\sqrt{\frac{2}{N} } \sum _{k}e^{-ik.l} a_{1k},\,\,\,a_{l:odd} =\sqrt{\frac{2}{N} } \sum _{k}e^{-ik.l} a_{2k}
\\&
b_{l:even} =\sqrt{\frac{2}{N} } \sum _{k}e^{ik.l} b_{1k},\,\,\,b_{l:odd} =\sqrt{\frac{2}{N} } \sum _{k}e^{ik.l} b_{2k}
\end{split}
\end{equation} 
where the wave vector $k$ in each expression takes $N/2$ values and belong to the reduced Brillouin zone $BZ'=\left[-\frac{\pi }{2} ;\frac{\pi }{2} \right]$,  the Hamiltonian given in Eq.~\eqref{4} can be rewritten in momentum space and under the complex quadratic form as
\begin{equation} \label{5} 
\begin{split}
&H=H_{0} +\sum _{k}\varepsilon \left(a_{1 k}^{+} a_{1 k} +b_{1 k}^{+} b_{1 k} \right) +\sum _{k}\left(\lambda _{1 k}^{*} a_{1 k} b_{1 k} +\lambda _{1 k} a_{1 k}^{+} b_{1 k}^{+} \right) +g\mu _{B} B_{+} \sum _{k}\left(a_{1 k}^{+} a_{1 k} -b_{1 k}^{+} b_{1 k} \right)
\\&
+\sum _{k}\varepsilon \left(a_{2 k}^{+} a_{2 k} +b_{2 k}^{+} b_{2 k} \right) +\sum _{k}\left(\lambda _{2 k}^{*} a_{2 k} b_{2 k} +\lambda _{2 k} a_{2 k}^{+} b_{2 k}^{+} \right) +g\mu _{B} B_{-} \sum _{k}\left(a_{2 k}^{+} a_{2 k} -b_{2 k}^{+} b_{2 k} \right) 
\end{split}
\end{equation} 
where $\gamma _{k} =\frac{1}{M} \sum _{k}e^{ik\delta } $ is the Fourier transform coupling constant, $\varepsilon =g\mu _{B} B_{A} +2SD'+2\Delta JMS$, $\lambda _{1 k} =\gamma _{k} \left(2JMS-2E_{+} MSi\right)$, $\lambda _{2 k} =\gamma _{k} \left(2JMS-2E_{-} MSi\right)$ ,$\lambda _{1 k}^{*} $ and $\lambda _{2 k}^{*} $ are the complex conjugates of $\lambda _{1 k}$ and $\lambda _{2 k}$ respectively. 

\par From the complex quadratic form, the matrix form of the Hamiltonian is obtained as,
\begin{equation} \label{6} 
H=H_{0} +\sum _{k}\left(\frac{1}{2} X^{+} TX-2\varepsilon \right) -g\mu _{B} \sum _{k}\frac{1}{2} X^{+} T' X 
\end{equation} 

where
\[X=\left(X^{+} \right)^{+} =\left(\begin{array}{c} {\begin{array}{c} {a_{1 k} } \\ {b_{1 k}^{+} } \end{array}} \\ {\begin{array}{c} {b_{1 k} } \\ {a_{1 k}^{+} } \end{array}} \\ {\begin{array}{c} {a_{2 k} } \\ {b_{2 k}^{+} } \end{array}} \\ {\begin{array}{c} {b_{2 k} } \\ {a_{2 k}^{+} } \end{array}} \end{array}\right),\,\,\,T=\left(\begin{array}{cccc} {\begin{array}{cc} {\varepsilon } & {\lambda _{1 k} } \\ {\lambda _{1 k}^{*} } & {\varepsilon } \end{array}} & {\begin{array}{cc} {0} & {0} \\ {0} & {0} \end{array}} & {\begin{array}{cc} {0} & {0} \\ {0} & {0} \end{array}} & {\begin{array}{cc} {0} & {0} \\ {0} & {0} \end{array}} \\ {\begin{array}{cc} {0} & {0} \\ {0} & {0} \end{array}} & {\begin{array}{cc} {\varepsilon } & {\lambda _{1 k} } \\ {\lambda _{1 k}^{*} } & {\varepsilon } \end{array}} & {\begin{array}{cc} {0} & {0} \\ {0} & {0} \end{array}} & {\begin{array}{cc} {0} & {0} \\ {0} & {0} \end{array}} \\ {\begin{array}{cc} {0} & {0} \\ {0} & {0} \end{array}} & {\begin{array}{cc} {0} & {0} \\ {0} & {0} \end{array}} & {\begin{array}{cc} {\varepsilon } & {\lambda _{2 k} } \\ {\lambda _{2 k}^{*} } & {\varepsilon } \end{array}} & {\begin{array}{cc} {0} & {0} \\ {0} & {0} \end{array}} \\ {\begin{array}{cc} {0} & {0} \\ {0} & {0} \end{array}} & {\begin{array}{cc} {0} & {0} \\ {0} & {0} \end{array}} & {\begin{array}{cc} {0} & {0} \\ {0} & {0} \end{array}} & {\begin{array}{cc} {\varepsilon } & {\lambda _{2 k} } \\ {\lambda _{2 k}^{*} } & {\varepsilon } \end{array}} \end{array}\right), \] 

and
\[T'=\left(\begin{array}{cccc} {\begin{array}{cc} {-B_{+} } & {0} \\ {0} & {B_{+} } \end{array}} & {\begin{array}{cc} {0} & {0} \\ {0} & {0} \end{array}} & {\begin{array}{cc} {0} & {0} \\ {0} & {0} \end{array}} & {\begin{array}{cc} {0} & {0} \\ {0} & {0} \end{array}} \\ {\begin{array}{cc} {0} & {0} \\ {0} & {0} \end{array}} & {\begin{array}{cc} {B_{+} } & {0} \\ {0} & {-B_{+} } \end{array}} & {\begin{array}{cc} {0} & {0} \\ {0} & {0} \end{array}} & {\begin{array}{cc} {0} & {0} \\ {0} & {0} \end{array}} \\ {\begin{array}{cc} {0} & {0} \\ {0} & {0} \end{array}} & {\begin{array}{cc} {0} & {0} \\ {0} & {0} \end{array}} & {\begin{array}{cc} {-B_{-} } & {0} \\ {0} & {B_{-} } \end{array}} & {\begin{array}{cc} {0} & {0} \\ {0} & {0} \end{array}} \\ {\begin{array}{cc} {0} & {0} \\ {0} & {0} \end{array}} & {\begin{array}{cc} {0} & {0} \\ {0} & {0} \end{array}} & {\begin{array}{cc} {0} & {0} \\ {0} & {0} \end{array}} & {\begin{array}{cc} {B_{-} } & {0} \\ {0} & {-B_{-} } \end{array}} \end{array}\right).\] 

Finally, the Hamiltonian in Eq.~\eqref{6} can be diagonalized by following the

Bogoliubov transformations \cite{44} given by,
\begin{equation} \label{7} 
\left\{\begin{split}
&a_{m k} =u_{m k} \alpha _{m k} +v_{m k} \beta _{m k}^{+} 
\\&
b_{m k} =v_{m k} \beta _{m k} +u_{m k} \alpha _{m k}^{+}
\end{split}
\right.  
\end{equation} 

where the coefficients $u_{m k} $  and $v_{m k} $ are complex numbers and throughout this work $m=\left\{1;2\right\}$ are used to differ the parameters denoting the even atom sites from those denoting the odd atoms sites. It is important to note that the new operators $\alpha _{m k} $ and  $\beta _{m k} $ obey the boson commutation rules, thus, leading to the following constraint $\left|u_{m k} \right|^{2} -\left|v_{m k} \right|^{2} =1$. Therefore, $u_{m k} $  and$v_{m k} $ can be parameterized as $u_{m k} =\cosh \theta _{m k} $ and $v_{m k} =e^{i\varphi _{m} } \sinh \theta _{m k} $ with $\varphi _{m} =\varphi ^{v_{m k} } -\varphi ^{u_{m k} } $. From the above transformations, the vectors $X$ and $X^{+} $ are given by $X=\vartheta \cdot \phi $ and $X^{+} =\phi ^{+} \cdot \vartheta ^{+} $ where $\phi =\left(\phi ^{+} \right)^{+} =\left(\begin{array}{c} {\begin{array}{c} {\alpha _{1 k} } \\ {\beta _{1 k}^{+} } \end{array}} \\ {\begin{array}{c} {\beta _{1 k} } \\ {\alpha _{1 k}^{+} } \end{array}} \\ {\begin{array}{c} {\alpha _{2 k} } \\ {\beta _{2 k}^{+} } \end{array}} \\ {\begin{array}{c} {\beta _{2 k} } \\ {\alpha _{2 k}^{+} } \end{array}} \end{array}\right)$
and 
\[\vartheta =\left(\vartheta ^{+} \right)^{+} =\left(\begin{array}{cccc} {\begin{array}{cc} {u_{1 k} } & {v_{1 k} } \\ {v_{1 k}^{*} } & {u_{1 k}^{*} } \end{array}} & {\begin{array}{cc} {0} & {0} \\ {0} & {0} \end{array}} & {\begin{array}{cc} {0} & {0} \\ {0} & {0} \end{array}} & {\begin{array}{cc} {0} & {0} \\ {0} & {0} \end{array}} \\ {\begin{array}{cc} {0} & {0} \\ {0} & {0} \end{array}} & {\begin{array}{cc} {u_{1 k} } & {v_{1 k} } \\ {v_{1 k}^{*} } & {u_{1 k}^{*} } \end{array}} & {\begin{array}{cc} {0} & {0} \\ {0} & {0} \end{array}} & {\begin{array}{cc} {0} & {0} \\ {0} & {0} \end{array}} \\ {\begin{array}{cc} {0} & {0} \\ {0} & {0} \end{array}} & {\begin{array}{cc} {0} & {0} \\ {0} & {0} \end{array}} & {\begin{array}{cc} {u_{2 k} } & {v_{2 k} } \\ {v_{2 k}^{*} } & {u_{2 k}^{*} } \end{array}} & {\begin{array}{cc} {0} & {0} \\ {0} & {0} \end{array}} \\ {\begin{array}{cc} {0} & {0} \\ {0} & {0} \end{array}} & {\begin{array}{cc} {0} & {0} \\ {0} & {0} \end{array}} & {\begin{array}{cc} {0} & {0} \\ {0} & {0} \end{array}} & {\begin{array}{cc} {u_{2 k} } & {v_{2 k} } \\ {v_{2 k}^{*} } & {u_{2 k}^{*} } \end{array}} \end{array}\right).\] 

Substituting these vectors in Eq.~\eqref{6}  and once the calculations are performed, the resulting Hamiltonian is written as 
\begin{equation} \label{8} 
H=H_{0} +\sum _{k}\left(\frac{1}{2} \phi ^{+} D\phi -2\varepsilon \right) ,                      
\end{equation} 

Where D is a diagonal matrix
\[D=\left(\begin{array}{cccc} {\begin{array}{cc} {\omega _{1 k}^{\left(+\right)} } & {0} \\ {0} & {\omega _{1 k}^{\left(-\right)} } \end{array}} & {\begin{array}{cc} {0} & {0} \\ {0} & {0} \end{array}} & {\begin{array}{cc} {0} & {0} \\ {0} & {0} \end{array}} & {\begin{array}{cc} {0} & {0} \\ {0} & {0} \end{array}} \\ {\begin{array}{cc} {0} & {0} \\ {0} & {0} \end{array}} & {\begin{array}{cc} {\omega _{1 k}^{\left(-\right)} } & {0} \\ {0} & {\omega _{1 k}^{\left(+\right)} } \end{array}} & {\begin{array}{cc} {0} & {0} \\ {0} & {0} \end{array}} & {\begin{array}{cc} {0} & {0} \\ {0} & {0} \end{array}} \\ {\begin{array}{cc} {0} & {0} \\ {0} & {0} \end{array}} & {\begin{array}{cc} {0} & {0} \\ {0} & {0} \end{array}} & {\begin{array}{cc} {\omega _{2 k}^{\left(+\right)} } & {0} \\ {0} & {\omega _{2 k}^{\left(-\right)} } \end{array}} & {\begin{array}{cc} {0} & {0} \\ {0} & {0} \end{array}} \\ {\begin{array}{cc} {0} & {0} \\ {0} & {0} \end{array}} & {\begin{array}{cc} {0} & {0} \\ {0} & {0} \end{array}} & {\begin{array}{cc} {0} & {0} \\ {0} & {0} \end{array}} & {\begin{array}{cc} {\omega _{2 k}^{\left(-\right)} } & {0} \\ {0} & {\omega _{2 k}^{\left(+\right)} } \end{array}} \end{array}\right).\] 
Thus, we obtain the so-called diagonalized form for the Hamiltonian as
\begin{equation} \label{9} 
\begin{split}
&
H=H'_{0} +\sum _{k}\left[\omega _{1 k}^{\left(+\right)} \left(\alpha _{1 k}^{+} \alpha _{1 k} +\frac{1}{2} \right)+\omega _{1 k}^{\left(-\right)} \left(\beta _{1 k}^{+} \beta _{1 k} +\frac{1}{2} \right)\right]
\\&
+\sum _{k}\left[\omega _{2 k}^{\left(+\right)} \left(\alpha _{2 k}^{+} \alpha _{2 k} +\frac{1}{2} \right)+\omega _{2 k}^{\left(-\right)} \left(\beta _{2 k}^{+} \beta _{2 k} +\frac{1}{2} \right)\right] 
\end{split}
\end{equation} 
where $H'_{0} =H_{0} -2\varepsilon N$. From the above Hamiltonian, each frequency of the magnon at the symmetric positions obtained in the case of uniform fields \cite{25} is divided into two pairs of frequencies in this case with site-dependent magnetic and electric field, which can be easily derived as
\begin{equation} \label{10} 
\omega _{1 k}^{\left(\pm \right)} =\tilde{\varepsilon }_{1} \pm g\mu _{B} B_{+} ,  \,\,\, \omega _{2 k}^{\left(\pm \right)} =\tilde{\varepsilon }_{2} \pm g\mu _{B} B_{-}  
\end{equation} 

with $\tilde{\varepsilon }_{m} =\sqrt{\varepsilon ^{2} -\left|\lambda _{m} \right|^{2} } $. It is important to remark that the equilibrium state of the system is ensured by the following constraint $\varepsilon >\left|\lambda _{m} \right|$. In contrast to the case of uniform magnetic field where only one equilibrium critical point has been observed \cite{25}, here there are two equilibrium critical points corresponding to the two pairs of frequencies and which can be obtained by taking $\omega _{m k}^{\left(-\right)} =0$ and $k=0$,
\begin{equation} \label{11} 
B_{c 1} =\frac{2JMS}{g\mu _{B} } \sqrt{\left(\frac{D'}{JM} +\Delta +\frac{g\mu _{B} B_{A} }{2JMS} \right)^{2} -\left(\frac{E_{+}^{2} }{J^{2} } +1\right)} ,
\end{equation}
\begin{equation}\label{12}
 B_{c 2} =\frac{2JMS}{g\mu _{B} } \sqrt{\left(\frac{D'}{JM} +\Delta +\frac{g\mu _{B} B_{A} }{2JMS} \right)^{2} -\left(\frac{E_{-}^{2} }{J^{2} } +1\right)} .        
\end{equation} 

 Once the Hamiltonian of the system is diagonalized, the statistical sum is found as
\begin{equation} \label{13} 
Z=\Tr\left(e^{-\beta H} \right) 
\end{equation} 
Where $\beta =1/k_{B} T$ with $k_{B} $ as the Boltzmann constant, $T$ as the absolute temperature and $H$ the diagonalized Hamiltonian in Eq.~\eqref{9}.
The Helmholtz free energy of the system is related to the statistical sum by the following relation:
\begin{equation} \label{14} 
F=-T\ln Z.                                                              
\end{equation} 
Thus, after calculation, the free energy is given by,
\begin{equation} \label{15} 
\begin{split}
&
F=H'_{0} +\frac{1}{2} \sum _{k}\left(\omega _{1 k}^{\left(+\right)} +\omega _{2 k}^{\left(+\right)} \right) +T\sum _{k}\ln \left(1-e^{-\frac{1}{T} \omega _{1 k}^{\left(+\right)} } \right) +T\sum _{k}\ln \left(1-e^{-\frac{1}{T} \omega _{2 k}^{\left(+\right)} } \right) 
\\&
+\frac{1}{2} \sum _{k}\left(\omega _{1 k}^{\left(-\right)} +\omega _{2 k}^{\left(-\right)} \right) +T\sum _{k}\ln \left(1-e^{-\frac{1}{T} \omega _{1 k}^{\left(-\right)} } \right) +T\sum _{k}\ln \left(1-e^{-\frac{1}{T} \omega _{2 k}^{\left(-\right)} } \right)
\end{split}
\end{equation} 
where the frequencies obtained in Eq.~\eqref{10} are rewritten in terms of the system parameters as follow, 
\[\omega _{1 k}^{\left(\pm \right)} =2JMS\sqrt{\left(\frac{D'}{JM} +\Delta +\frac{g\mu _{B} B_{A} }{2JMS} \right)^{2} +\left(\frac{E_{+}^{2} }{J^{2} } +1\right)\left(\frac{2k^{2} l^{2} }{M} -1\right)} \pm g\mu _{B} B_{+} ,\] 
\[\omega _{2 k}^{\left(\pm \right)} =2JMS\sqrt{\left(\frac{D'}{JM} +\Delta +\frac{g\mu _{B} B_{A} }{2JMS} \right)^{2} +\left(\frac{E_{-}^{2} }{J^{2} } +1\right)\left(\frac{2k^{2} l^{2} }{M} -1\right)} \pm g\mu _{B} B_{-} \] 
with  $l$ as the length sides of the cubic primitive cell of each sublattice.

From the free energy, some magnetoelectric properties \cite{45} such as the magnetization, the electric polarization, and the magnetoelectric polarizability are derived \cite{R13},
\begin{equation} \label{16} 
\begin{split}
M&=-\mu _{0} \frac{\partial F}{\partial B}
\\&
=-g\mu _{0} \mu _{B} \left[\frac{N}{2} +\sum _{k}\left(\frac{e^{-\frac{1}{T} \omega _{1 k}^{\left(+\right)} } }{1-e^{-\frac{1}{T} \omega _{1\, k}^{\left(+\right)} } } +\frac{e^{-\frac{1}{T} \omega _{2 k}^{\left(+\right)} } }{1-e^{-\frac{1}{T} \omega _{2 k}^{\left(+\right)} } } \right) \right]
\\&
+g\mu _{0} \mu _{B} \left[\frac{N}{2} +\sum _{k}\left(\frac{e^{-\frac{1}{T} \omega _{1 k}^{\left(-\right)} } }{1-e^{-\frac{1}{T} \omega _{1\, k}^{\left(-\right)} } } +\frac{e^{-\frac{1}{T} \omega _{2 k}^{\left(-\right)} } }{1-e^{-\frac{1}{T} \omega _{2 k}^{\left(-\right)} } } \right) \right]
\end{split}
\end{equation} 
\begin{equation}\label{17}
\begin{split}
P&=-\frac{\partial F}{\partial E}
\\&
=\frac{2MS}{J} \sum _{k}\gamma _{k}^{2} \left(\frac{E_{+} }{\xi _{+}^{1/2} } +\frac{E_{-} }{\xi _{-}^{1/2} } \right) +\frac{2MS}{J} \sum _{k}\gamma _{k}^{2} \left[\frac{E_{+} e^{-\frac{1}{T} \omega _{1 k}^{\left(+\right)} } }{\xi _{+}^{1/2} \left(1-e^{-\frac{1}{T} \omega _{1 k}^{\left(+\right)} } \right)} +\frac{E_{-} e^{-\frac{1}{T} \omega _{2 k}^{\left(+\right)} } }{\xi _{-}^{1/2} \left(1-e^{-\frac{1}{T} \omega _{2 k}^{\left(+\right)} } \right)} \right] 
\\&
+\frac{2MS}{J} \sum _{k}\gamma _{k}^{2} \left[\frac{E_{+} e^{-\frac{1}{T} \omega _{1 k}^{\left(-\right)} } }{\xi _{+}^{1/2} \left(1-e^{-\frac{1}{T} \omega _{1 k}^{\left(-\right)} } \right)} +\frac{E_{-} e^{-\frac{1}{T} \omega _{2 k}^{\left(-\right)} } }{\xi _{-}^{1/2} \left(1-e^{-\frac{1}{T} \omega _{2 k}^{\left(-\right)} } \right)} \right] 
\end{split}
\end{equation}
\begin{equation}\label{18}
\begin{split}
\alpha& =-\mu _{0} \frac{\partial F}{\partial B\partial E}
\\&
=-\frac{2g\mu _{0} \mu _{B} MS}{JT}\left\lbrace 
\begin{split}
&\sum _{k}\left[\frac{E_{+} \gamma _{k}^{2} e^{-\frac{1}{T} \omega _{1 k}^{\left(+\right)} } }{\xi _{+}^{1/2} \left(1-e^{-\frac{1}{T} \omega _{1 k}^{\left(+\right)} } \right)^{2} } +\frac{E_{-} \gamma _{k}^{2} e^{-\frac{1}{T} \omega _{2 k}^{\left(+\right)} } }{\xi _{-}^{1/2} \left(1-e^{-\frac{1}{T} \omega _{2 k}^{\left(+\right)} } \right)^{2} } \right] 
\\&
-\sum _{k}\left[\frac{E_{+} \gamma _{k}^{2} e^{-\frac{1}{T} \omega _{1 k}^{\left(-\right)} } }{\xi _{+}^{1/2} \left(1-e^{-\frac{1}{T} \omega _{1 k}^{\left(-\right)} } \right)^{2} } +\frac{E_{-} \gamma _{k}^{2} e^{-\frac{1}{T} \omega _{2 k}^{\left(-\right)} } }{\xi _{-}^{1/2} \left(1-e^{-\frac{1}{T} \omega _{2 k}^{\left(-\right)} } \right)^{2} } \right]
\end{split}
\right\rbrace 
\end{split}
\end{equation}
respectively, where 
\begin{equation*}
\xi _{+} =\left(\frac{g\mu _{B} B_{A} }{2JMS} +\Delta +\frac{D'}{JM} \right)^{2} +\left(\frac{E_{+}^{2} }{J^{2} } +1\right)\left(\frac{2k^{2} l^{2} }{M} -1\right),
\end{equation*}
\begin{equation*}
\xi _{-} =\left(\frac{g\mu _{B} B_{A} }{2JMS} +\Delta +\frac{D'}{JM} \right)^{2} +\left(\frac{E_{-}^{2} }{J^{2} } +1\right)\left(\frac{2k^{2} l^{2} }{M} -1\right).
\end{equation*}

\section{Entropy and specific heat capacity}\label{S3}
\subsection{Entropy}
The Boltzmann entropy which quantified the rate of the disorder in the system is derived from the free energy as,
\begin{equation} \label{19} 
S=-\frac{\partial F}{\partial T}
\end{equation} 
Thus, after performing calculation the entropy is given by,
\begin{equation} \label{20}
\begin{split}
&S=-\left[\sum _{k}\ln \left(1-e^{-\frac{1}{T} \omega _{1 k}^{\left(+\right)} } \right)\left(1-e^{-\frac{1}{T} \omega _{2 k}^{\left(+\right)} } \right) +\sum _{k}\ln \left(1-e^{-\frac{1}{T} \omega _{1 k}^{\left(-\right)} } \right)\left(1-e^{-\frac{1}{T} \omega _{2 k}^{\left(-\right)} } \right) \right]
\\&
+\frac{1}{T} \left[\sum _{k}\frac{\omega _{1 k}^{\left(+\right)} e^{-\frac{1}{T} \omega _{1 k}^{\left(+\right)} } }{\left(1-e^{-\frac{1}{T} \omega _{1 k}^{\left(+\right)} } \right)}  +\sum _{k}\frac{\omega _{2 k}^{\left(+\right)} e^{-\frac{1}{T} \omega _{2 k}^{\left(+\right)} } }{\left(1-e^{-\frac{1}{T} \omega _{2 k}^{\left(+\right)} } \right)}  \right]+\frac{1}{T} \left[\sum _{k}\frac{\omega _{1 k}^{\left(-\right)} e^{-\frac{1}{T} \omega _{1 k}^{\left(-\right)} } }{\left(1-e^{-\frac{1}{T} \omega _{1 k}^{\left(-\right)} } \right)}  +\sum _{k}\frac{\omega _{2 k}^{\left(-\right)} e^{-\frac{1}{T} \omega _{2 k}^{\left(-\right)} } }{\left(1-e^{-\frac{1}{T} \omega _{2 k}^{\left(-\right)} } \right)}\right]
\end{split}
\end{equation} 
First of all, we note that the numerical results presented in the current section and  next section are obtained using the following parameters: $B_{A} =4.14 T , D'=5.6 , \Delta =0.5 , J=1$ and $M=6$.
\par
\begin{figure}[!htb]
\centerline{
\subfigure[]{\includegraphics[width=0.35\textwidth]{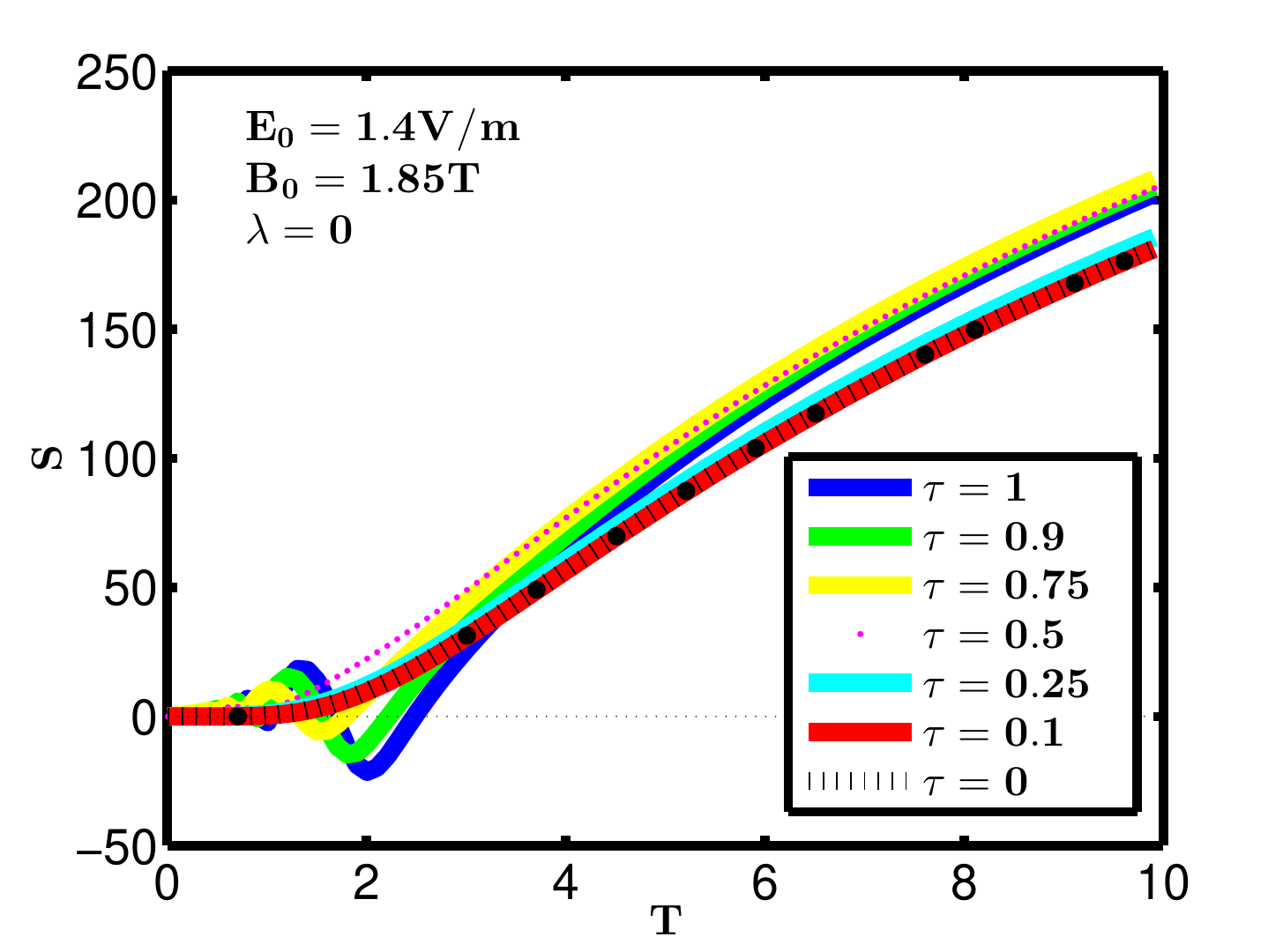}}
\subfigure[]{\includegraphics[width=0.35\textwidth]{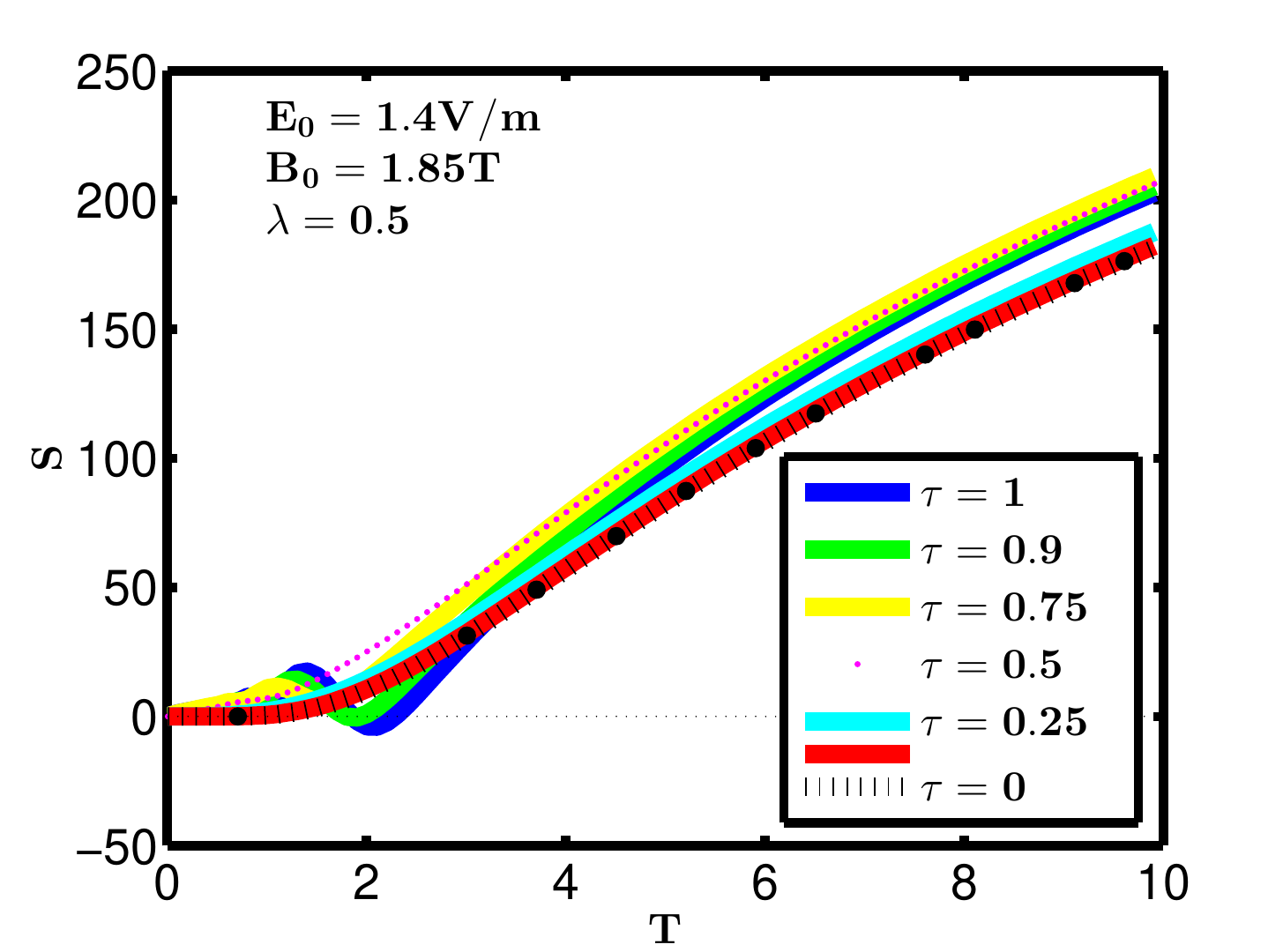}}
\subfigure[]{\includegraphics[width=0.35\textwidth]{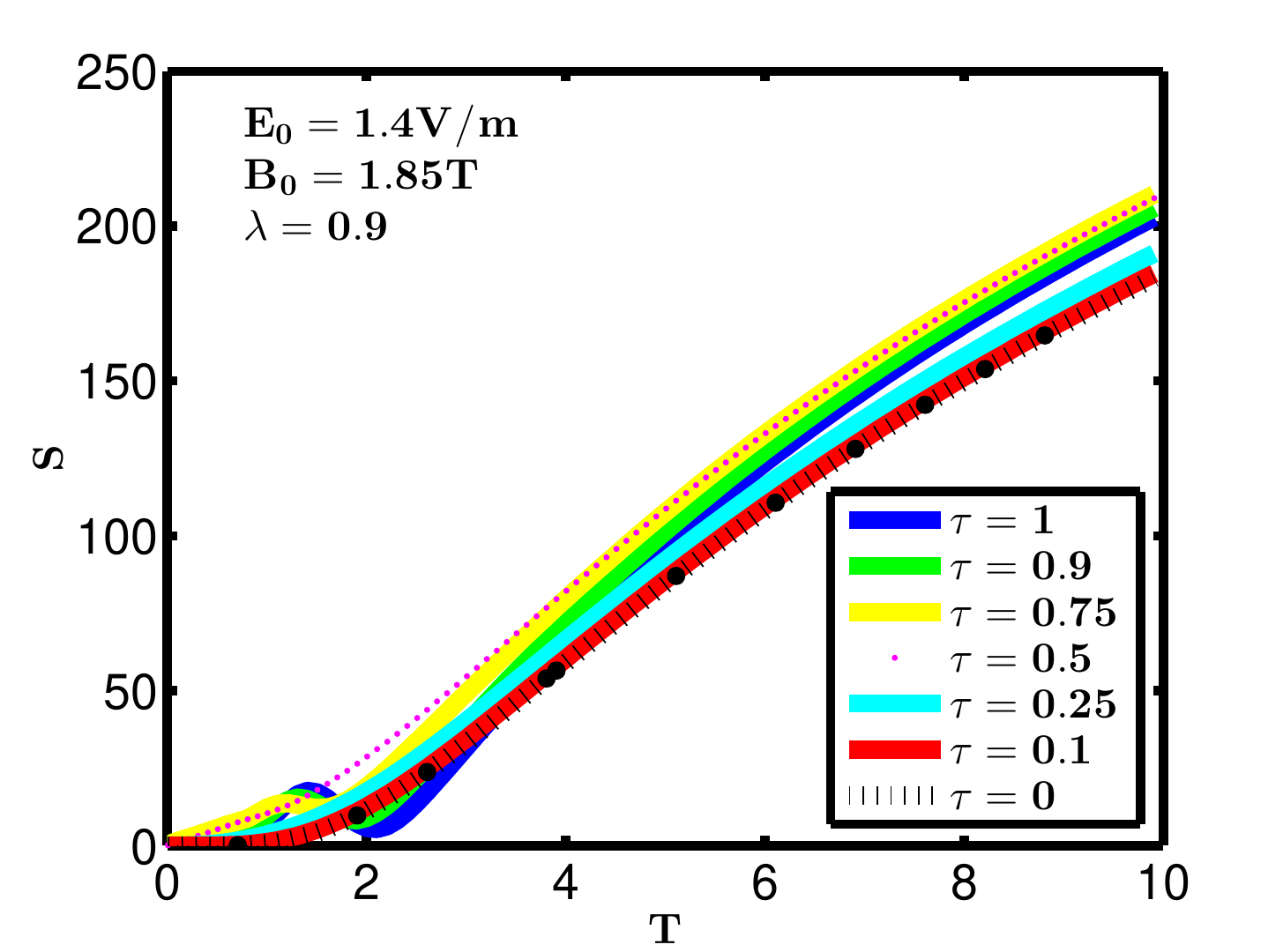}}  
}
\centerline{
\subfigure[]{\includegraphics[width=0.35\textwidth]{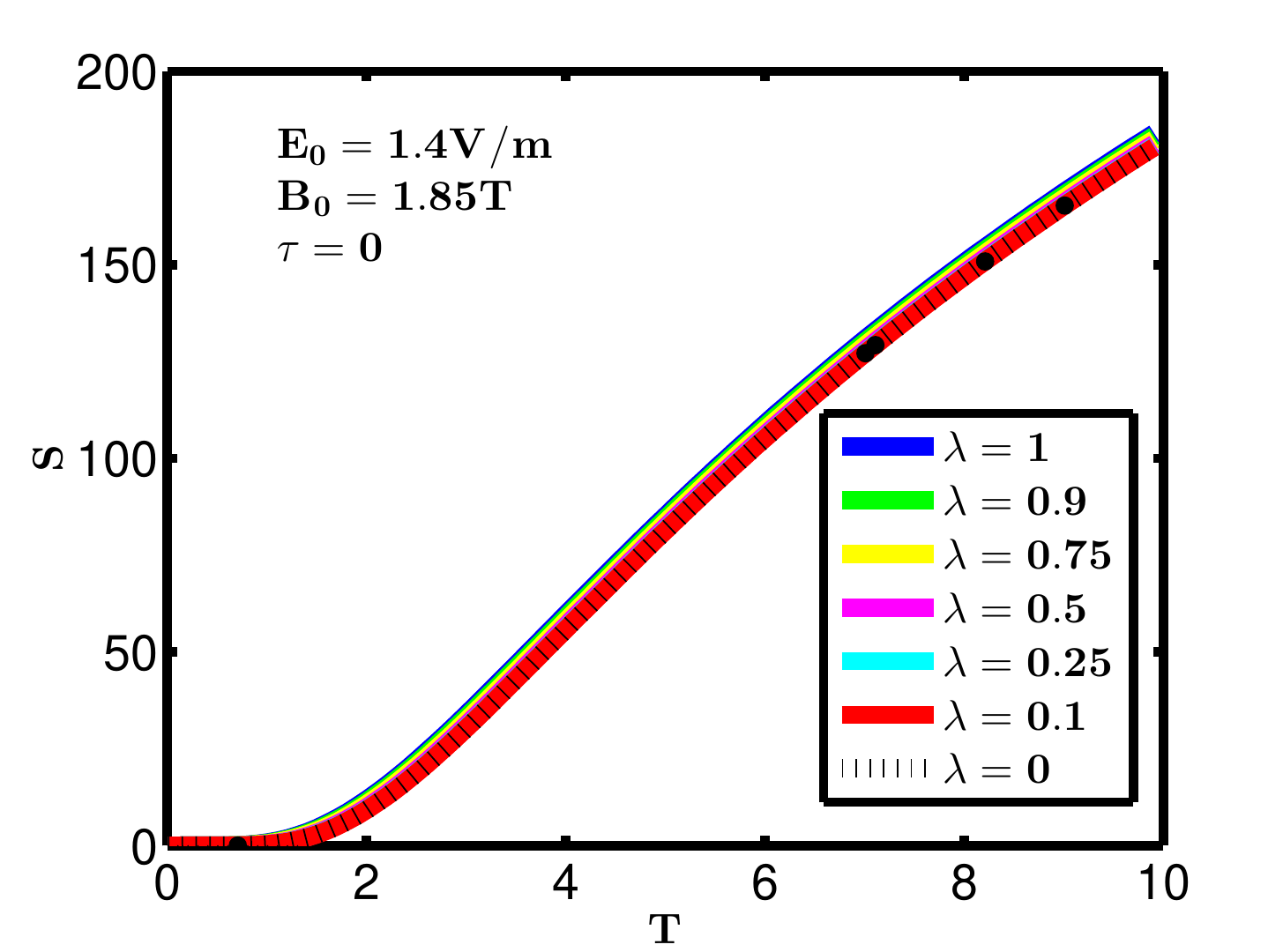}}
\subfigure[]{\includegraphics[width=0.35\textwidth]{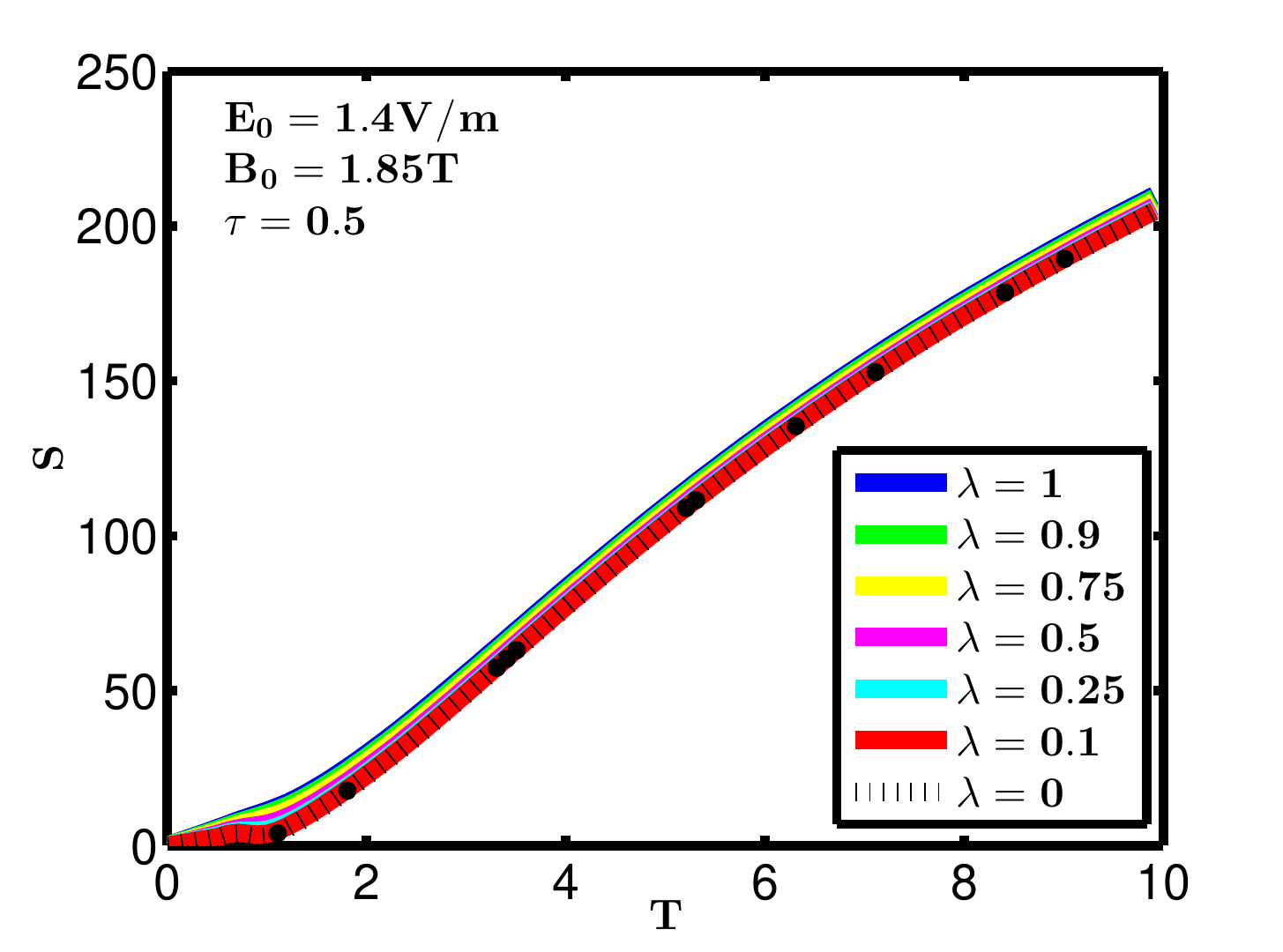}}
\subfigure[]{\includegraphics[width=0.35\textwidth]{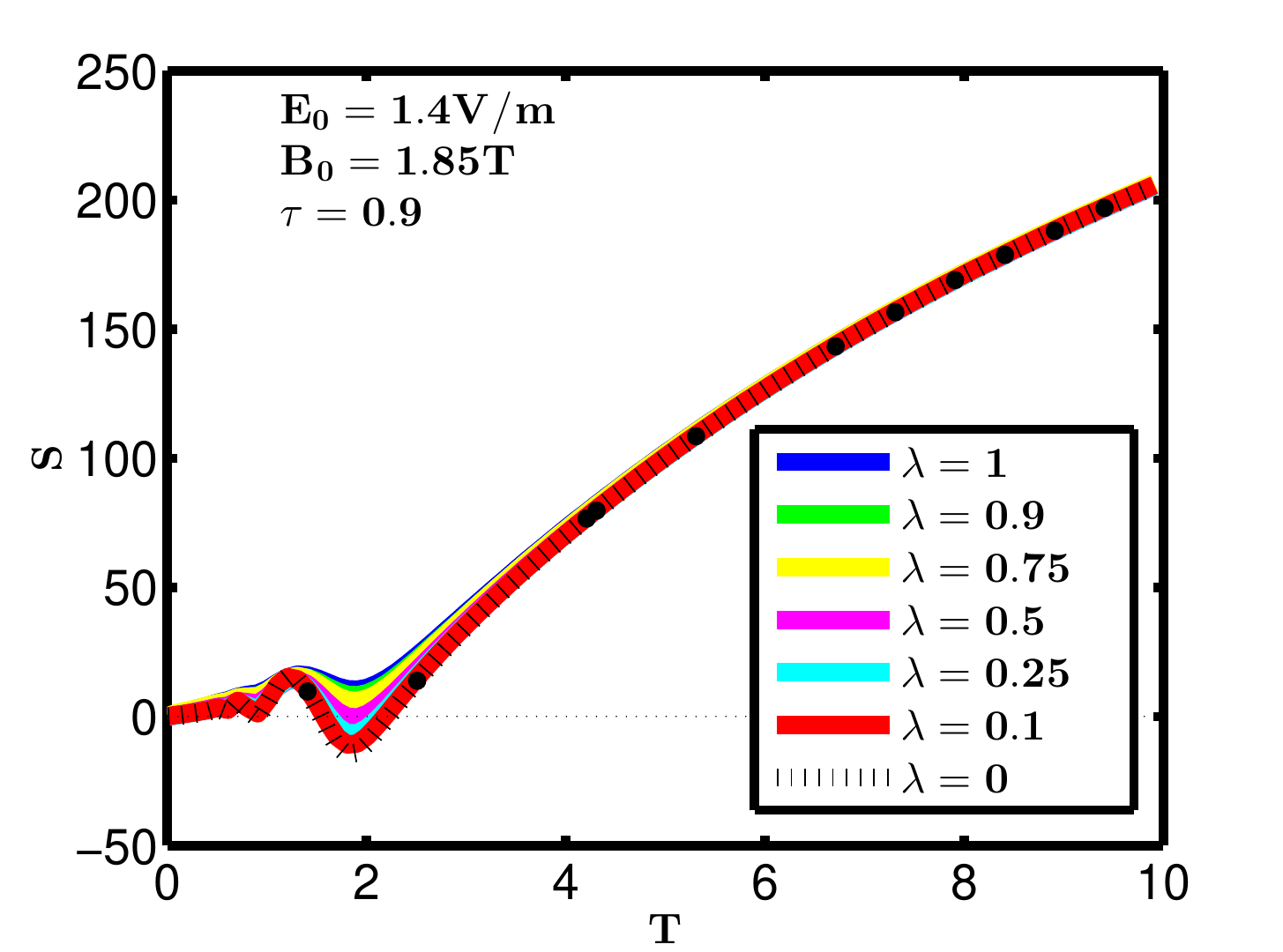}}
}
\caption{the upper panels show the evolution of the entropy of the system for different values of the electric site-dependent parameters and for three values of the magnetic site-dependent parameter namely 0 (a), 0.5 (b) and 0.9 (c). In the lower panels we plotted the evolution of the entropy for different values of the magnetic site-dependent parameters and for three values of the electric site-dependent parameter namely 0 (d), 0.5 (e) and 0.9 (f). }
\label{F21}
\end{figure}

 
The temperature dependence of the entropy for different values of the electric (upper panels) and magnetic (lower panels) site-dependent parameters is depicted in Fig.~\eqref{F21}. By varying the electric site-dependent parameter, we observe that when this parameter is greater than $0.5$ the curves exhibit the anomalous oscillating-like behavior with a negative part at low temperature ($T\mathrm{<} 2.5 K$) but the amplitude of the negative part decreases with the increasing of the magnetic site-dependent parameter and completely disappear when $\lambda =0.9$. That negative part of the entropy can be interpreting as evidence of the cooling of the system (inverse magneto/electrocaloric effect).  In addition, by varying the magnetic site-dependent parameter, it is observed that the site-dependent magnetic field affects the entropy at the oscillating-like part. Such an entropy behavior clearly demonstrates the violation of the second law of thermodynamic at low temperatures. 

\par
\begin{figure}[!htb]
\centerline{
\subfigure[]{\includegraphics[width=0.35\textwidth]{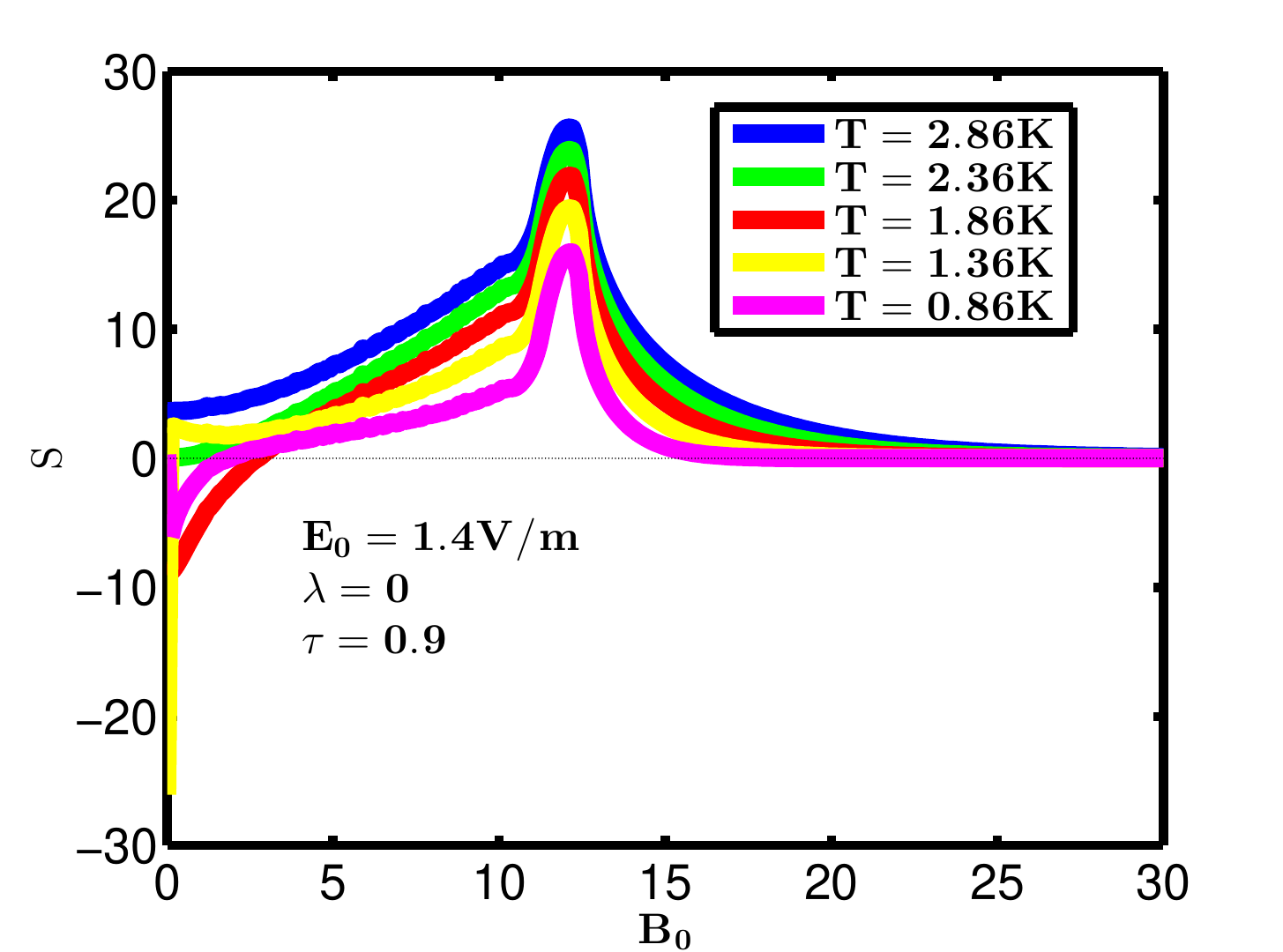}}
\subfigure[]{\includegraphics[width=0.35\textwidth]{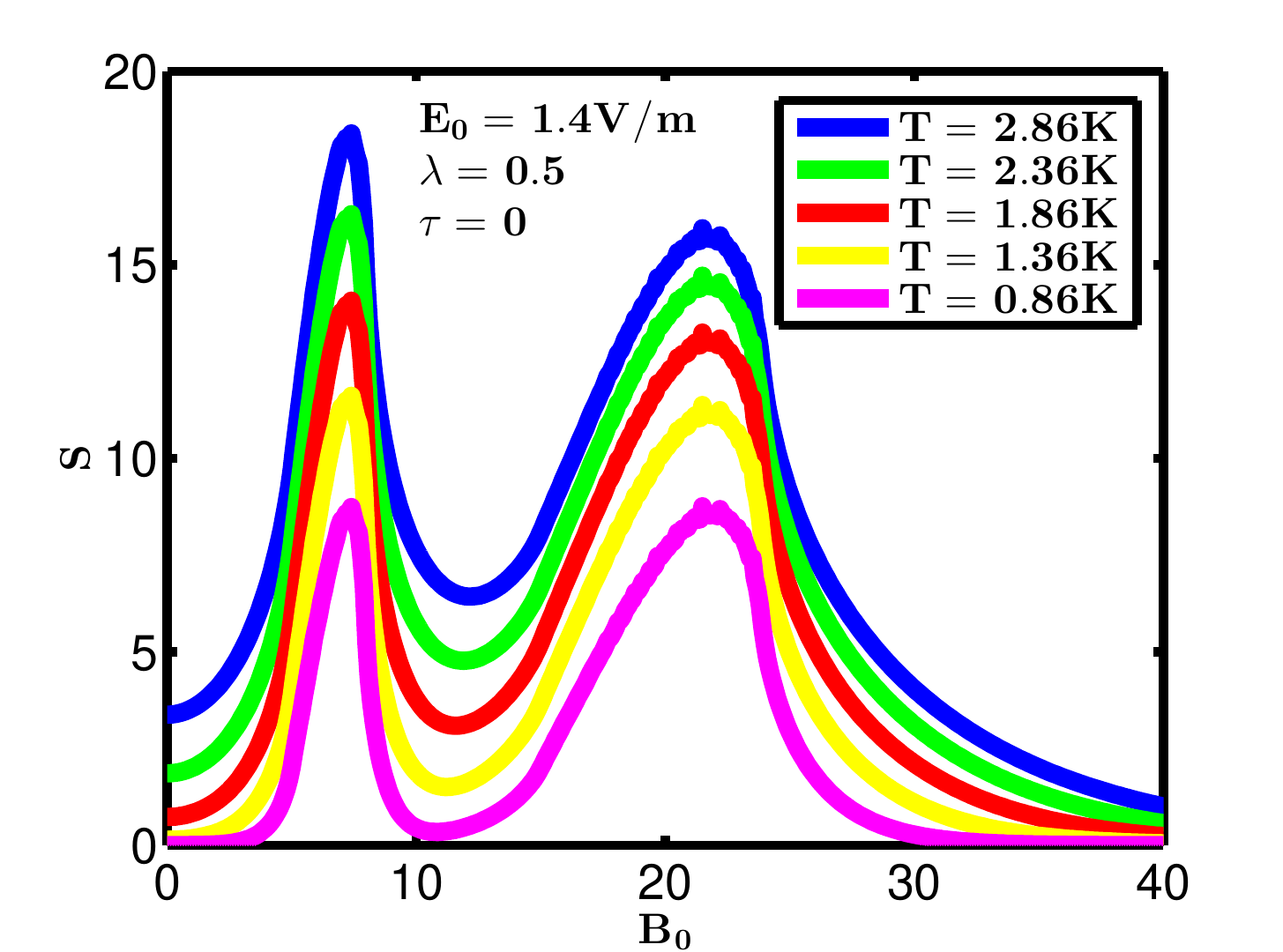}}
}
\centerline{
\subfigure[]{\includegraphics[width=0.35\textwidth]{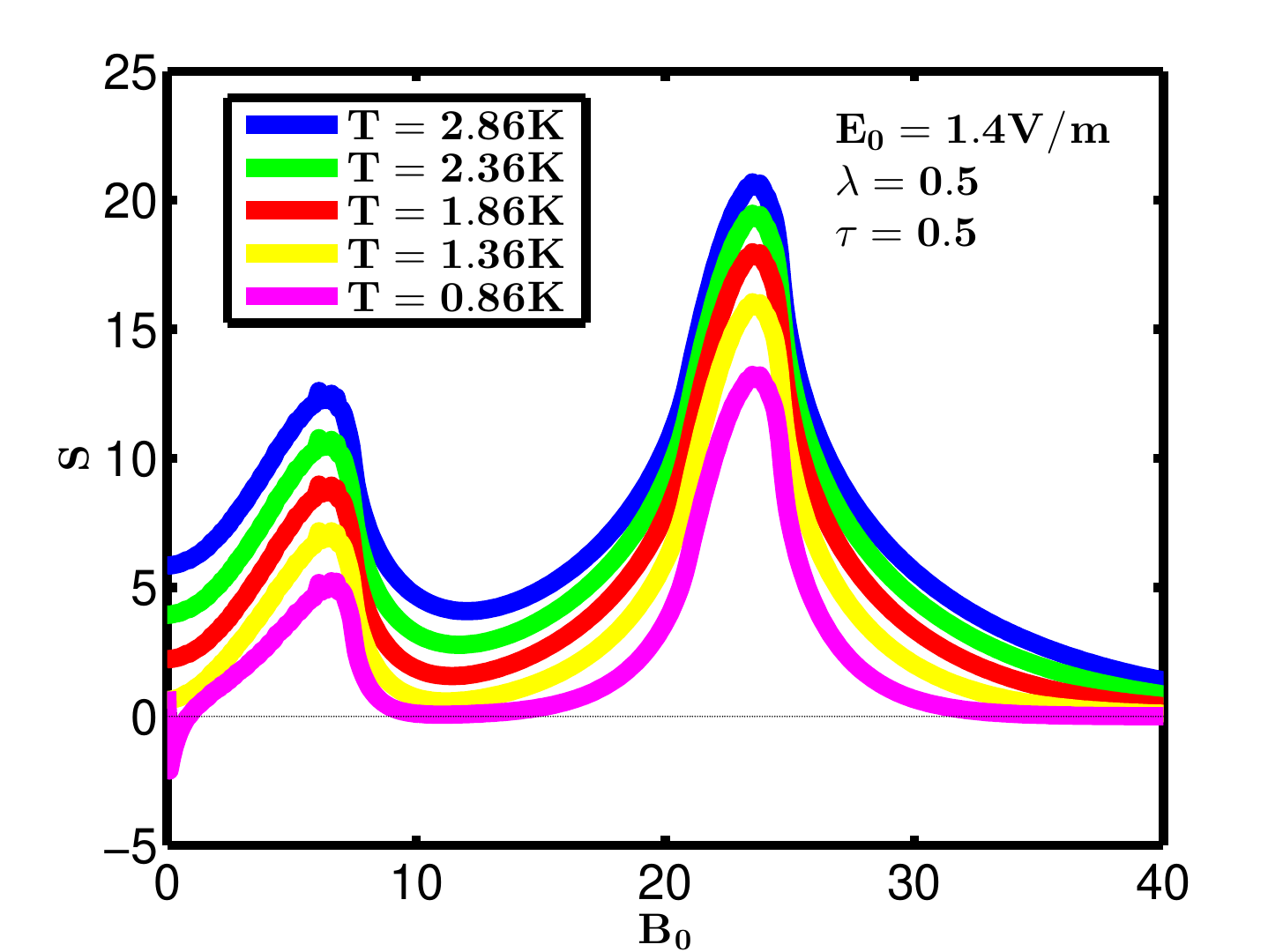}}
\subfigure[]{\includegraphics[width=0.35\textwidth]{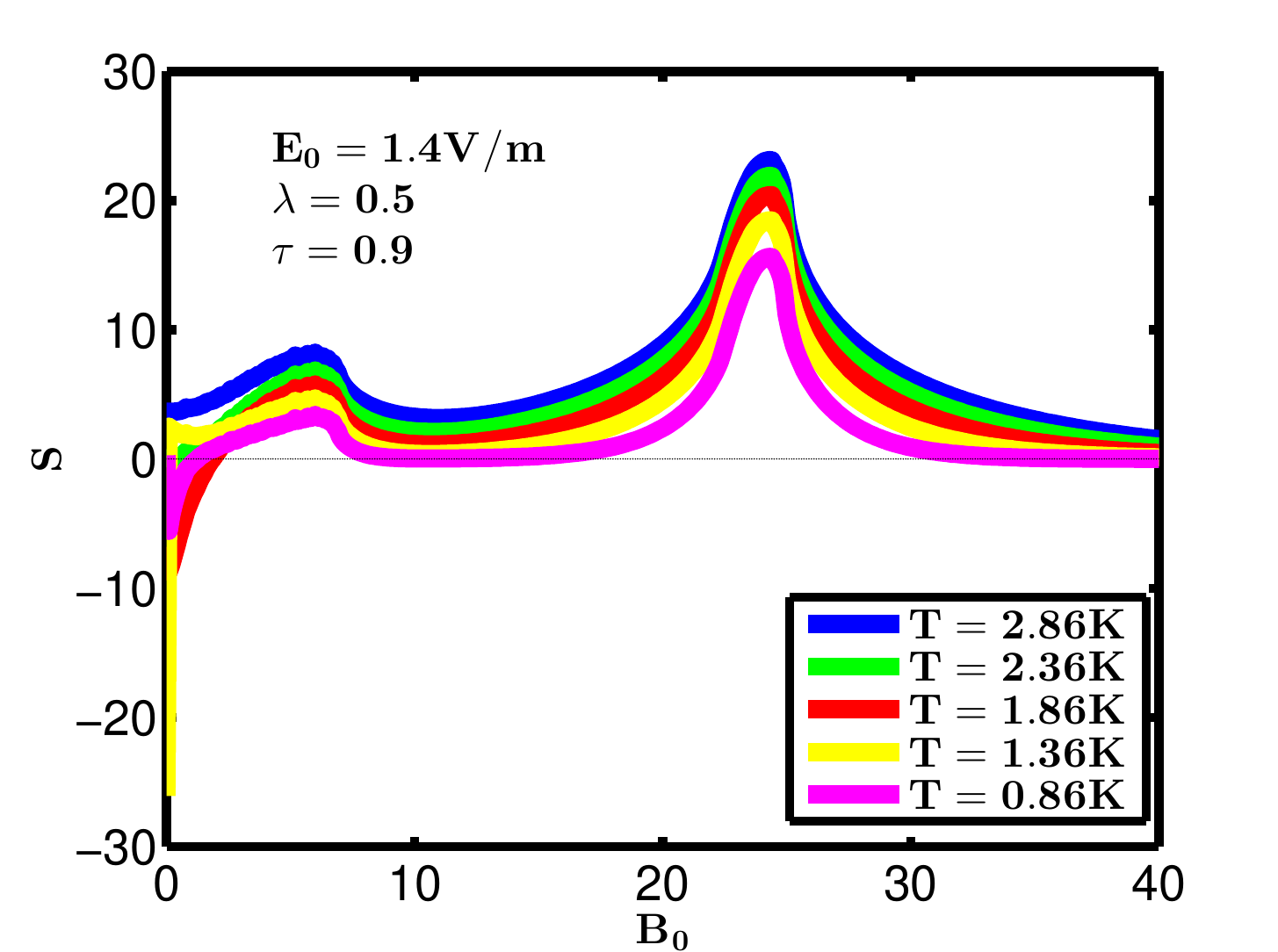}}
\subfigure[]{\includegraphics[width=0.35\textwidth]{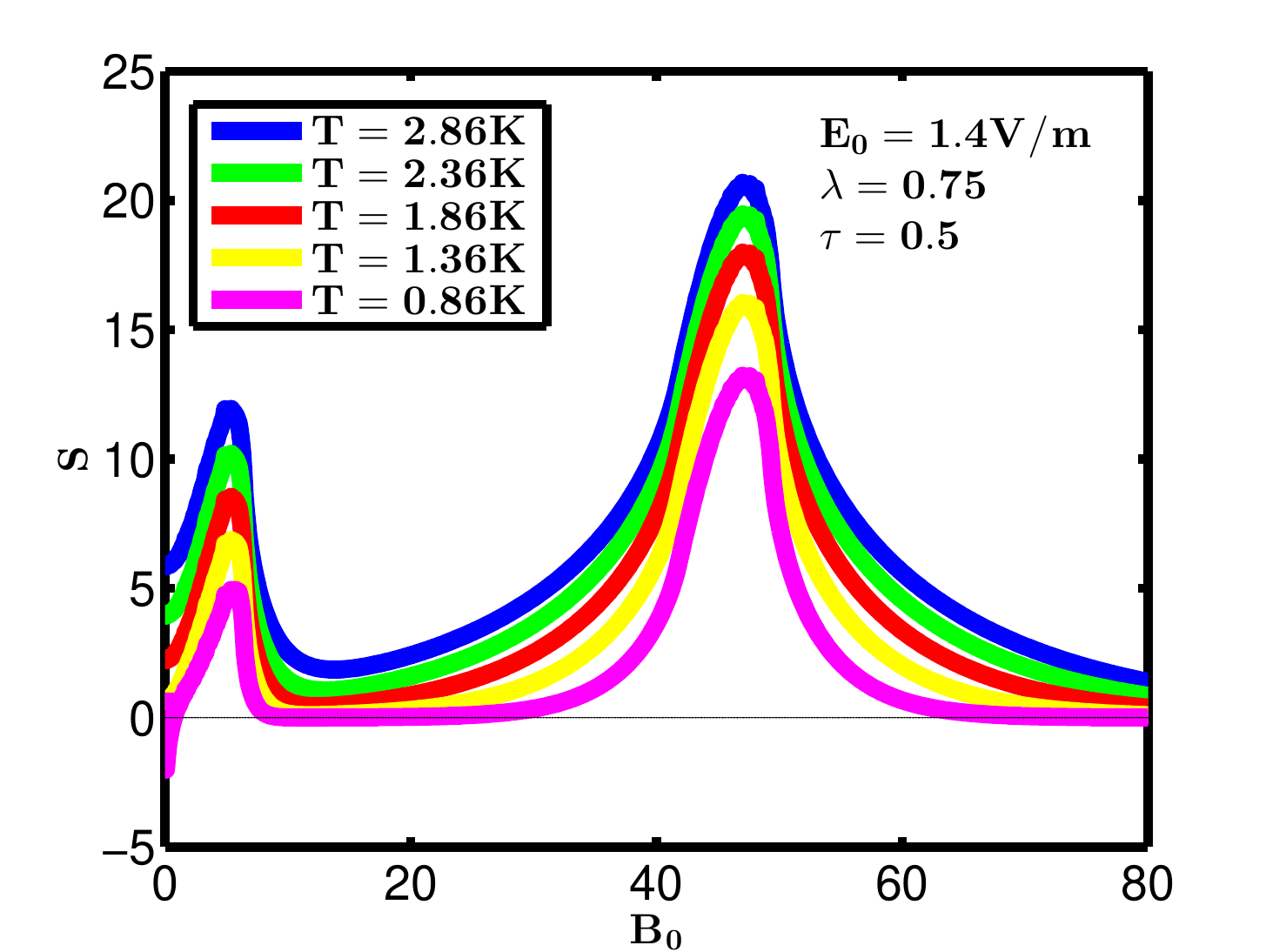}}
}
\caption{magnetic field dependence of  entropy by varying the temperature with the following magnetic and electric site-dependent parameters: $\lambda = 0$, $\tau =0.9$ (a); $\lambda = 0.5$, $\tau =0$ (b); $\lambda = 0.5$, $\tau =0.5$ (c); $\lambda = 0.5$, $\tau =0.9$ (d) ; $\lambda = 0.75$, $\tau =0.5$ (e).}
\label{F22}
\end{figure}

 Moreover, the plots in Fig.~\eqref{F22} highlight the magnetic response of the entropy for different values of temperature. It is showed that one peak-like point occurs in the case of the uniform magnetic field ($\lambda =0$) as observed in the previous work \cite{25}. However, two peaks-like points occur in the case of a site-dependent magnetic field ($\lambda \mathrm{\neq} 0$) and this confirms the two values of the critical magnetic fields obtained in the Eqs.~\eqref{11} and \eqref{12}. When the magnetic site-dependent parameter is equal to $0.5$, the values of the first and second critical magnetic fields are approximately $B_{c1}=7.5 T$ and $B_{c2}=22.5 T$ respectively, both for a uniform and site-dependent electric field. It is observed that the parameter $\tau $ considerable affects the amplitudes of the peaks whereas the parameter $\lambda $ affects the position of theses peaks (critical magnetic fields). Indeed, when $\lambda $ increase the first critical point is shifted into the left meanwhile and the second critical point is shifted into the right. This implies that by increasing the parameter $\lambda $ one can effectively decrease and increase the values of the critical magnetic fields $B_{c1}$ and $B_{c2}$ respectively. Notice that the two peaks observed when the system is under the influence of the site-dependent magnetic field indicate the existence of the intermediate phase between the order and the disordered phase as observed in reference \cite{46}.
  
 \par
\begin{figure}[!htb]
\centerline{
\subfigure[]{\includegraphics[width=0.35\textwidth]{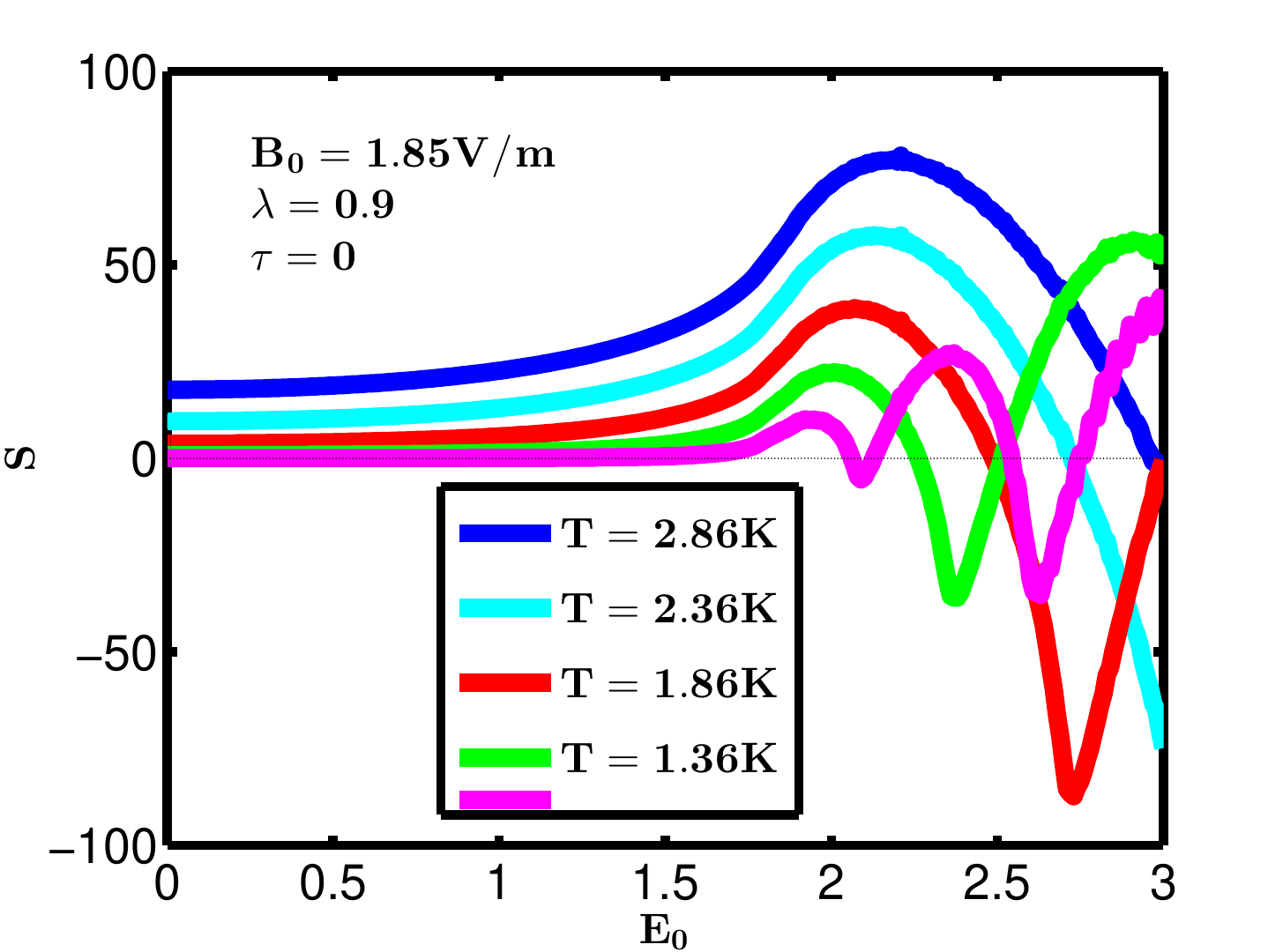}}
\subfigure[]{\includegraphics[width=0.35\textwidth]{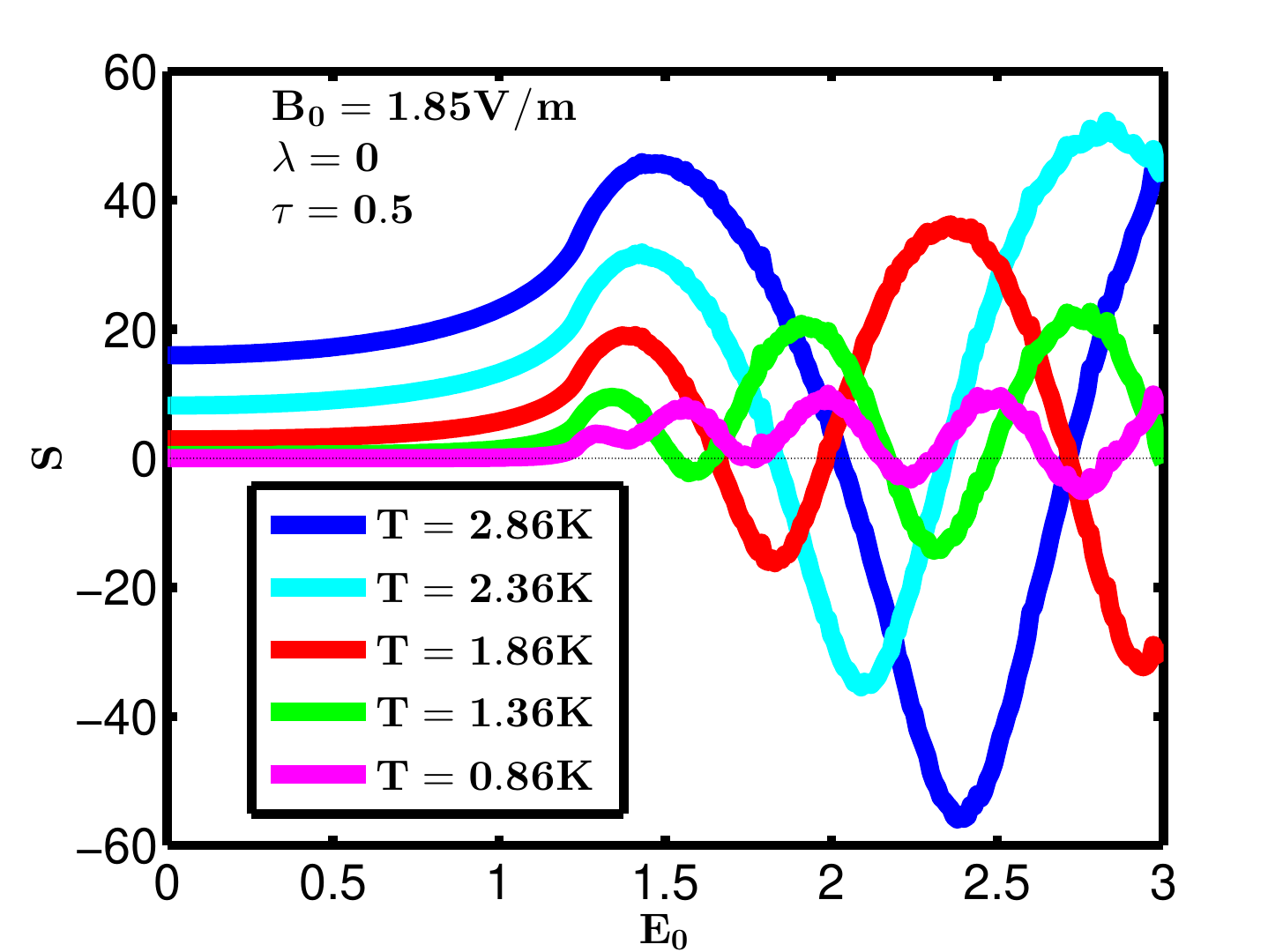}}
}
\centerline{
\subfigure[]{\includegraphics[width=0.35\textwidth]{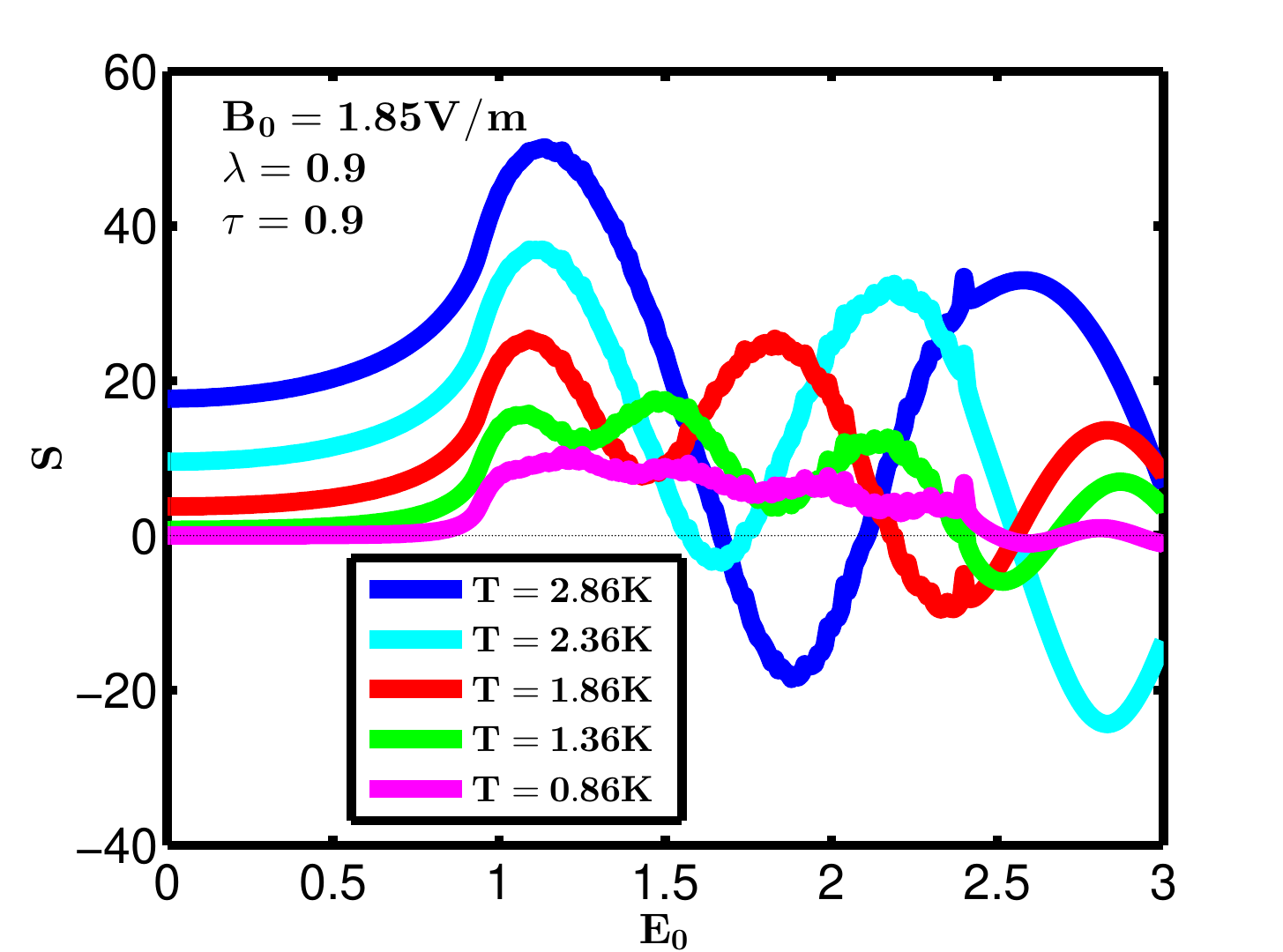}}
\subfigure[]{\includegraphics[width=0.35\textwidth]{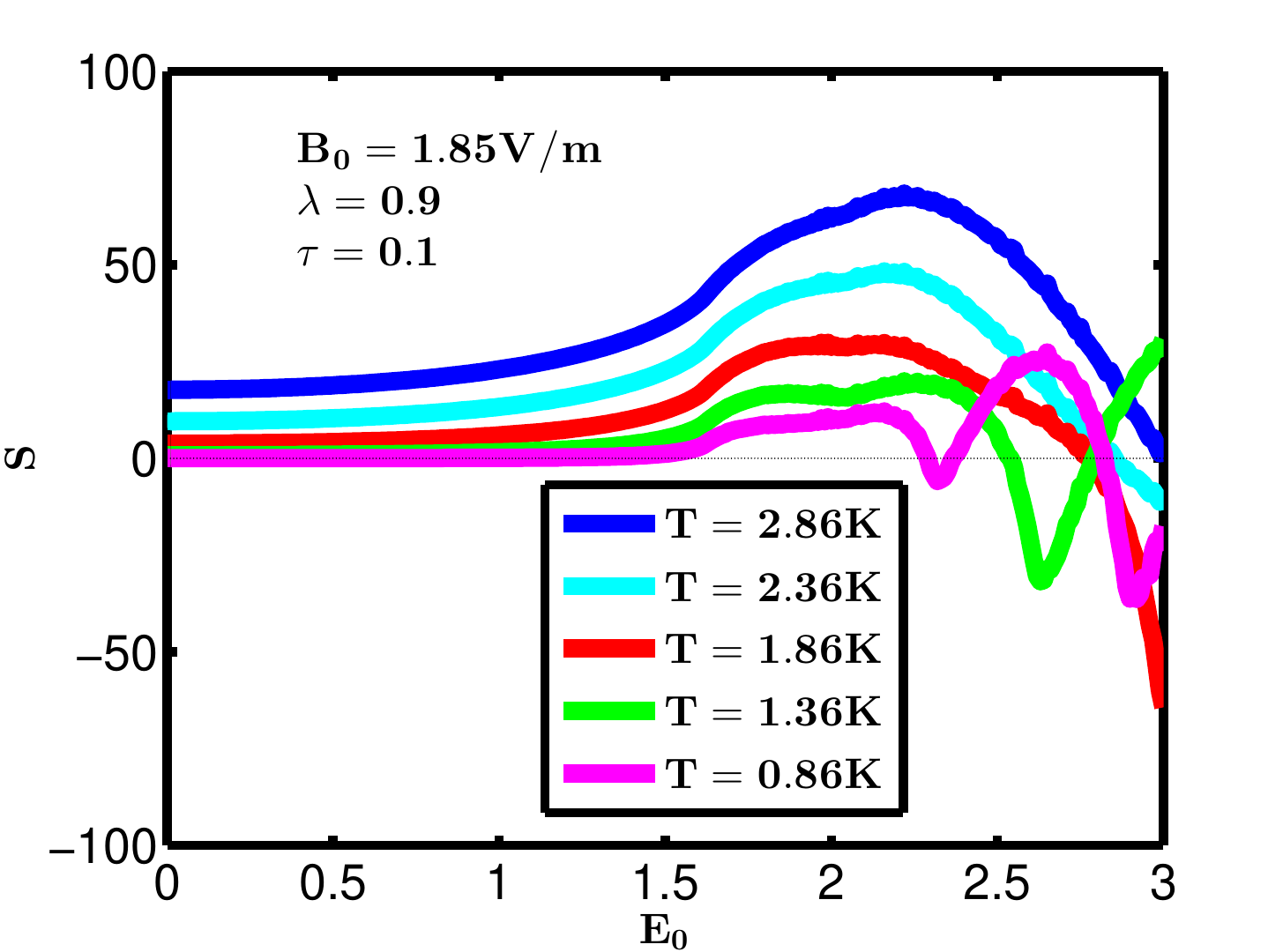}}
\subfigure[]{\includegraphics[width=0.35\textwidth]{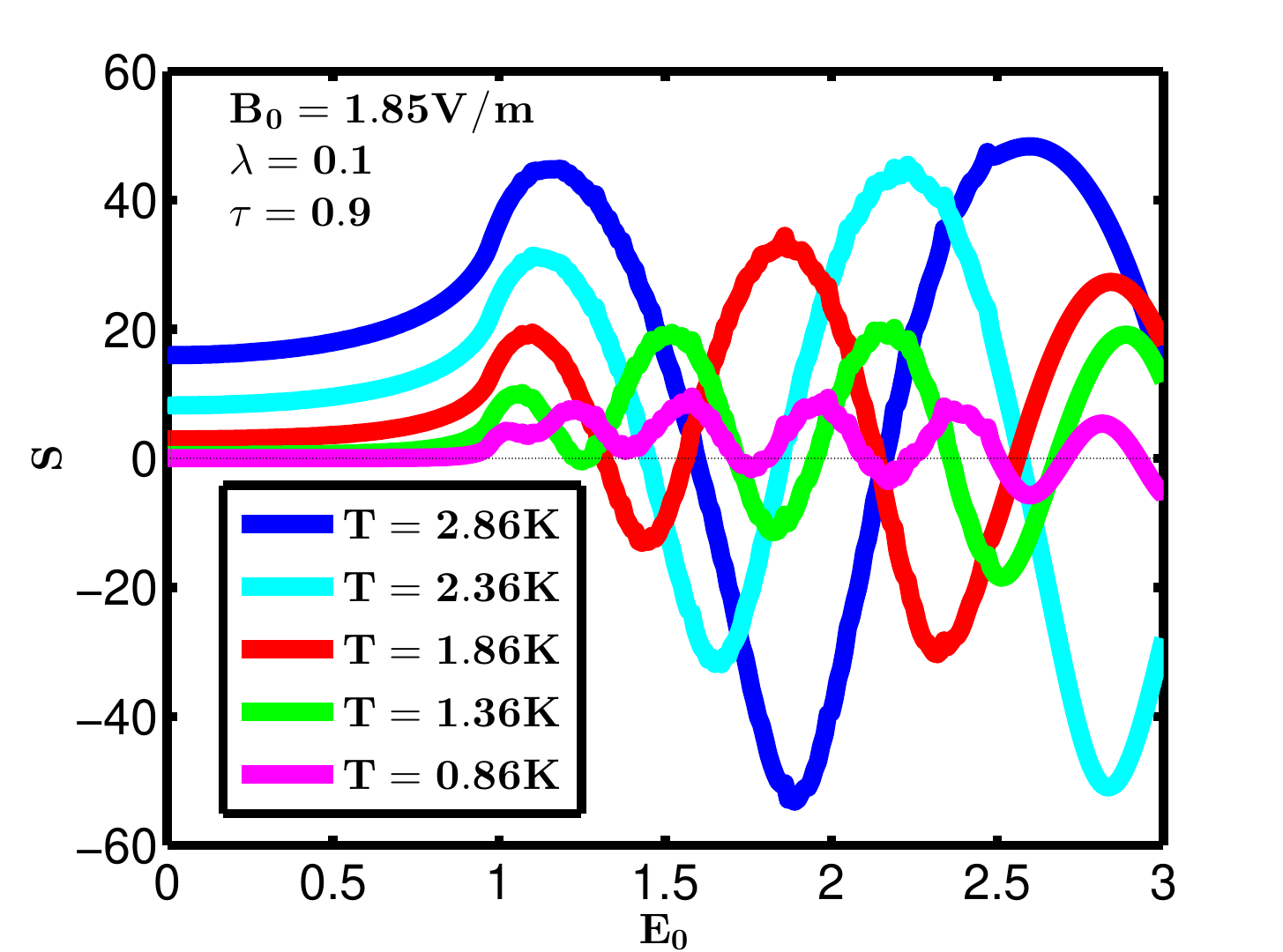}}
}
\caption{electric field dependence of  entropy by varying the temperature with the following magnetic and electric site-dependent parameters: $\lambda = 0.9$, $\tau =0$ (a); $\lambda = 0$, $\tau =0.5$ (b); $\lambda = 0.9$, $\tau =0.9$ (c); $\lambda = 0.9$, $\tau =0.1$ (d) ; $\lambda = 0.1$, $\tau =0.9$ (e).}
\label{F23}
\end{figure}

\begin{figure}[!htb]
\centerline{
\subfigure[]{\includegraphics[width=0.35\textwidth]{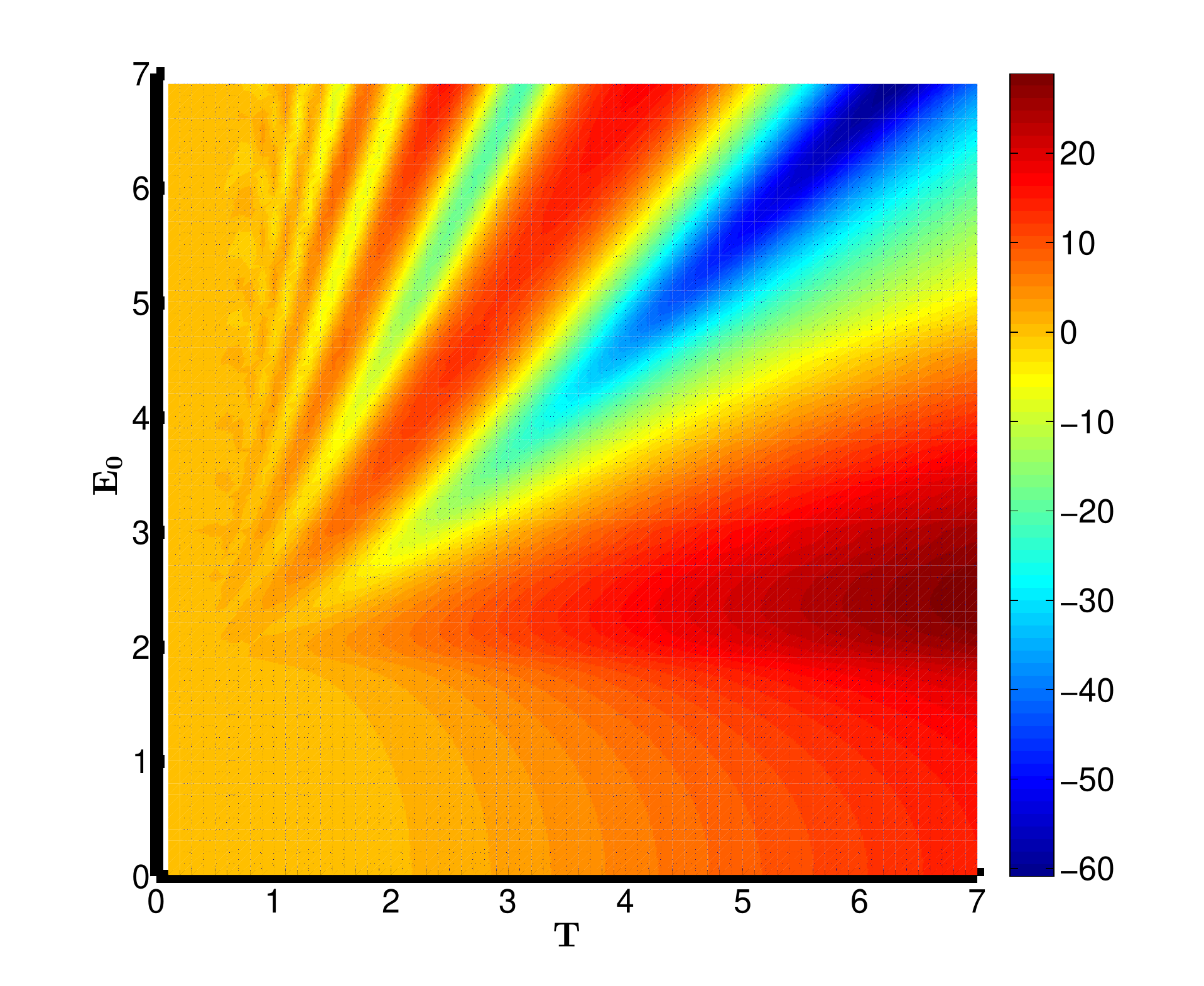}}
\subfigure[]{\includegraphics[width=0.35\textwidth]{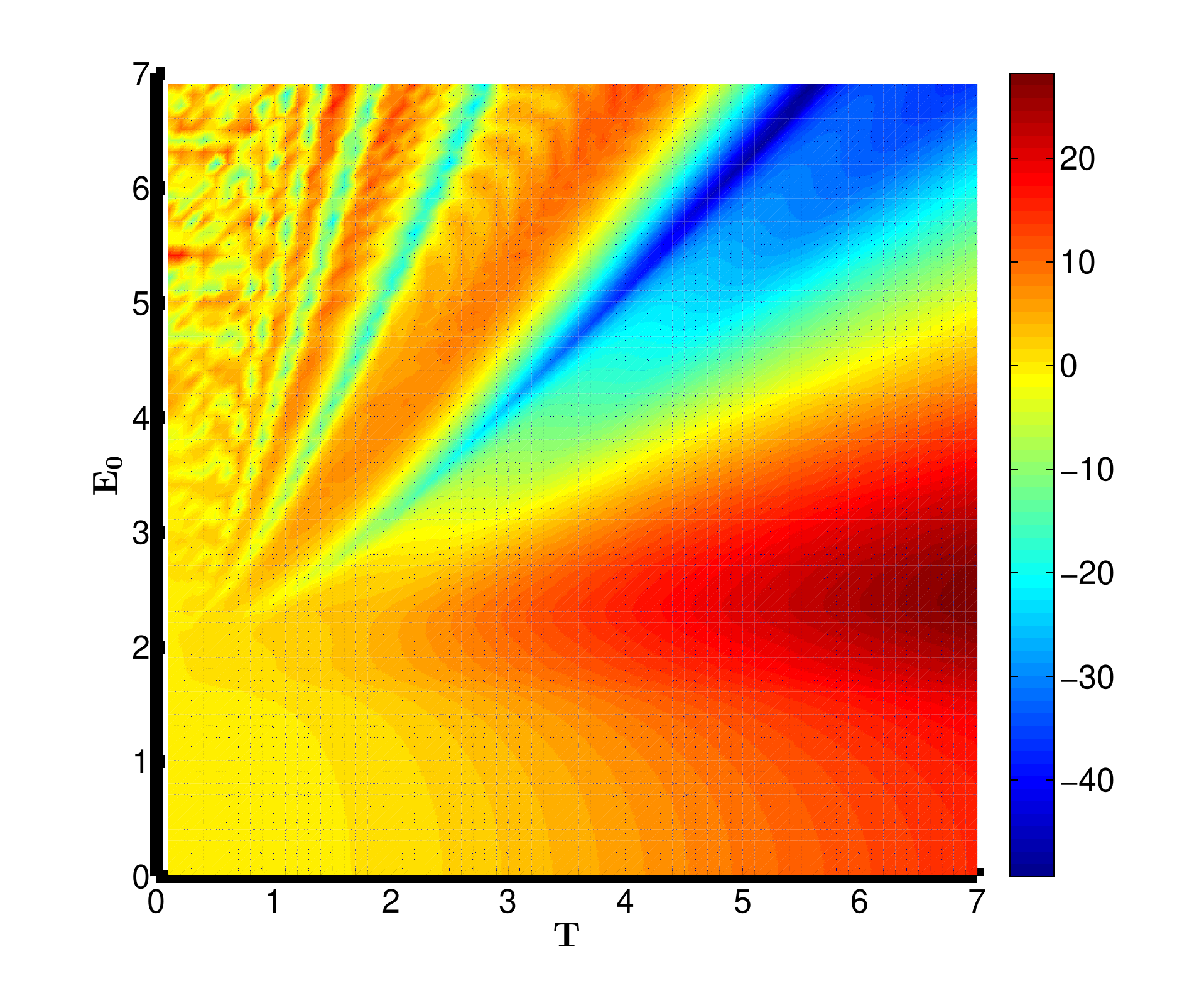}}
\subfigure[]{\includegraphics[width=0.35\textwidth]{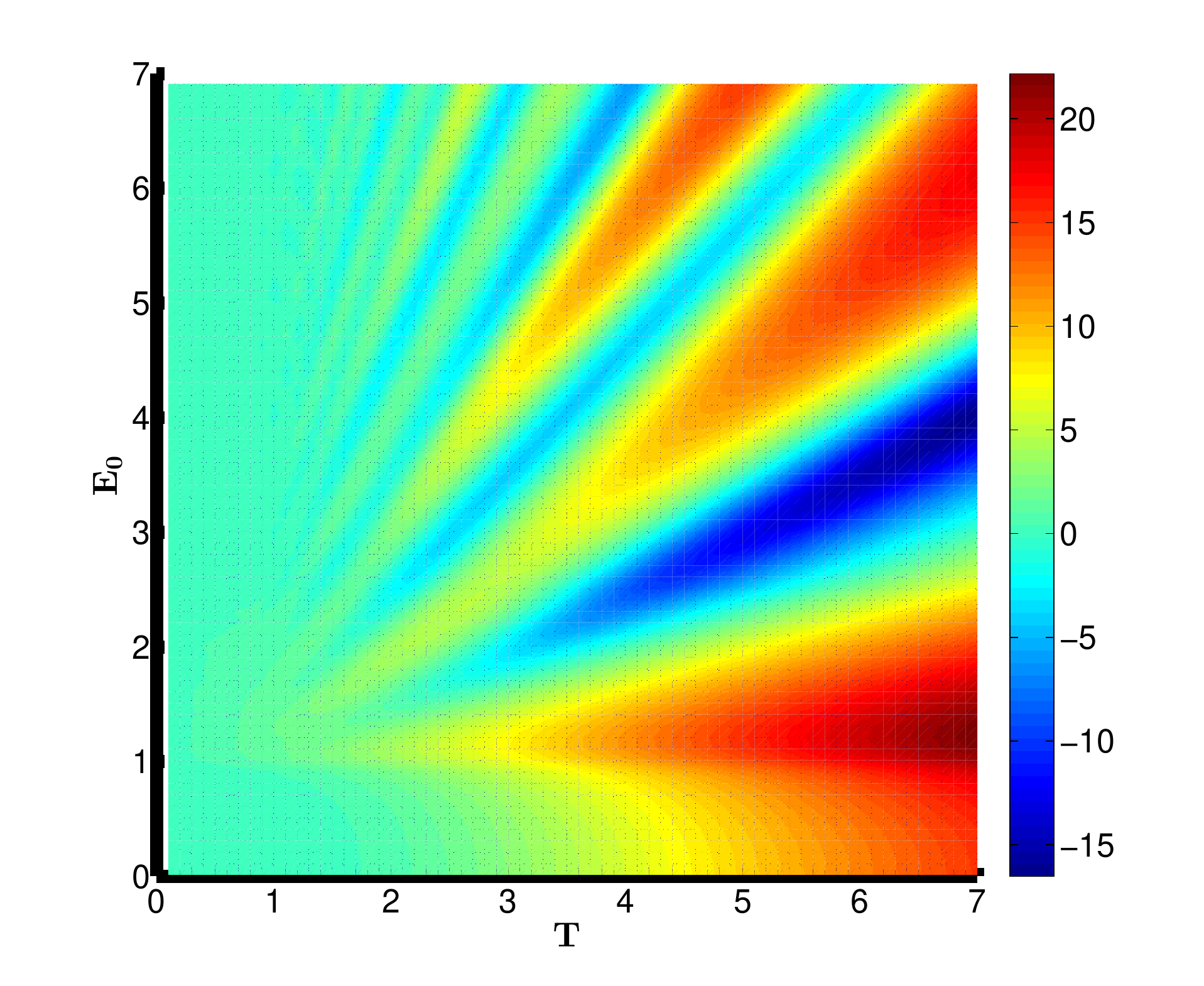}}
}
\caption{the surface plot of entropy against electric field and temperature with the following magnetic field, magnetic and electric site-dependent parameters :(a)$B_0 =1.85 T, \tau = \lambda = 0$, (b) $ B_0 =1.85 T, \lambda =0.9, \tau =0.1$, (c) $B_0 =1.85 T, \lambda =0.9, \tau =   0.9$.  }
\label{FO0}
\end{figure}

 
 On the other hand, in Fig.~\eqref{F23}, it is observed that the entropy increases with the electric field and reaches a relative maximum point from which it shows an oscillating-like behavior both in the case of uniform and site-dependent fields. Also, increasing the temperature as well as the magnetic site-dependent parameter enhances the value of the electric field from which appears the first maximum point (critical electric field) whereas the opposite situation occurs when the electric site-dependent parameter increases. Note that the critical electric field is always approximately localized between $1 V/m$ and $2.5 V/m$, no matter the values of the parameters of the system considered. It is worth noticing that beyond the critical electric field, the system violates the second law of thermodynamics. Indeed, according to the second law of thermodynamics, the entropy of an isolated system always increase but hereby the entropy displays an oscillatory behavior from the critical electric field. In addition, that oscillatory behaviour of the entropy with negative part is due to the heat dissipation for the system under external electric field and thermal condition. Note that the site-dependent electric field enhance such a behaviour of the entropy also observed in Fig.~\eqref{F21}. Furthermore, the phase diagram in term of the entropy obtained within the $E_0 - T$ plane, in  Fig.~\eqref{FO0}, shows three regions as function of approximated values of critical electric field and temperature: a region where the entropy varies slightly with the electric field ($T\mathrm{<} 1 K$), a region where the entropy increases with both the electric field and the temperature ($E_0 \mathrm{<} 1 V/m$) and a region where the entropy oscillates both with the electric field and the temperature ($E_0 \mathrm{>} 1 V/m$ and $T \mathrm{>} 1 K$ ). Note that the critical electric and the critical temperature are tunable by the magnetic and electric site-dependent parameters.    

\par
\par
\par
\subsection{specific heat capacity}
\begin{figure}[!htb]
\centerline{
\subfigure[]{\includegraphics[width=0.35\textwidth]{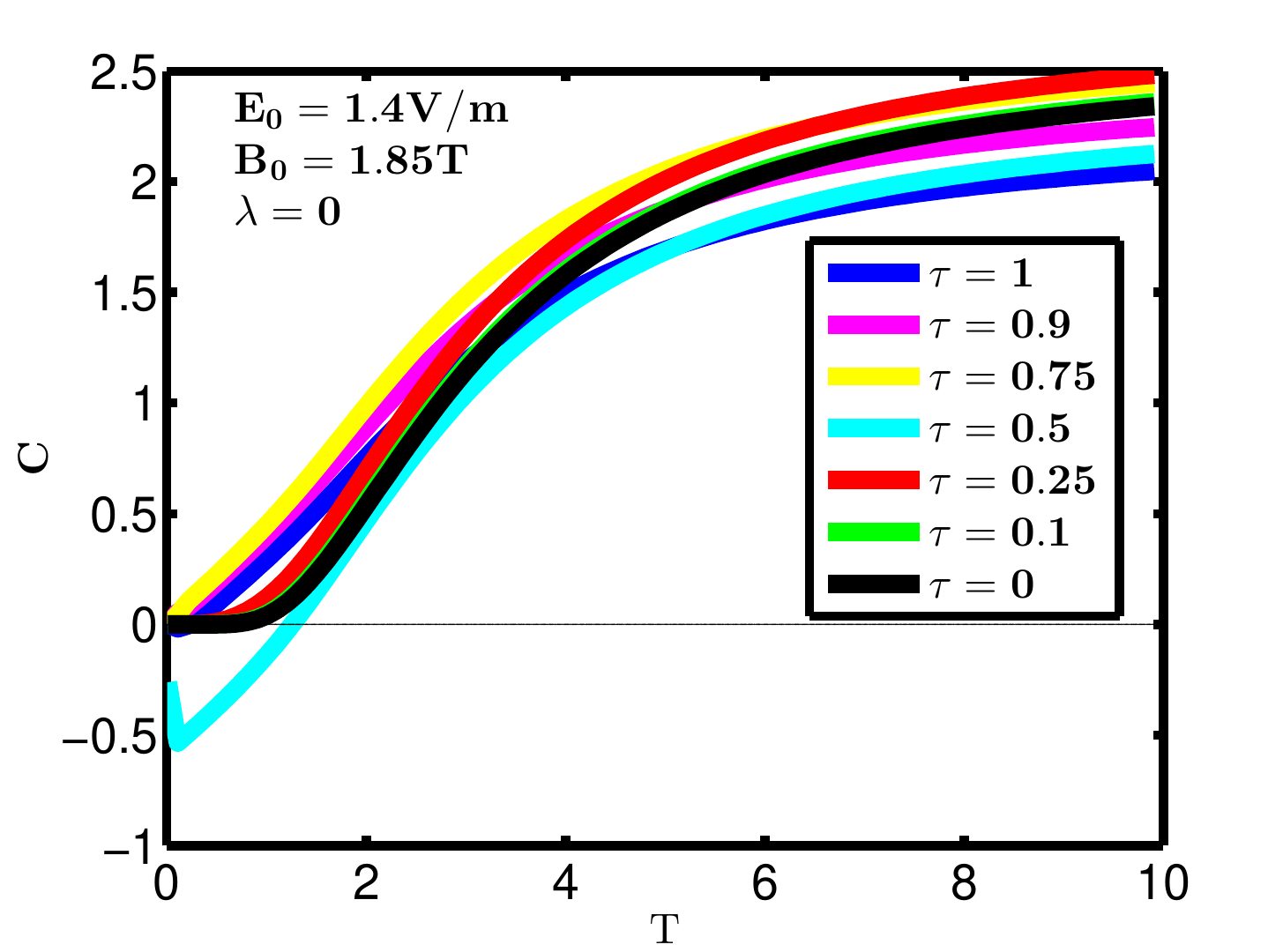}}
\subfigure[]{\includegraphics[width=0.35\textwidth]{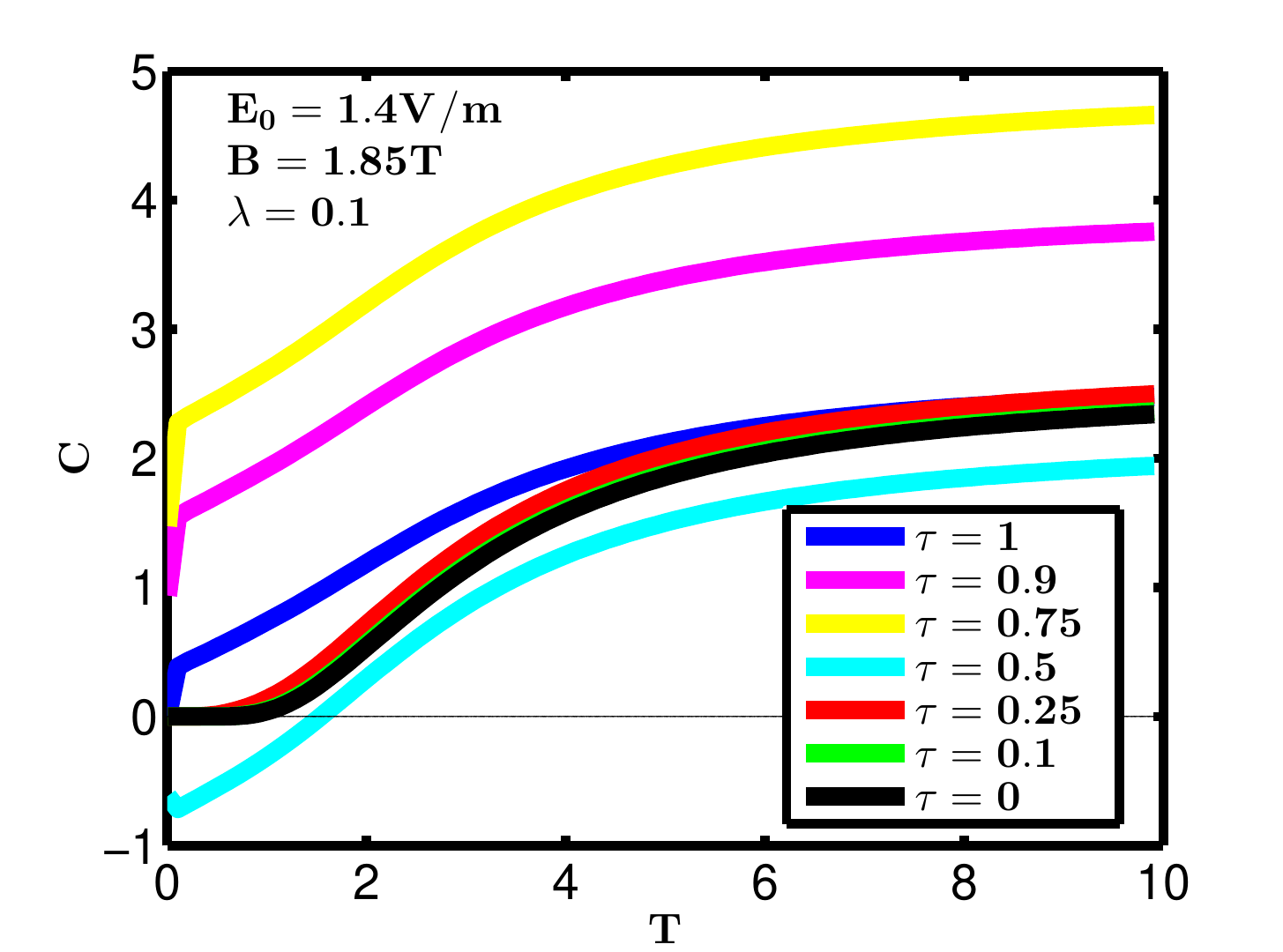}}
\subfigure[]{\includegraphics[width=0.35\textwidth]{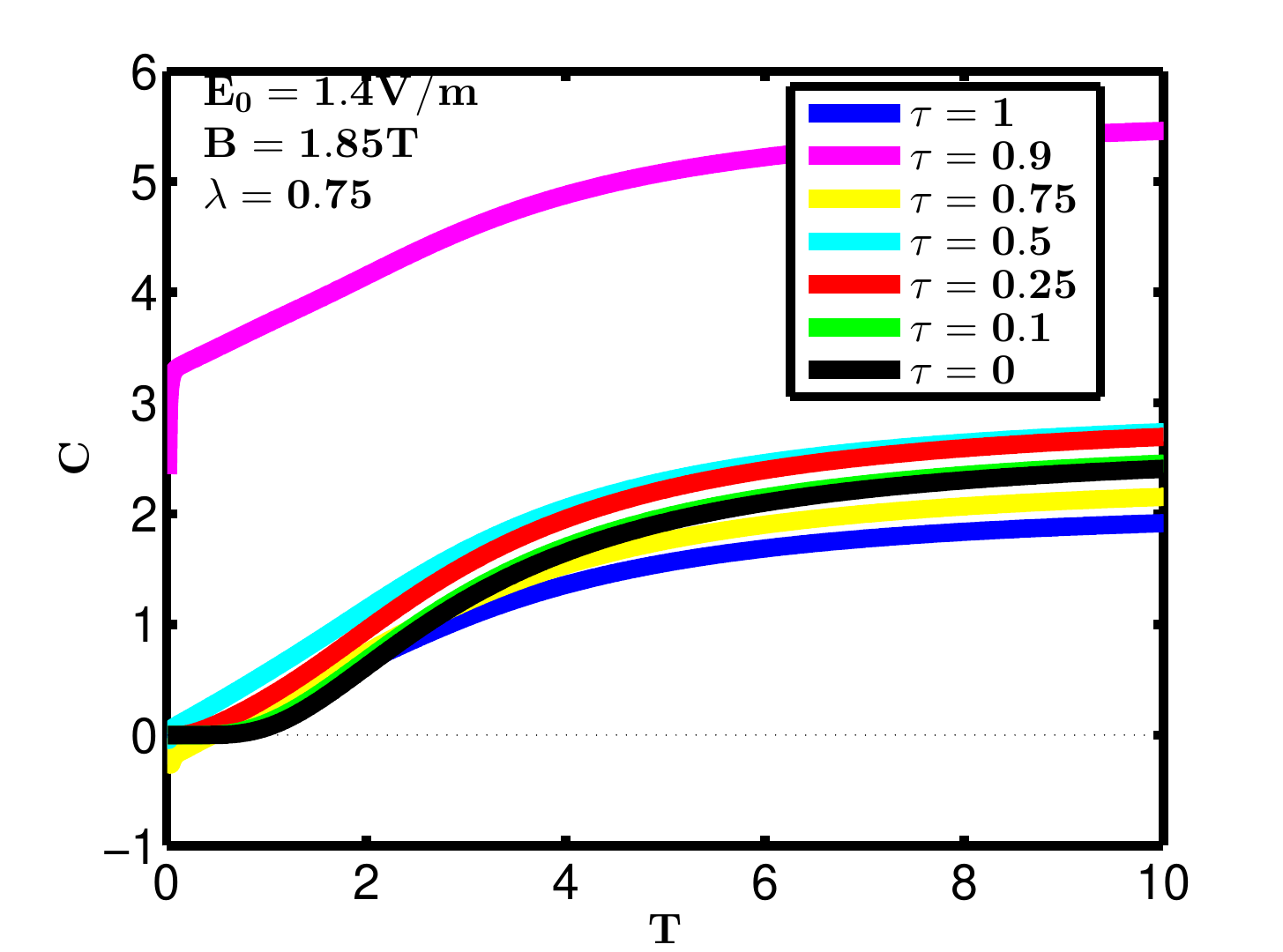}}  
}
\centerline{\subfigure[]{\includegraphics[width=0.35\textwidth]{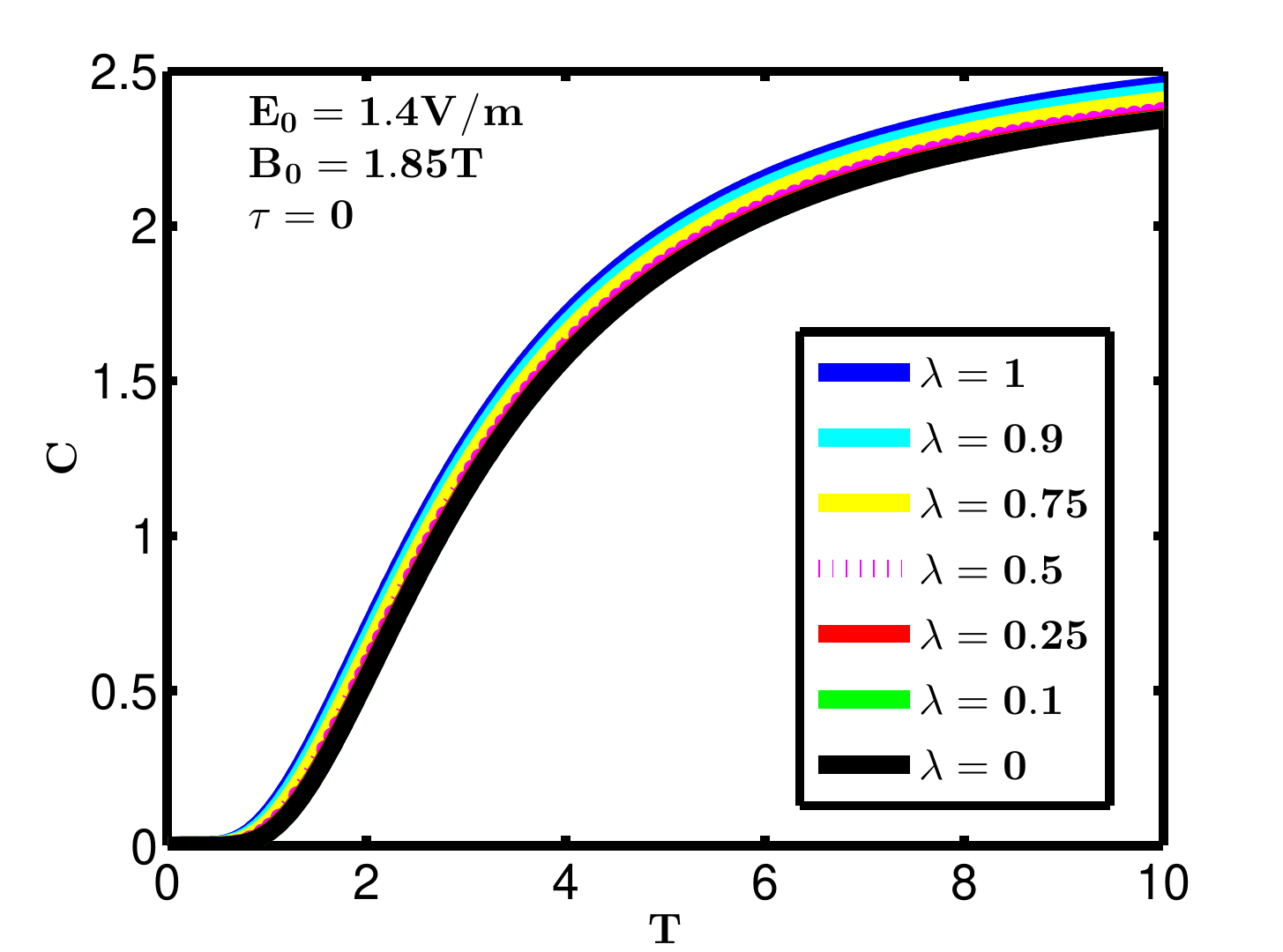}}
\subfigure[]{\includegraphics[width=0.35\textwidth]{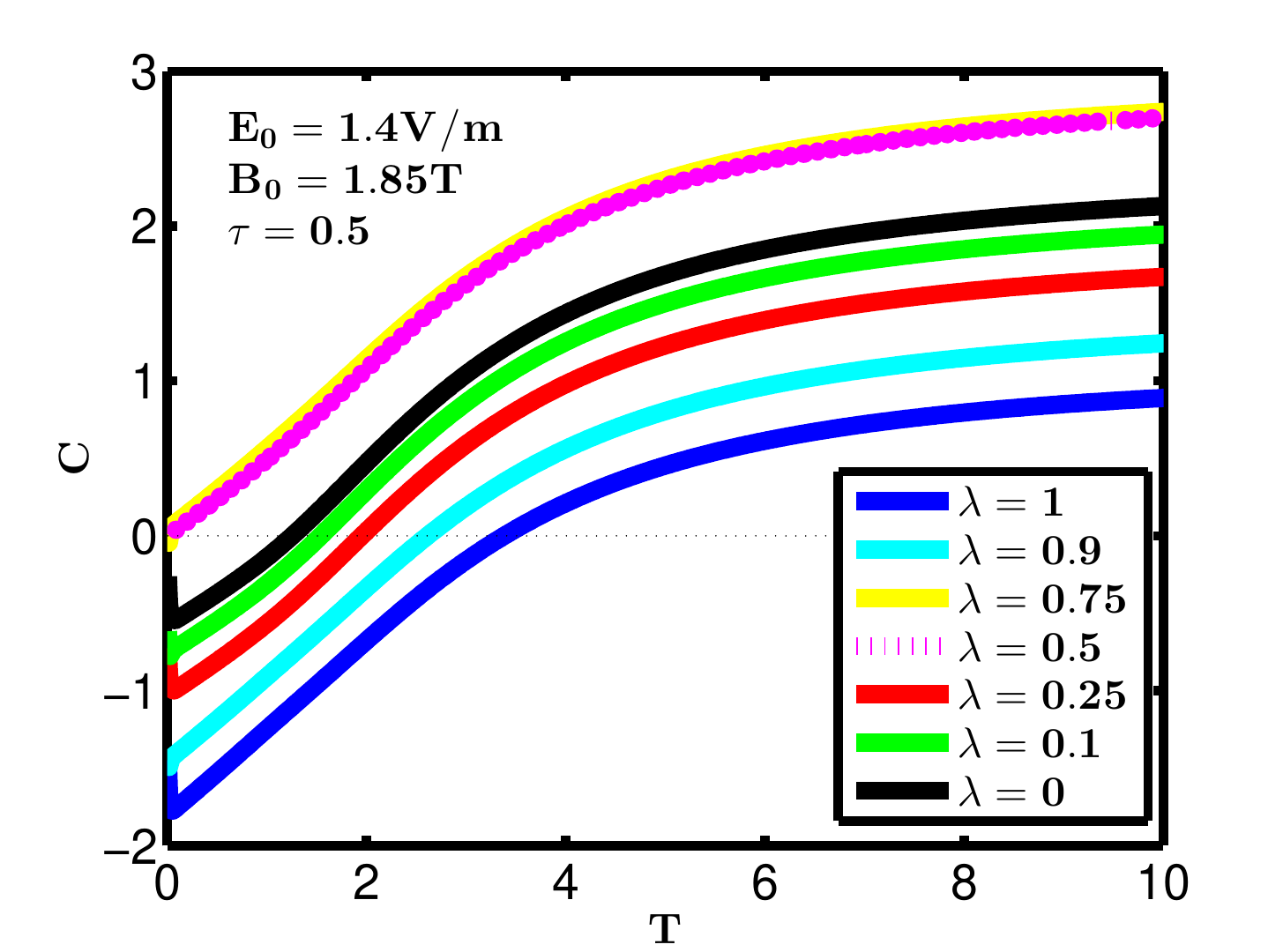}}
\subfigure[]{\includegraphics[width=0.35\textwidth]{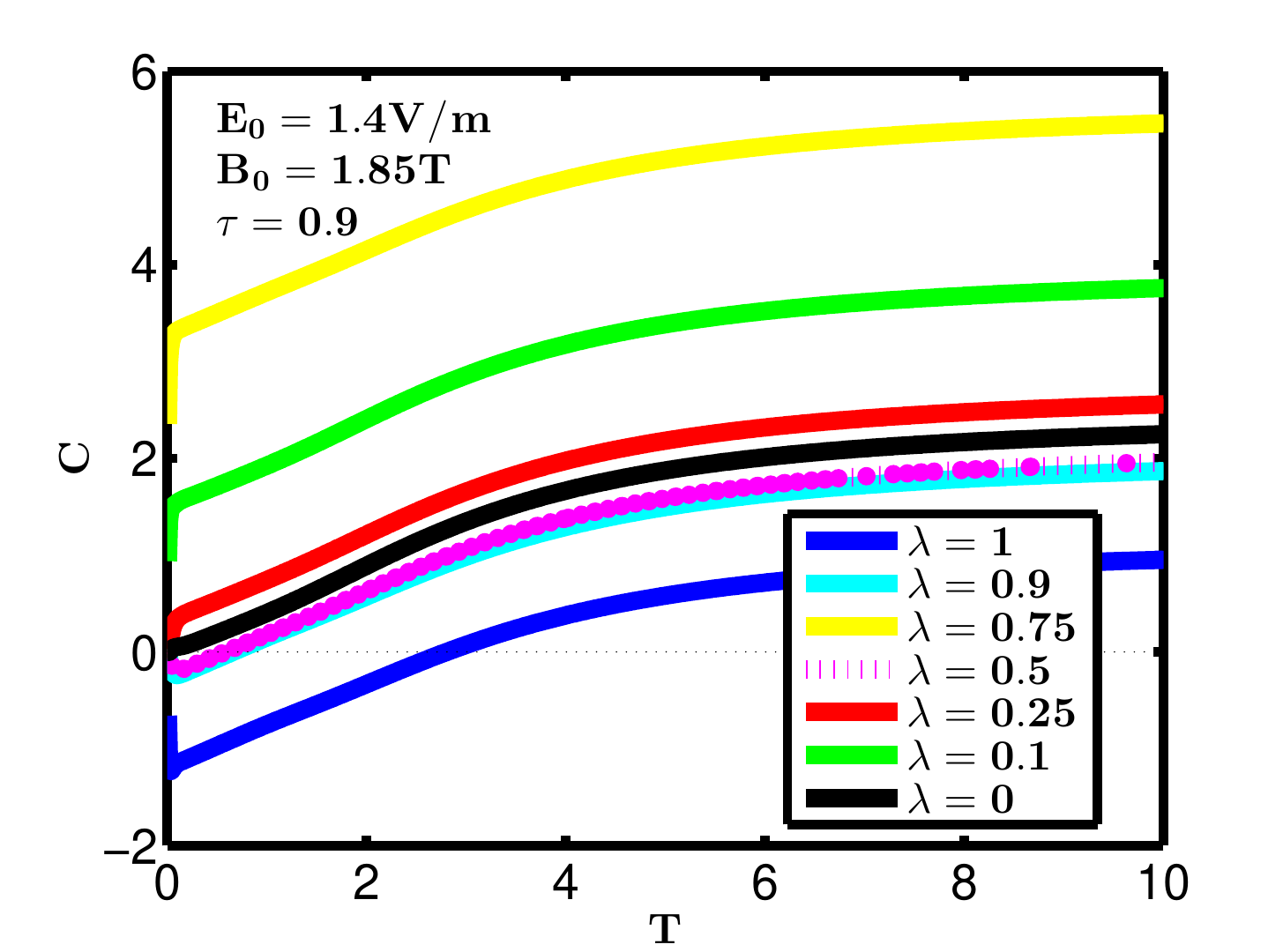}}
}
\caption{the upper panels show the evolution of the specific heat capacity of the system for different values of the electric site-dependent parameters and for three values of the magnetic site-dependent parameter namely 0 (a), 0.1 (b) and 0.75 (c). In the lower panels we plotted the evolution of the specific heat capacity for different values of the magnetic site-dependent parameters and for three values of the electric site-dependent parameter namely 0 (d), 0.5 (e) and 0.9 (f). }
\label{F24}
\end{figure}

 From the entropy obtained in Eq.~\eqref{20}, the specific heat capacity at constant magnetic and electric fields is obtained as
\begin{equation} \label{21} 
C=T\left(\frac{\partial S}{\partial T} \right) .                                  
\end{equation} 
Thus 
\begin{equation} \label{22} 
C=\frac{1}{T^{2} } \left[
\begin{split}
&\sum _{k}\frac{\omega _{1 k}^{\left(+\right)} e^{-\frac{1}{T} \omega _{1 k}^{\left(+\right)} } }{\left(1-e^{-\frac{1}{T} \omega _{1 k}^{\left(+\right)} } \right)^{2} }  +\sum _{k}\frac{\omega _{1 k}^{\left(-\right)} e^{-\frac{1}{T} \omega _{1 k}^{\left(-\right)} } }{\left(1-e^{-\frac{1}{T} \omega _{1 k}^{\left(-\right)} } \right)^{2} }  +\sum _{k}\frac{\omega _{2 k}^{\left(+\right)} e^{-\frac{1}{T} \omega _{2 k}^{\left(+\right)} } }{\left(1-e^{-\frac{1}{T} \omega _{2 k}^{\left(+\right)} } \right)^{2} } 
\\&
+\sum _{k}\frac{\omega _{2 k}^{\left(-\right)} e^{-\frac{1}{T} \omega _{2 k}^{\left(-\right)} } }{\left(1-e^{-\frac{1}{T} \omega _{2 k}^{\left(-\right)} } \right)^{2} }
\end{split} \right].      
\end{equation} 
The temperature dependence of the specific heat capacity is shown in Fig.~\eqref{F24}. It is observed that the specific heat capacity increases with the temperature and reaches an asymptotic value.  The value of the temperature beyond which the heat capacity becomes constant (critical temperature) together with the corresponding asymptotic value depend upon the values of the magnetic and electric site-dependent parameters considered. Also, in the absence of site-dependent-electric field ($\tau =0$), the specific heat capacity is only slightly affected by the site-dependent magnetic field. However, when the site-dependent-electric field switch on ($\tau \mathrm{\neq} 0$), the impacts of the site-dependent magnetic field becomes more prominent. Moreover, for certain values of magnetic and electric site-dependent parameters, the specific heat capacity exhibits an abrupt increase or decrease near-zero temperature, indicating the occurrence of metamagnetic and metaelectric transition in the system \cite{47}.

\par
\begin{figure}[!htb]
\centerline{
\subfigure[]{\includegraphics[width=0.35\textwidth]{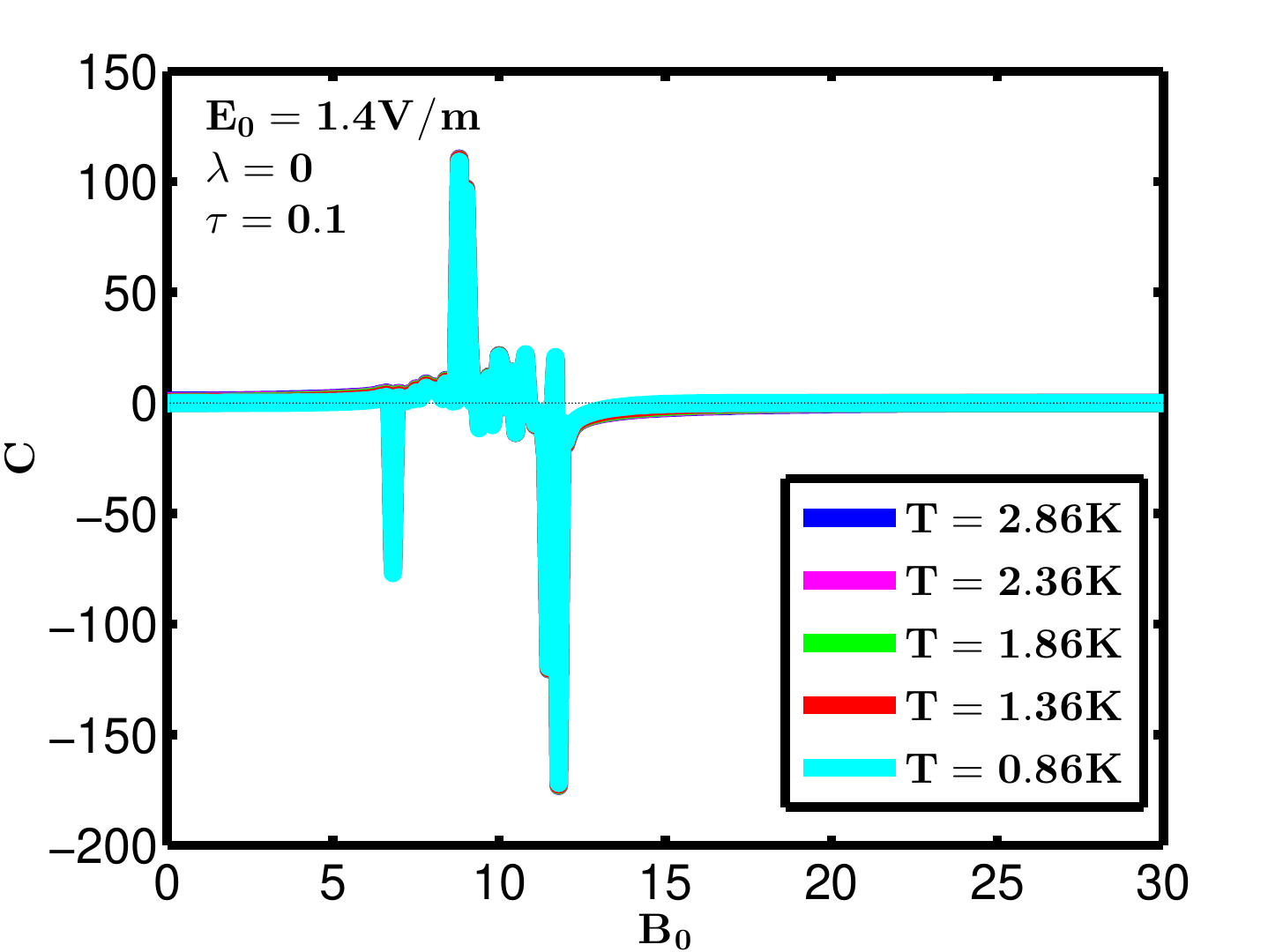}}
\subfigure[]{\includegraphics[width=0.35\textwidth]{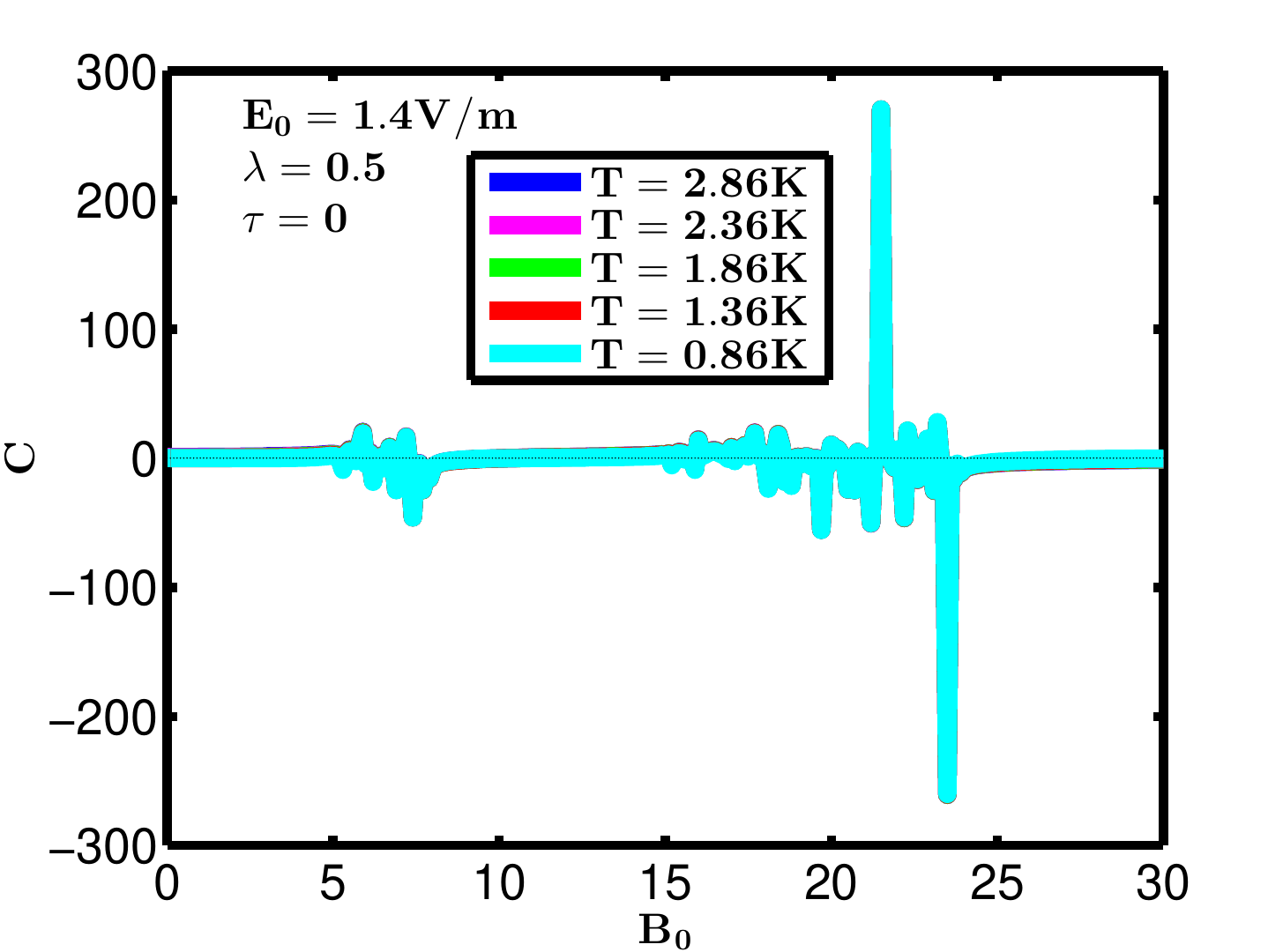}}
}
\centerline{\subfigure[]{\includegraphics[width=0.35\textwidth]{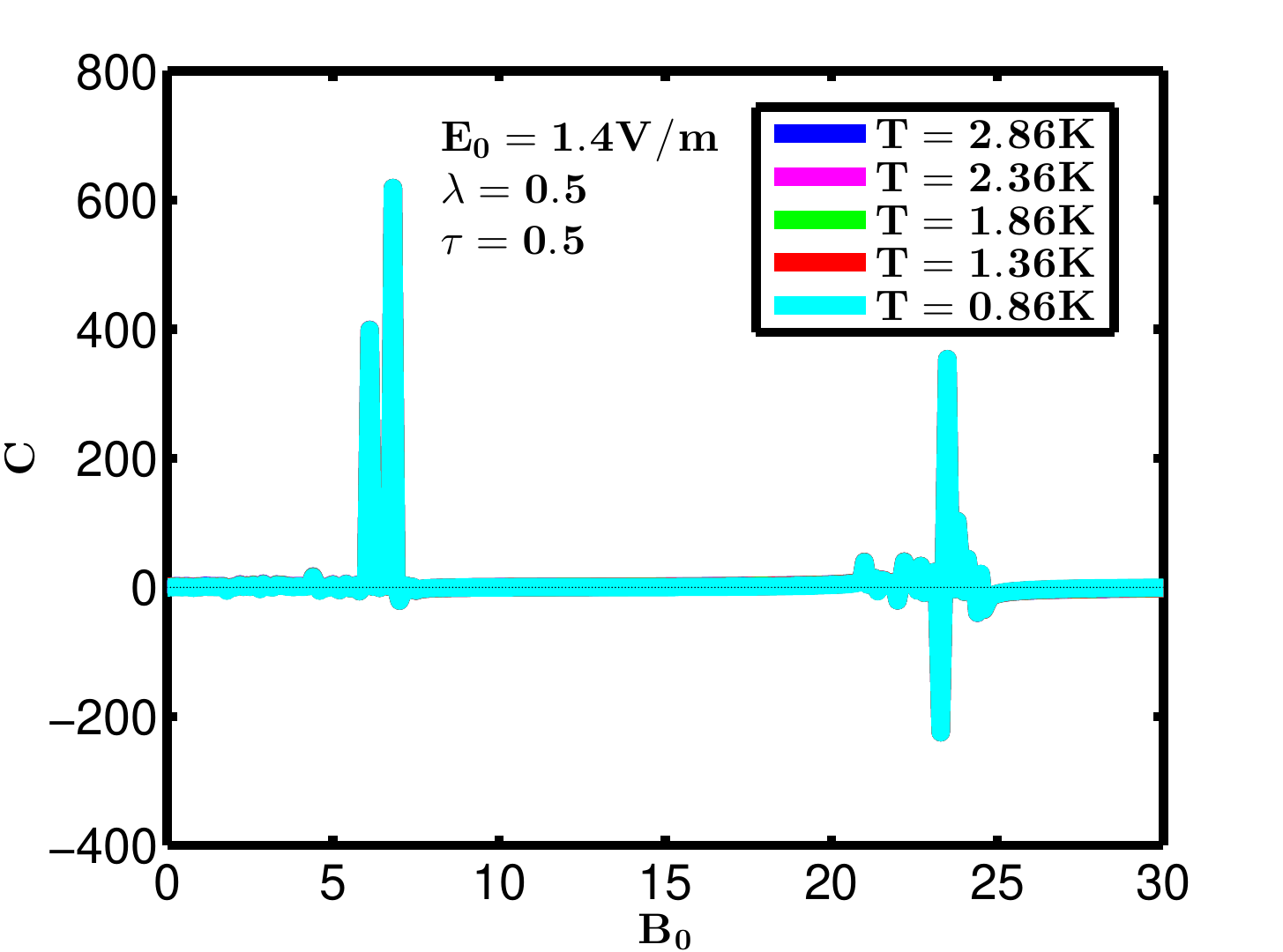}}
\subfigure[]{\includegraphics[width=0.35\textwidth]{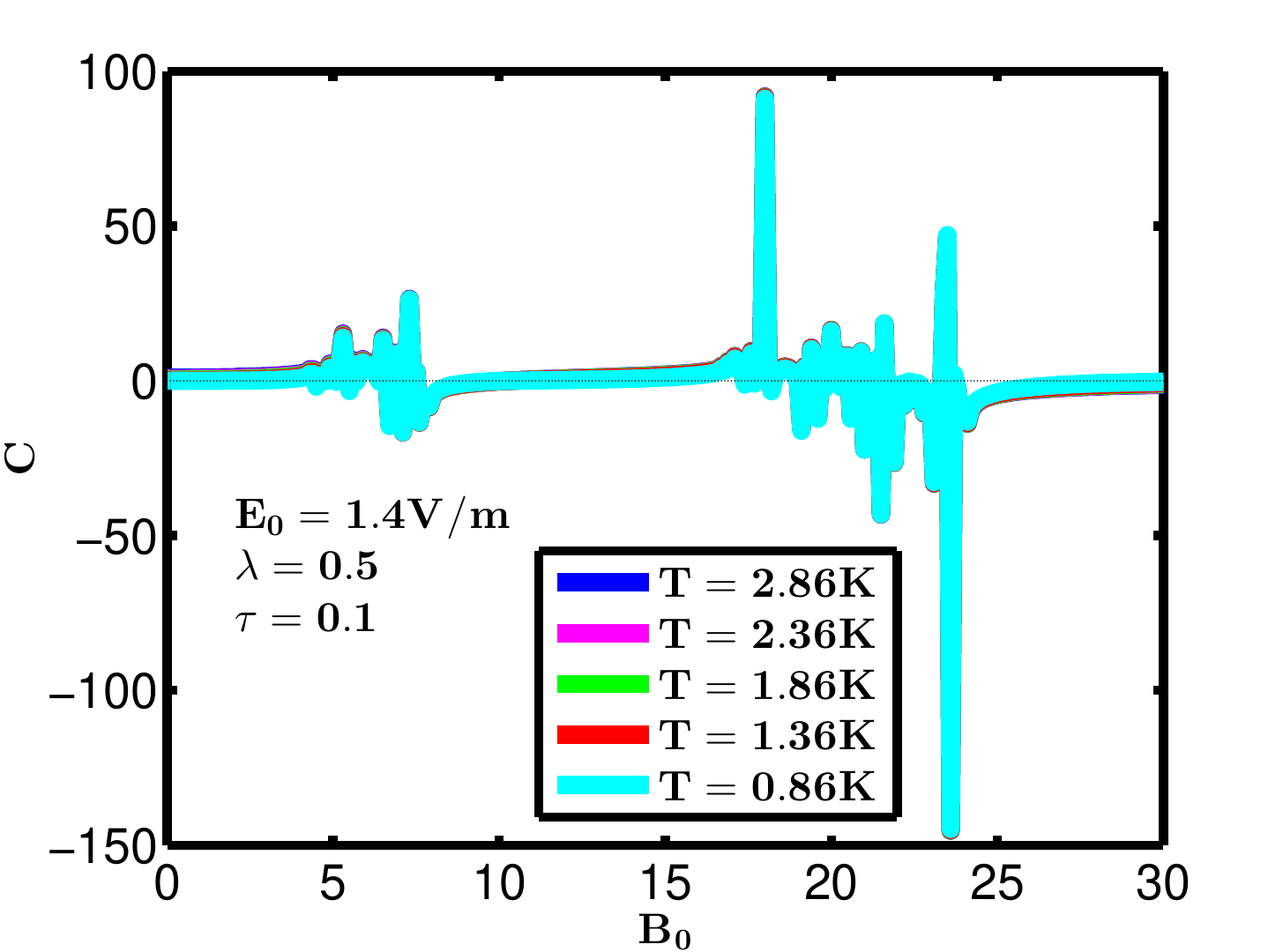}}
\subfigure[]{\includegraphics[width=0.35\textwidth]{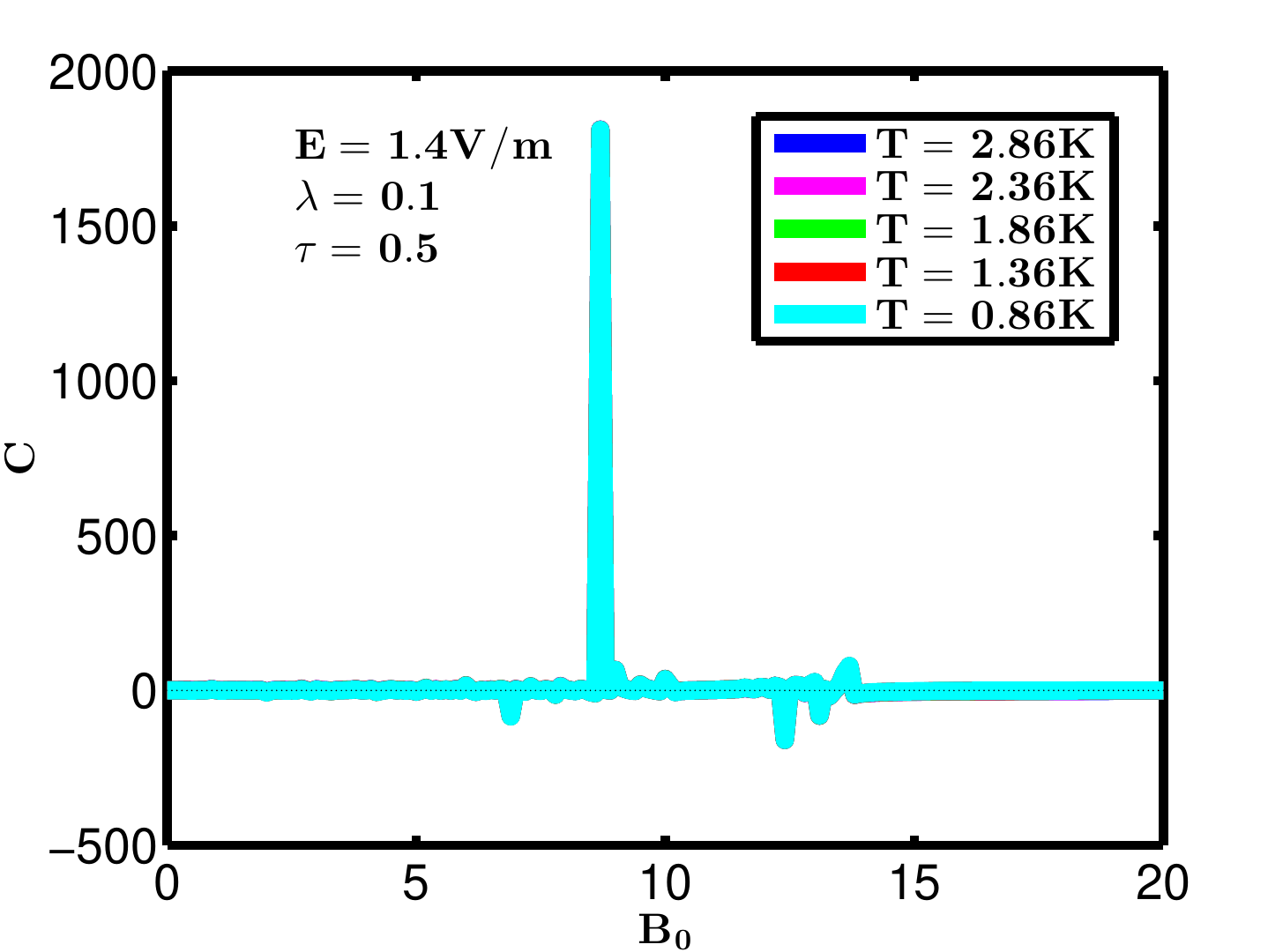}}
}
\caption{magnetic field dependence of the specific heat capacity by varying the temperature with the following magnetic and electric site-dependent parameters: $\lambda = 0$, $\tau =0.1$ (a); $\lambda = 0.5$, $\tau =0$ (b); $\lambda = 0.5$, $\tau =0.5$ (c); $\lambda = 0.5$, $\tau =0.1$ (d) ; $\lambda = 0.1$, $\tau =0.5$ (e).}
\label{F25}
\end{figure}

 
In Fig.~\eqref{F25}, the magnetic field dependence of the specific heat capacity with different values of the temperature displays a quasi-oscillating-like behavior accompanied by peak-like points which appear at one point in the case of zero magnetic site-dependent parameter ($\lambda =0$) and two different points corresponding to the critical magnetic fields in the case of nonzero magnetic site-dependent parameter ($\lambda \mathrm{\neq} 0$). 
\par
\begin{figure}[!htb]
\centerline{
\subfigure[]{\includegraphics[width=0.35\textwidth]{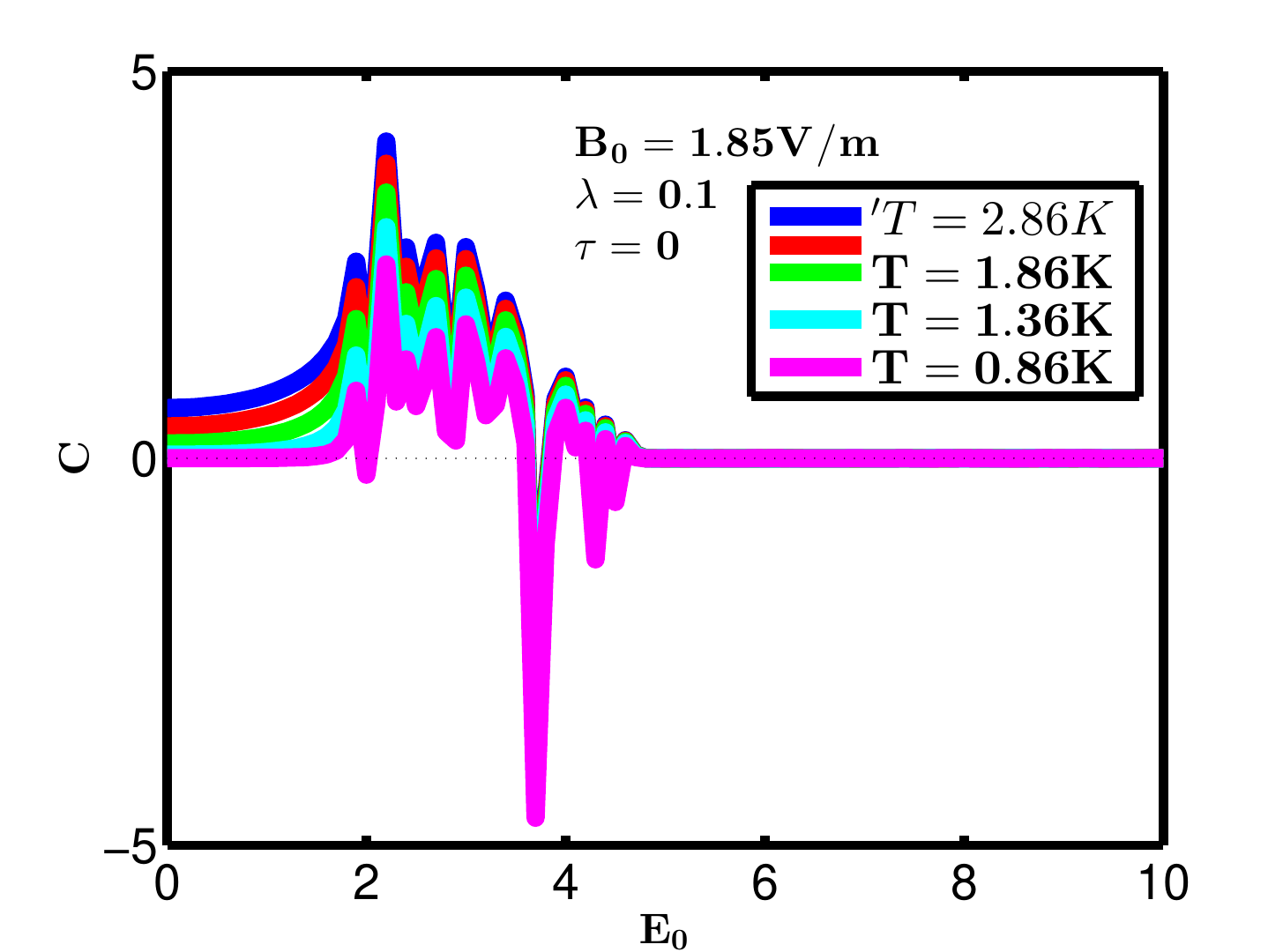}}
\subfigure[]{\includegraphics[width=0.35\textwidth]{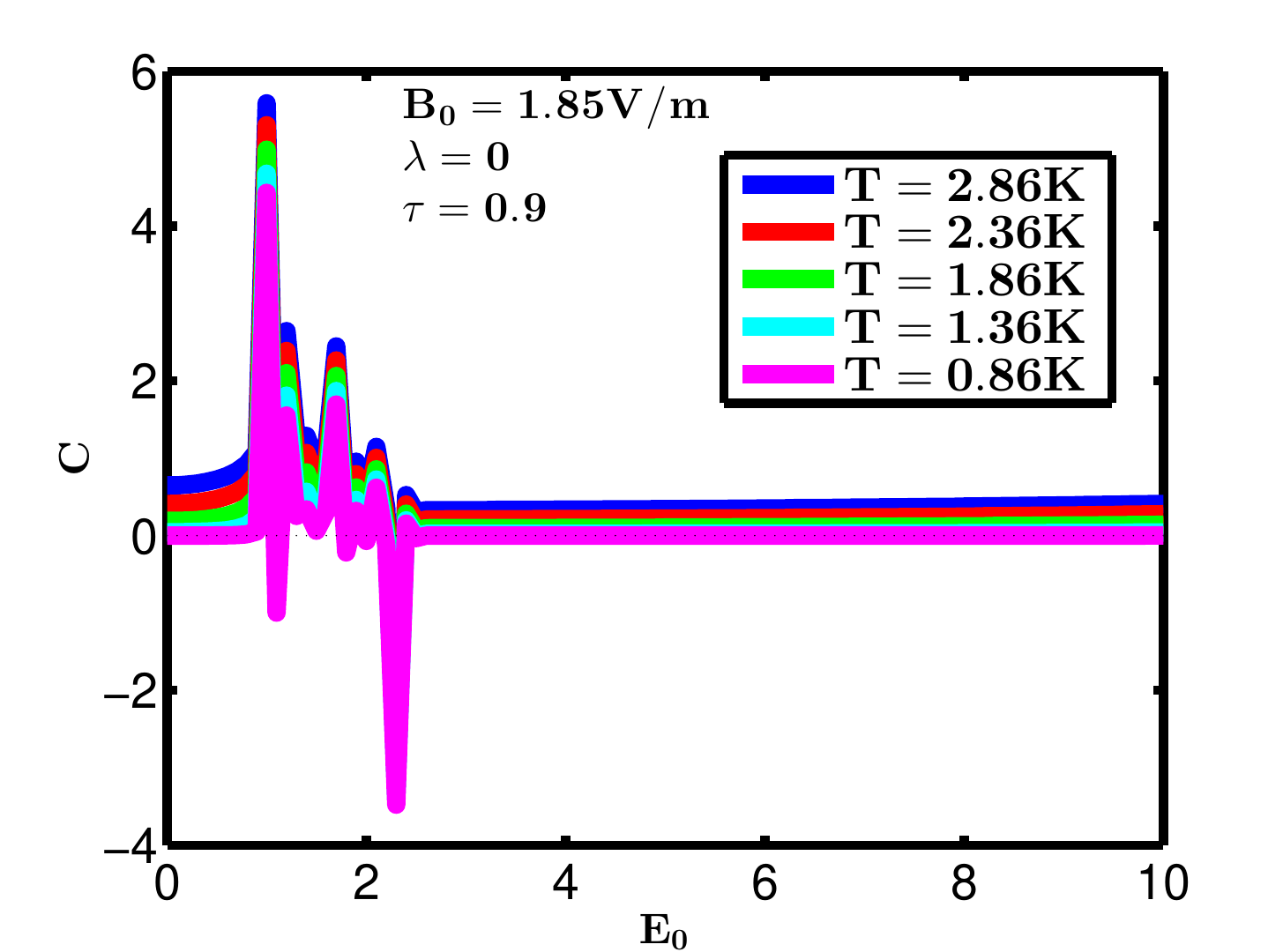}}
}
\centerline{\subfigure[]{\includegraphics[width=0.35\textwidth]{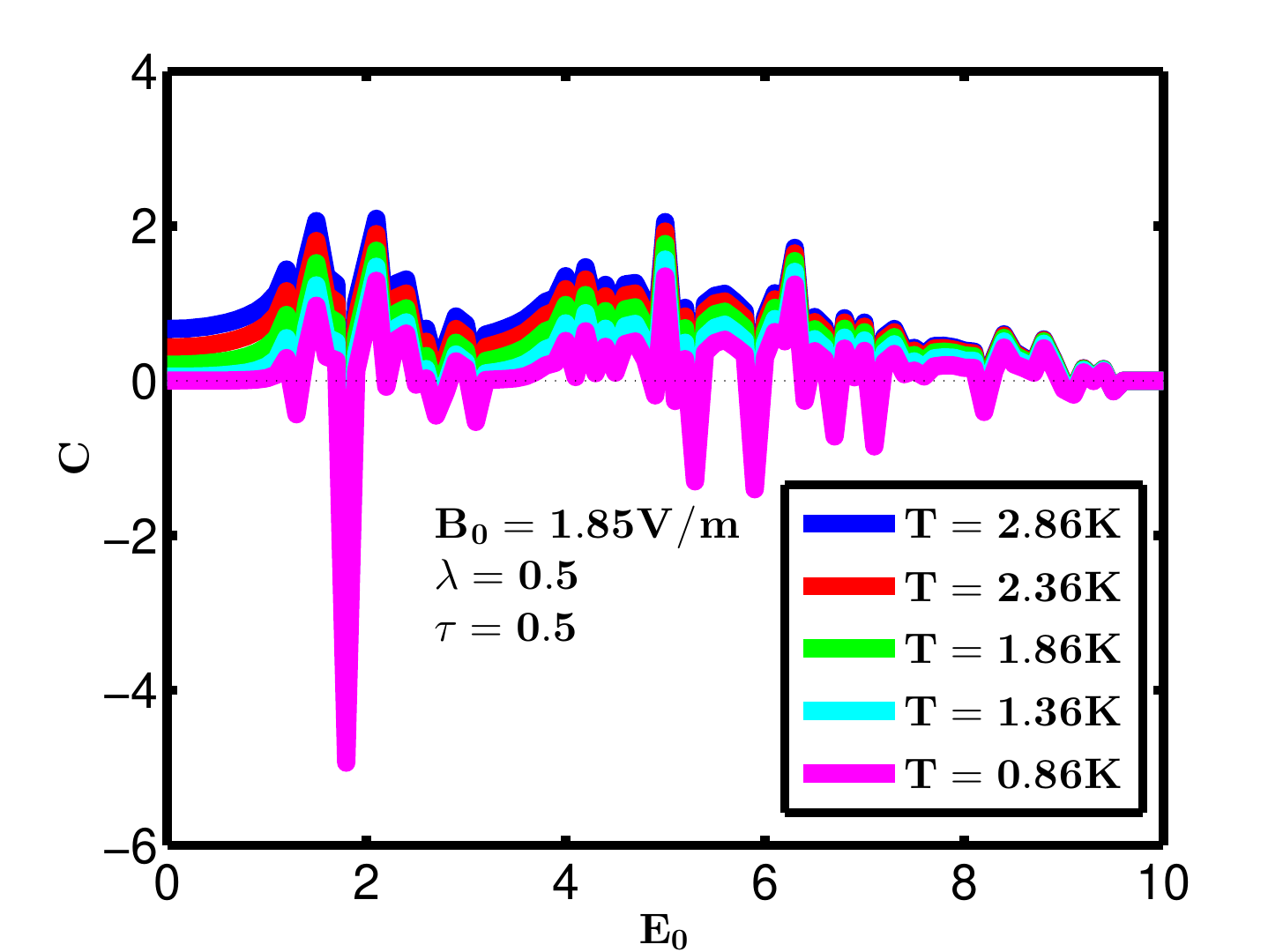}}
\subfigure[]{\includegraphics[width=0.35\textwidth]{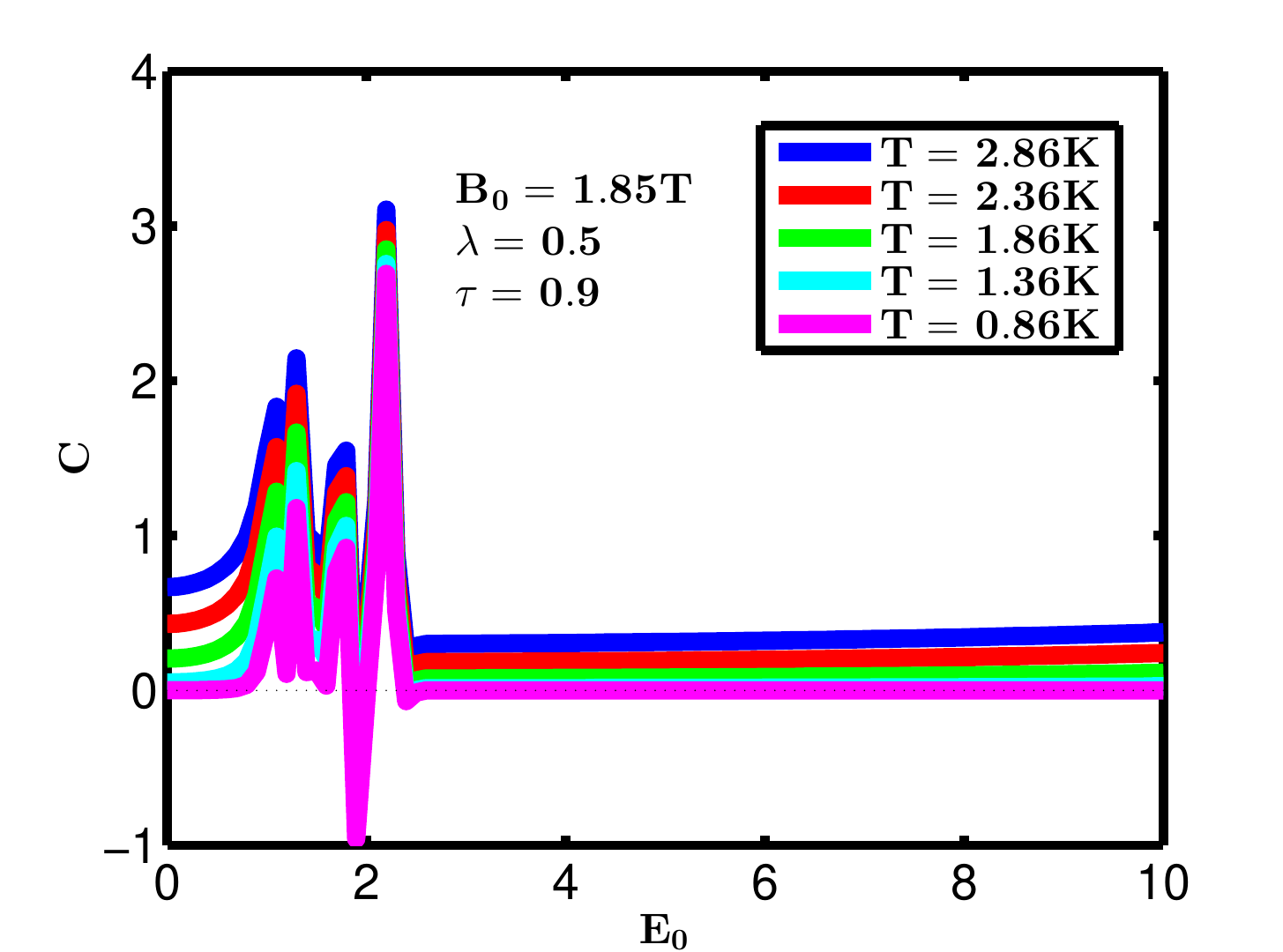}}
\subfigure[]{\includegraphics[width=0.35\textwidth]{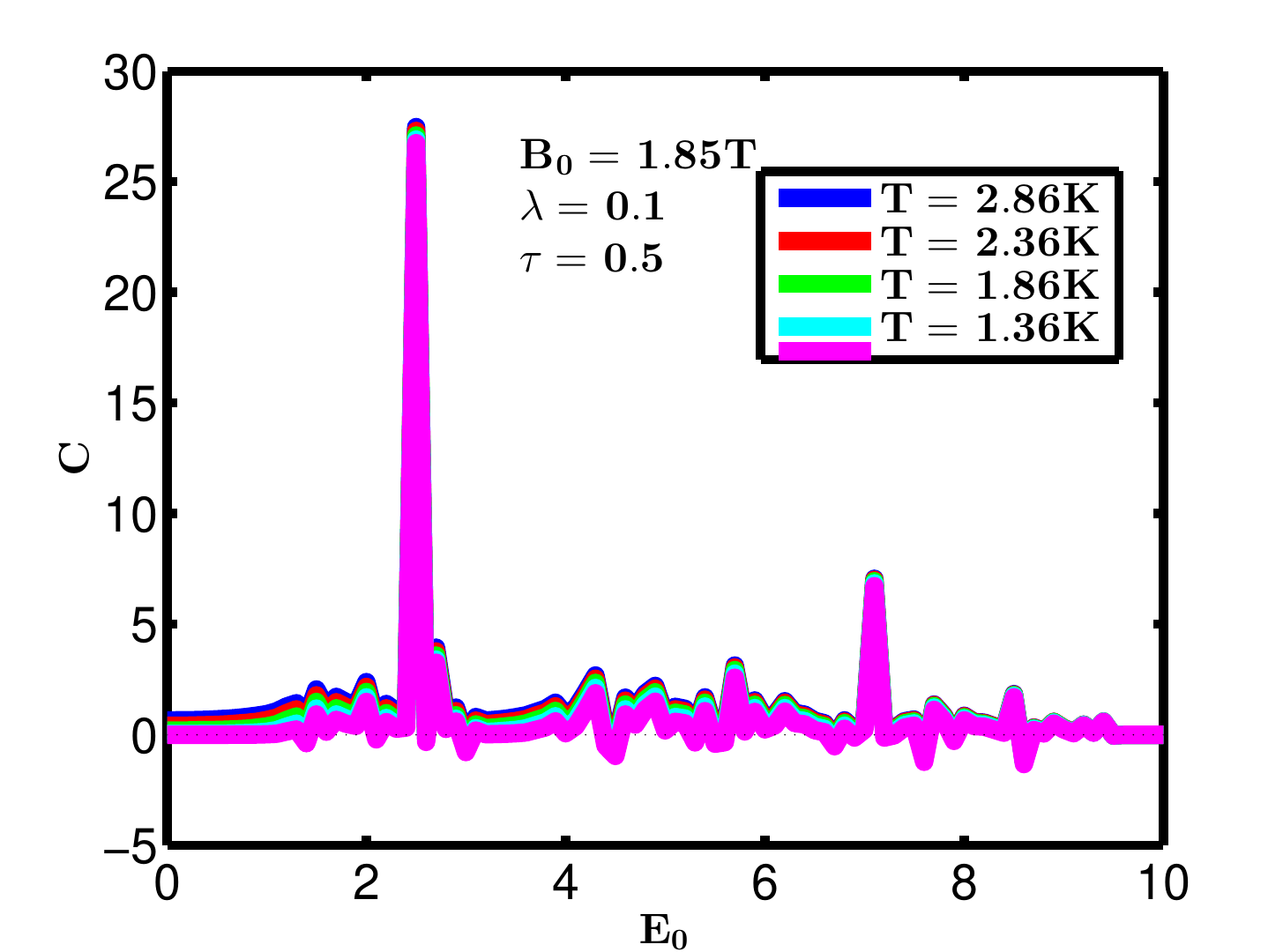}}
}
\caption{electric field dependence of  the specific heat capacity by varying the temperature with the following magnetic and electric site-dependent parameters: $\lambda = 0.1$, $\tau =0$ (a); $\lambda = 0$, $\tau =0.9$ (b); $\lambda = 0.5$, $\tau =0.5$ (c); $\lambda = 0.5$, $\tau =0.9$ (d) ; $\lambda = 0.1$, $\tau =0.5$ (e).}
\label{F26}
\end{figure}

 
Furthermore, the electric dependence of the specific heat capacity highlighted in Fig.~\eqref{F26} shows that the specific heat capacity firstly increases, then exhibits a quasi-oscillating-like behavior associated with peak points and finally freezes or becomes constant. This implies that the specific heat capacity of the system becomes constant when the electric field is strong enough, demonstrating that the system is in a thermal equilibrium state. It is worth noting that similar behaviors of specific heat capacity have been observed in Refs \cite{48,49}.       

\par
\par
\section{Adiabatic cooling rate}\label{S4}
In this section, the adiabatic cooling rates such as the adiabatic magnetic cooling rate, the adiabatic electric cooling rate, and adiabatic magnetoelectric cooling rate are evaluated. These parameters help to quantify the magnetocaloric effect, the electrocaloric effect, and the magneto-electrocaloric effect respectively. In fact, the adiabatic cooling rate can be defined clearly as an adiabatic temperature change of a system as a response of a suitable applied external field. From the Maxwell equations and the basic thermodynamic equations, the adiabatic cooling rate in a nonlinear media is generalized as \cite{50},
\begin{equation} \label{23} 
\Gamma =-\frac{T}{C} \frac{\partial X_{i} }{\partial T}  
\end{equation} 
where $X_{i} $ is generalized displacement (magnetization, electric polarization, strain, etc.). Thus, the adiabatic magnetic, electric, and magnetoelectric cooling rates are derived as
\begin{equation} \label{24} 
\begin{split}
\Gamma ^{m} &=\frac{\partial T}{\partial H} =-\frac{T}{C} \frac{\partial M}{\partial T}
\\&
=\frac{g\mu _{0} \mu _{B} }{TC} \left[
\begin{split}
&\sum _{k}\frac{\omega _{1 k}^{\left(+\right)} e^{-\frac{1}{T} \omega _{1 k}^{\left(+\right)} } }{\left(1-e^{-\frac{1}{T} \omega _{1 k}^{\left(+\right)} } \right)^{2} }  +\sum _{k}\frac{\omega _{2 k}^{\left(+\right)} e^{-\frac{1}{T} \omega _{2 k}^{\left(+\right)} } }{\left(1-e^{-\frac{1}{T} \omega _{2 k}^{\left(+\right)} } \right)^{2} }  -\sum _{k}\frac{\omega _{1 k}^{\left(-\right)} e^{-\frac{1}{T} \omega _{1 k}^{\left(-\right)} } }{\left(1-e^{-\frac{1}{T} \omega _{1 k}^{\left(-\right)} } \right)^{2} }  
\\&
-\sum _{k}\frac{\omega _{2 k}^{\left(-\right)} e^{-\frac{1}{T} \omega _{2 k}^{\left(-\right)} } }{\left(1-e^{-\frac{1}{T} \omega _{2 k}^{\left(-\right)} } \right)^{2} }
\end{split}  \right]
\end{split}
\end{equation} 
\begin{equation} \label{25} 
\begin{split}
&\Gamma ^{e} =\frac{\partial T}{\partial E} =-\frac{T}{C} \frac{\partial P}{\partial T}
\\&
=-\frac{2MS}{JTC} \left[
\begin{split}
&
\sum _{k}\frac{\gamma _{k}^{2} E_{+} \omega _{1 k}^{\left(+\right)} e^{-\frac{1}{T} \omega _{1 k}^{\left(+\right)} } }{\xi _{+}^{1/2} \left(1-e^{-\frac{1}{T} \omega _{1 k}^{\left(+\right)} } \right)^{2} }  +\sum _{k}\frac{\gamma _{k}^{2} E_{-} \omega _{2 k}^{\left(+\right)} e^{-\frac{1}{T} \omega _{2 k}^{\left(+\right)} } }{\xi _{-}^{1/2} \left(1-e^{-\frac{1}{T} \omega _{2 k}^{\left(+\right)} } \right)^{2} }  +\sum _{k}\frac{\gamma _{k}^{2} E_{+} \omega _{1 k}^{\left(-\right)} e^{-\frac{1}{T} \omega _{1 k}^{\left(-\right)} } }{\xi _{+}^{1/2} \left(1-e^{-\frac{1}{T} \omega _{1 k}^{\left(-\right)} } \right)^{2} }  
\\&
+\sum _{k}\frac{\gamma _{k}^{2} E_{-} \omega _{2 k}^{\left(-\right)} e^{-\frac{1}{T} \omega _{2 k}^{\left(-\right)} } }{\xi _{-}^{1/2} \left(1-e^{-\frac{1}{T} \omega _{2 k}^{\left(-\right)} } \right)^{2} } 
\end{split} \right]
\end{split}
\end{equation} 
and
\begin{equation} \label{26}
\begin{split}
&\Gamma ^{me} =\frac{\partial T}{\partial H\partial E} =-\frac{T}{C} \frac{\partial \alpha }{\partial T}
\\&
 =-\frac{2g\mu _{0} \mu _{B} MS}{JTC} \left[
\begin{split}
 &
 \sum _{k}\frac{\gamma _{k}^{2} E_{+} \omega _{1 k}^{\left(+\right)} e^{-\frac{1}{T} \omega _{1 k}^{\left(+\right)} } }{\xi _{+}^{1/2} \left(1-e^{-\frac{1}{T} \omega _{1 k}^{\left(+\right)} } \right)^{2} }  +\sum _{k}\frac{\gamma _{k}^{2} E_{-} \omega _{2 k}^{\left(+\right)} e^{-\frac{1}{T} \omega _{2 k}^{\left(+\right)} } }{\xi _{-}^{1/2} \left(1-e^{-\frac{1}{T} \omega _{2 k}^{\left(+\right)} } \right)^{2} }  
 \\&
 -\sum _{k}\frac{\gamma _{k}^{2} E_{+} \omega _{1 k}^{\left(-\right)} e^{-\frac{1}{T} \omega _{1 k}^{\left(-\right)} } }{\xi _{+}^{1/2} \left(1-e^{-\frac{1}{T} \omega _{1 k}^{\left(-\right)} } \right)^{2} }  -\sum _{k}\frac{\gamma _{k}^{2} E_{-} \omega _{2 k}^{\left(-\right)} e^{-\frac{1}{T} \omega _{2 k}^{\left(-\right)} } }{\xi _{-}^{1/2} \left(1-e^{-\frac{1}{T} \omega _{2 k}^{\left(-\right)} } \right)^{2} }
\end{split}  \right]
 \\&
 +\frac{2g\mu _{0} \mu _{B} MS}{JT^{2} C} \left[\sum _{k}\frac{\gamma _{k}^{2} E_{+} \omega _{1 k}^{\left(+\right)} e^{-\frac{1}{T} \omega _{1 k}^{\left(+\right)} } \left(1+e^{-\frac{1}{T} \omega _{1 k}^{\left(+\right)} } \right)}{\xi _{+}^{1/2} \left(1-e^{-\frac{1}{T} \omega _{1 k}^{\left(+\right)} } \right)^{3} }  +\sum _{k}\frac{\gamma _{k}^{2} E_{-} \omega _{2 k}^{\left(+\right)} e^{-\frac{1}{T} \omega _{2 k}^{\left(+\right)} } \left(1+e^{-\frac{1}{T} \omega _{2 k}^{\left(+\right)} } \right)}{\xi _{-}^{1/2} \left(1-e^{-\frac{1}{T} \omega _{2 k}^{\left(+\right)} } \right)^{3} }  \right]
 \\&
 -\frac{2g\mu _{0} \mu _{B} MS}{JT^{2} C} \left[\sum _{k}\frac{\gamma _{k}^{2} E_{+} \omega _{1 k}^{\left(-\right)} e^{-\frac{1}{T} \omega _{1 k}^{\left(-\right)} } \left(1+e^{-\frac{1}{T} \omega _{1 k}^{\left(-\right)} } \right)}{\xi _{+}^{1/2} \left(1-e^{-\frac{1}{T} \omega _{1 k}^{\left(-\right)} } \right)^{3} }  +\sum _{k}\frac{\gamma _{k}^{2} E_{-} \omega _{2 k}^{\left(-\right)} e^{-\frac{1}{T} \omega _{2 k}^{\left(-\right)} } \left(1+e^{-\frac{1}{T} \omega _{2 k}^{\left(-\right)} } \right)}{\xi _{-}^{1/2} \left(1-e^{-\frac{1}{T} \omega _{2 k}^{\left(-\right)} } \right)^{3} }  \right]
\end{split}
\end{equation}
respectively, where $C$ is the specific heat capacity at constant fields (Eq.~\eqref{22}), $M$ the magnetization (Eq.~\eqref{16}), $P$ the electric polarization (Eq.~\eqref{17}), $\alpha $ is the magnetoelectric polarizability (Eq.~\eqref{18}).
\subsection{Adiabatic magnetic and electric cooling rates}
\subsubsection{Adiabatic magnetic cooling rate}
\begin{figure}[!htb]
\centerline{
\subfigure[]{\includegraphics[width=0.35\textwidth]{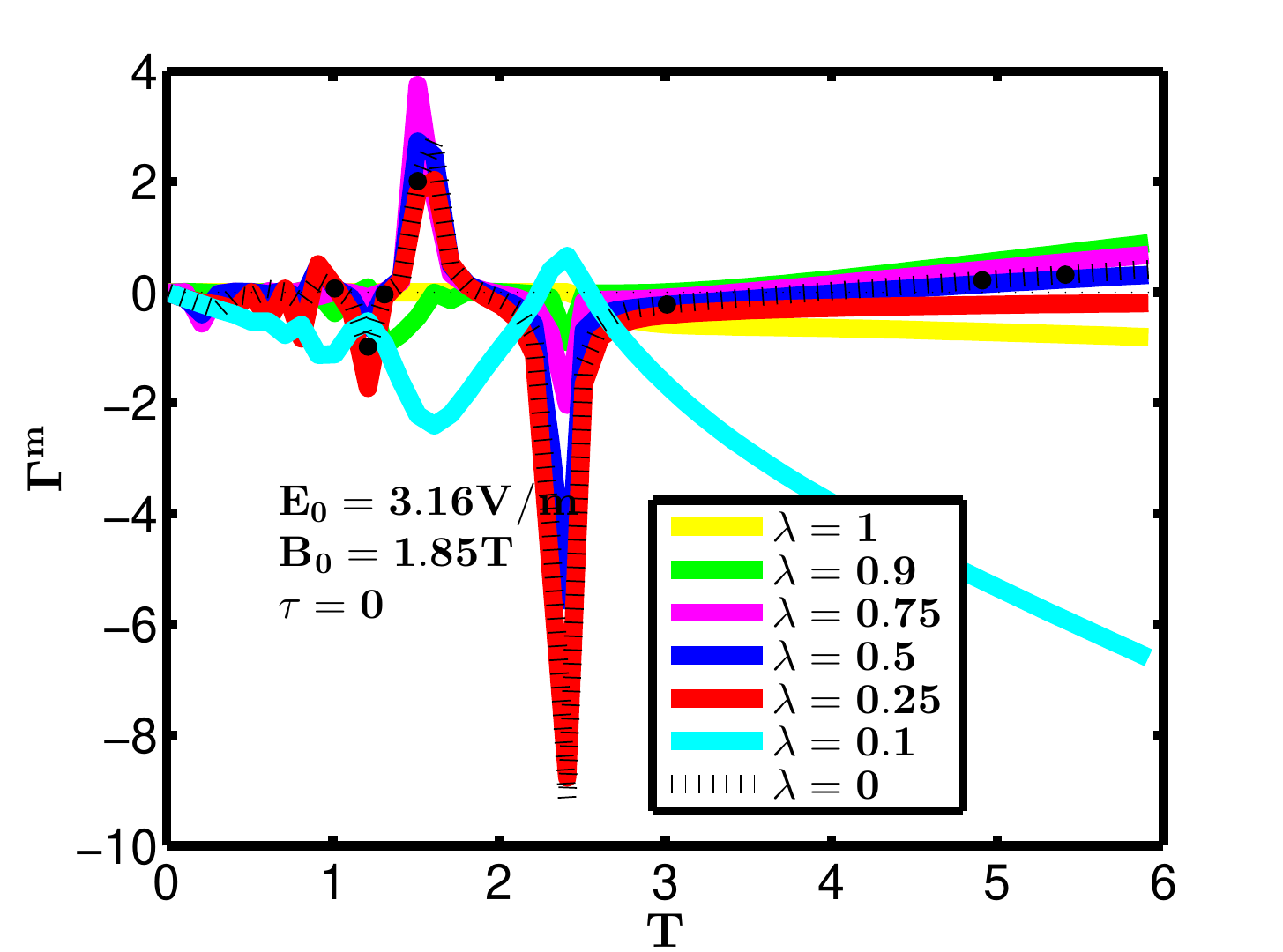}}
\subfigure[]{\includegraphics[width=0.35\textwidth]{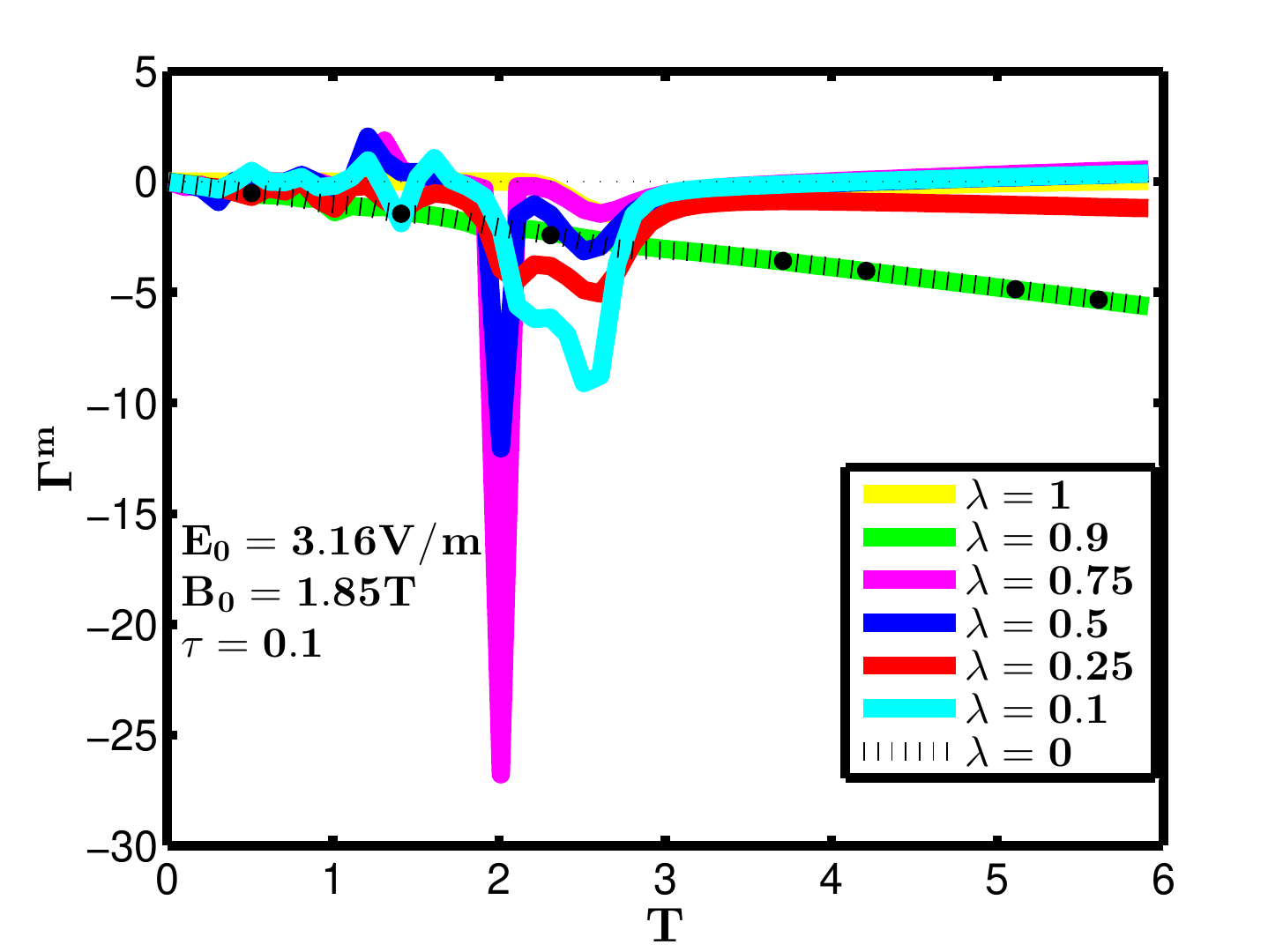}}
\subfigure[]{\includegraphics[width=0.35\textwidth]{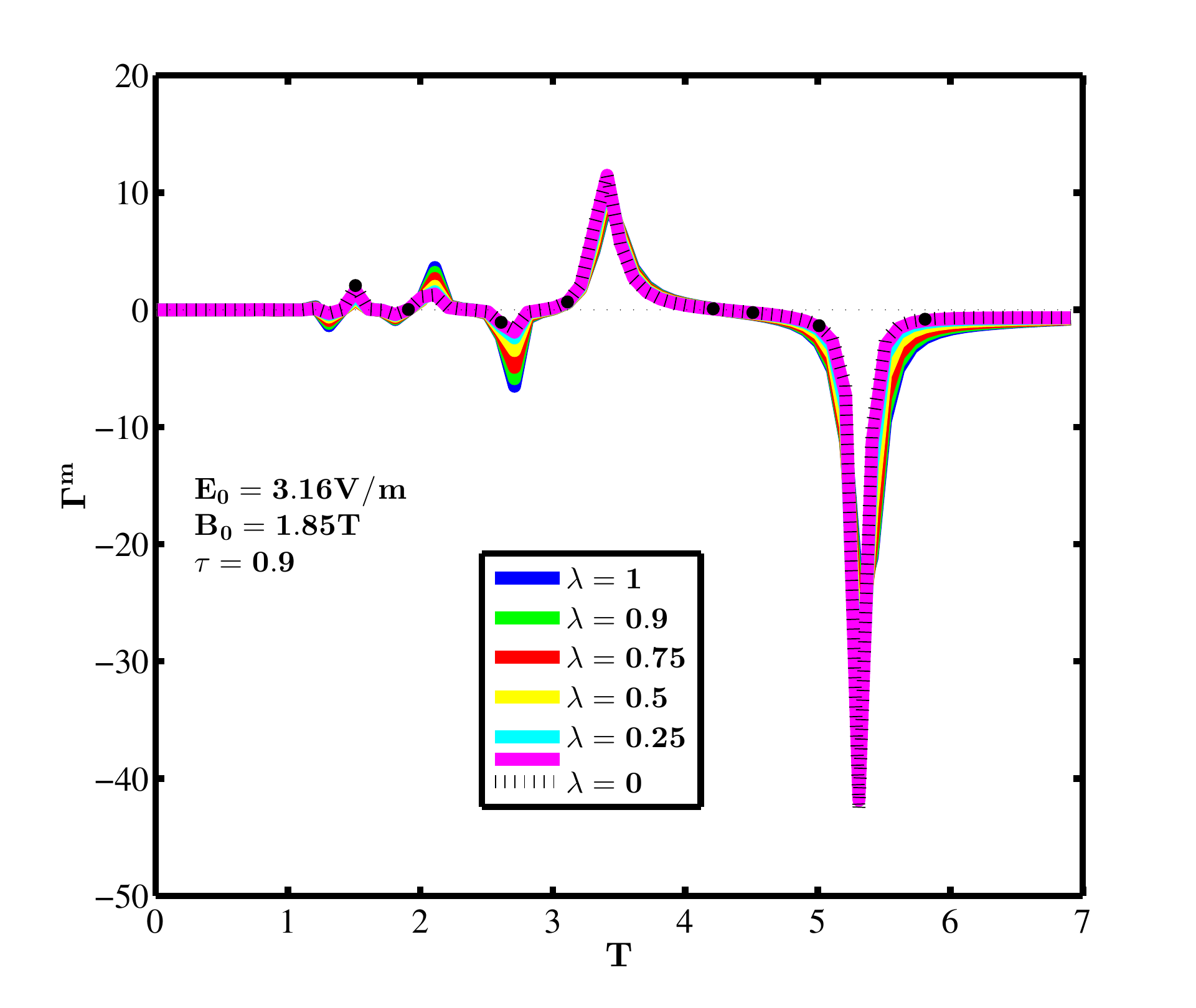}}  
}
\centerline{\subfigure[]{\includegraphics[width=0.35\textwidth]{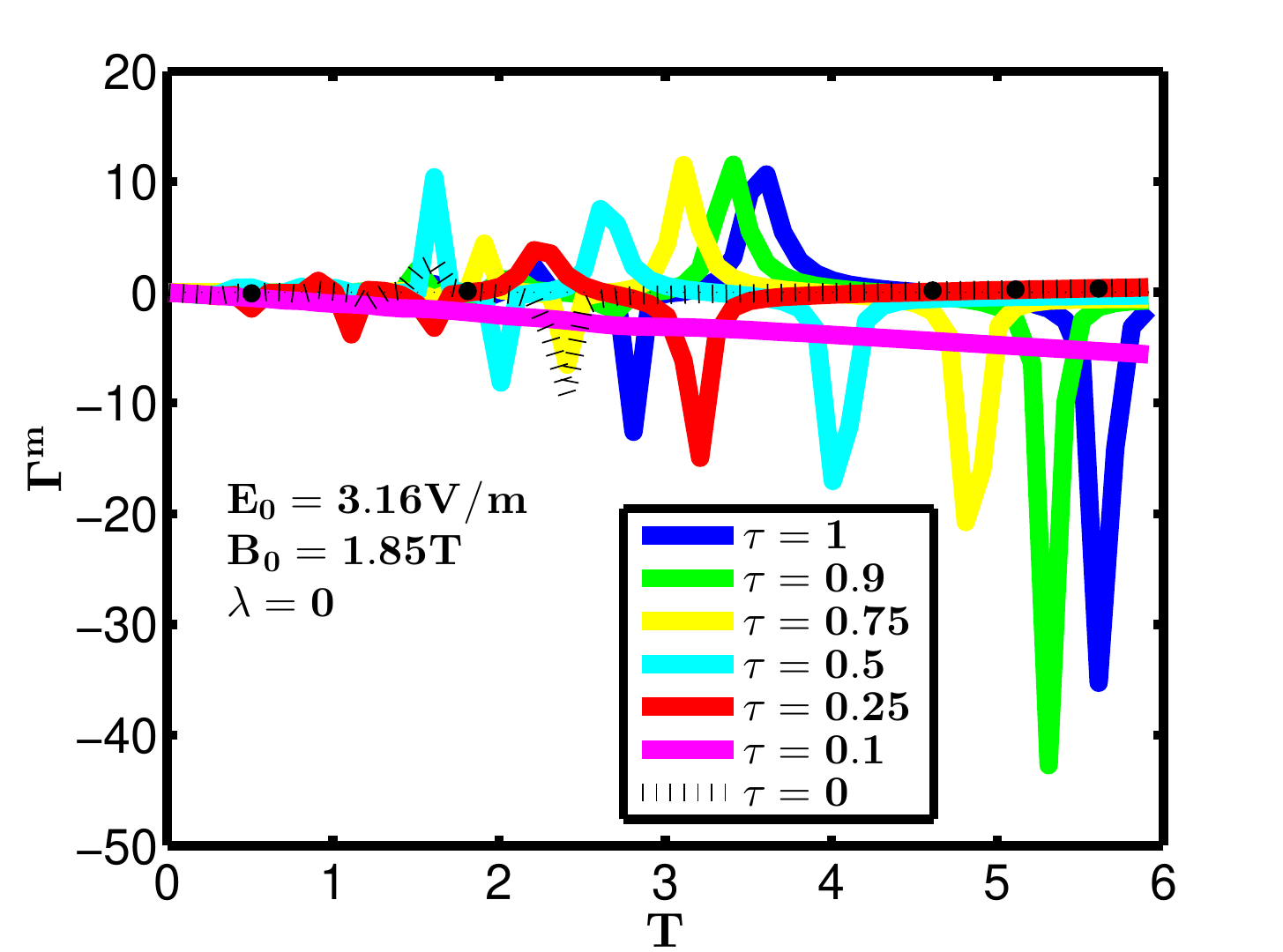}}
\subfigure[]{\includegraphics[width=0.35\textwidth]{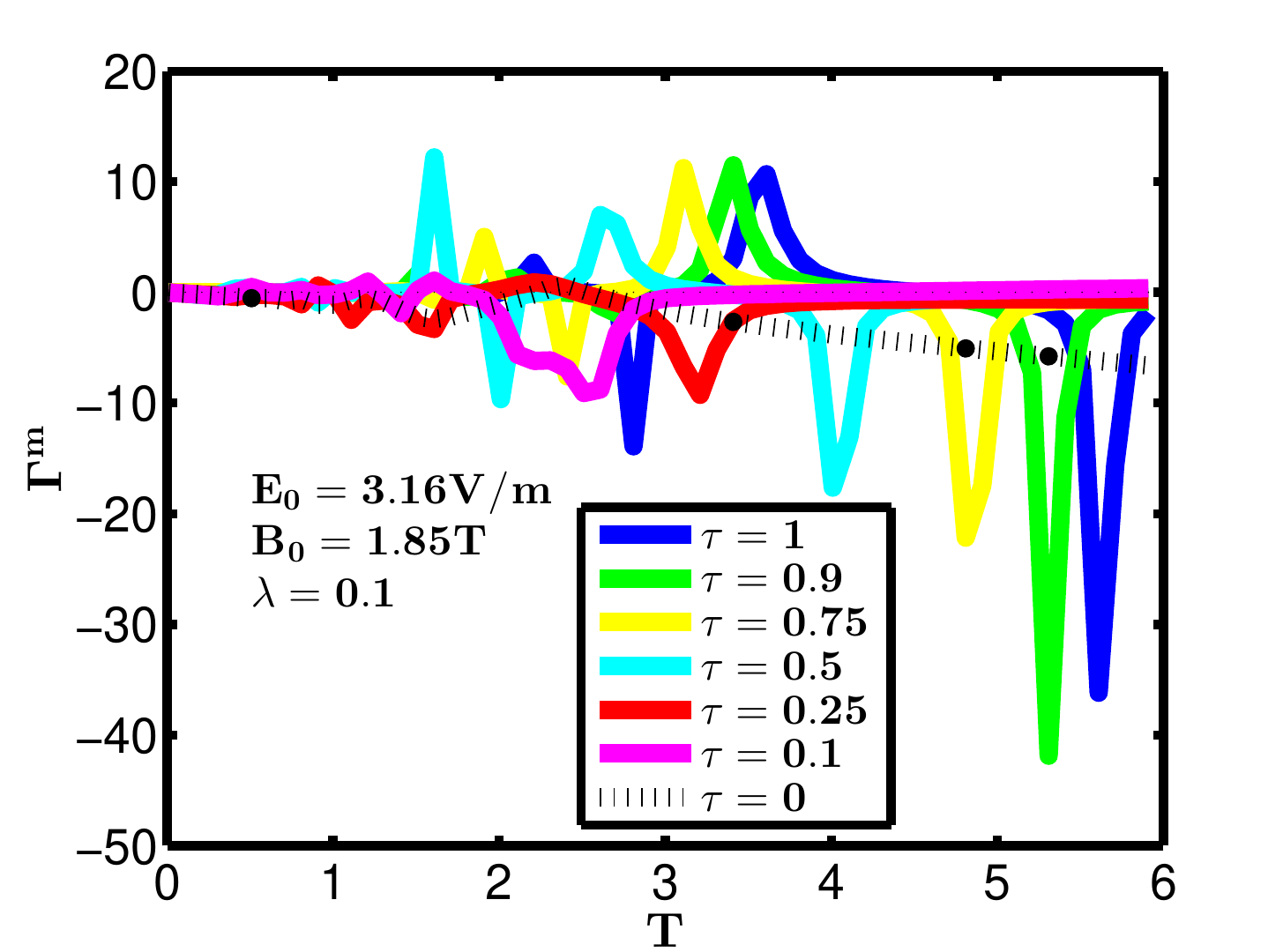}}
\subfigure[]{\includegraphics[width=0.35\textwidth]{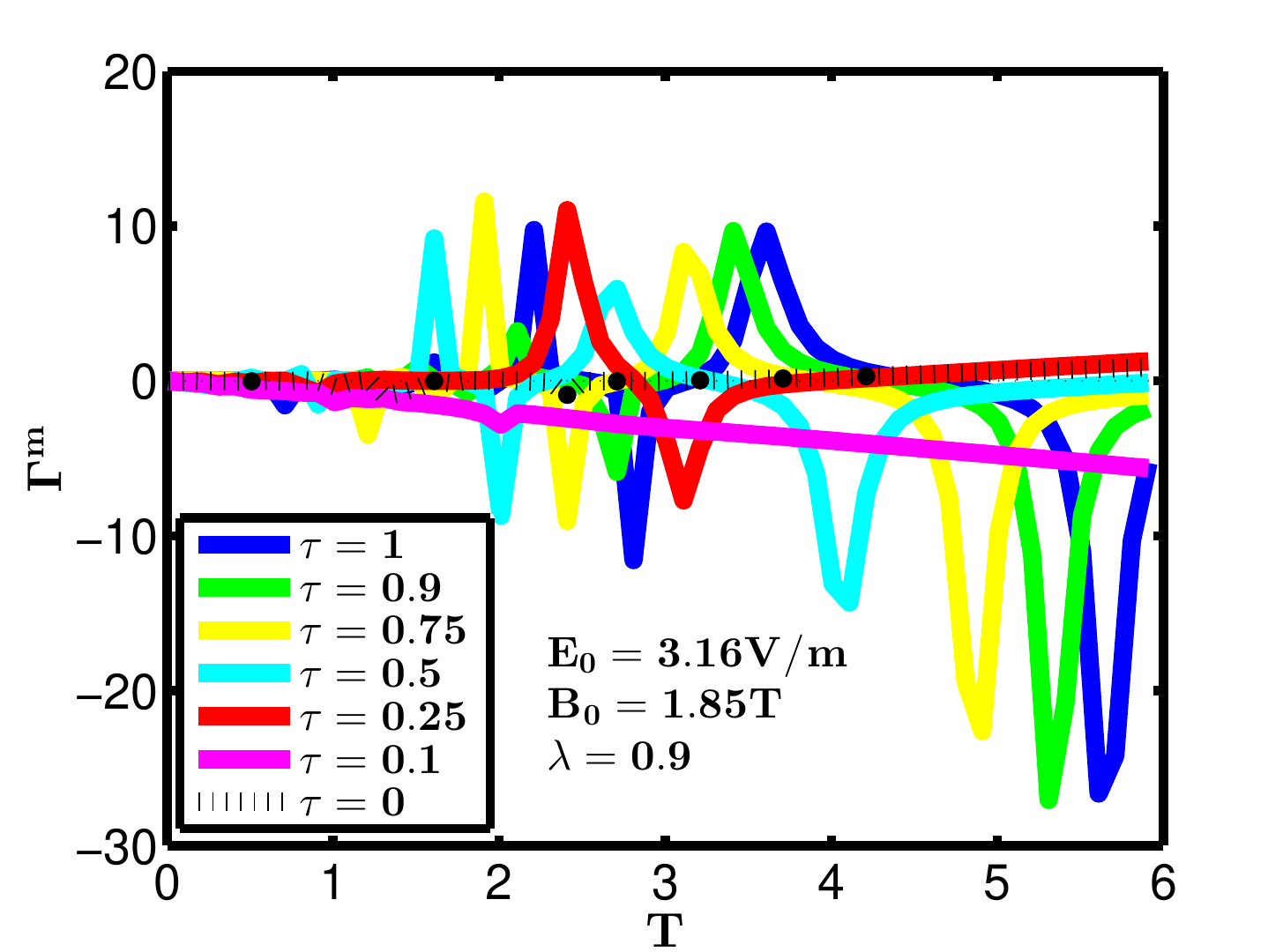}}
}
\caption{the upper panels show the evolution of adiabatic magnetic cooling rate of the system for different values of the magnetic site-dependent parameters and for three values of the electric site-dependent parameter namely 0 (a), 0.1 (b) and 0.9 (c). In the lower panels we plotted the evolution of adiabatic magnetic cooling rate for different values of the electric site-dependent parameters and for three values of the magnetic site-dependent parameter namely 0 (d), 0.5 (e) and 0.9 (f). }
\label{F31}
\end{figure}

 In Fig.~\eqref{F31}, the temperature response of the adiabatic magnetic cooling rate highlights that $\Gamma ^m$ exhibits alternating negative and positive peak points (in the region of low temperature) and then a linear increase or decrease with temperature (in the region of high temperature). It is observed that by varying the magnetic site-dependent parameter the peak points almost appear at the same values of temperature while they almost appear at different values of temperature when the electric site-dependent parameter varies. In the particular case in which the electric site-dependent parameter is equal to $0.9$ with different values of the magnetic site-dependent parameter, all the peaks with the same sign appear exactly at the same points, and the change points were $\Gamma ^{m} =0$ also occur at the same temperature see( Fig.~\eqref{F31}  panel (c)). Besides, for a certain combination of the magnetic and electric site-dependent parameters, there is a formation of two successive negative peaks (Fig.~\eqref{F31} panel (b)). These behaviors are in good accordance with the characteristic behaviors of magnetocaloric effects as obtained in the following experimental works \cite{51}. So it is important to notice that the positive peaks of the magnetic cooling rate demonstrates the heating of the system (normal magnetocaloric effect) whereas the negative ones signals the cooling of the system (inverse magneto caloric effect). 
\par
\begin{figure}[!htb]
\centerline{
\subfigure[]{\includegraphics[width=0.35\textwidth]{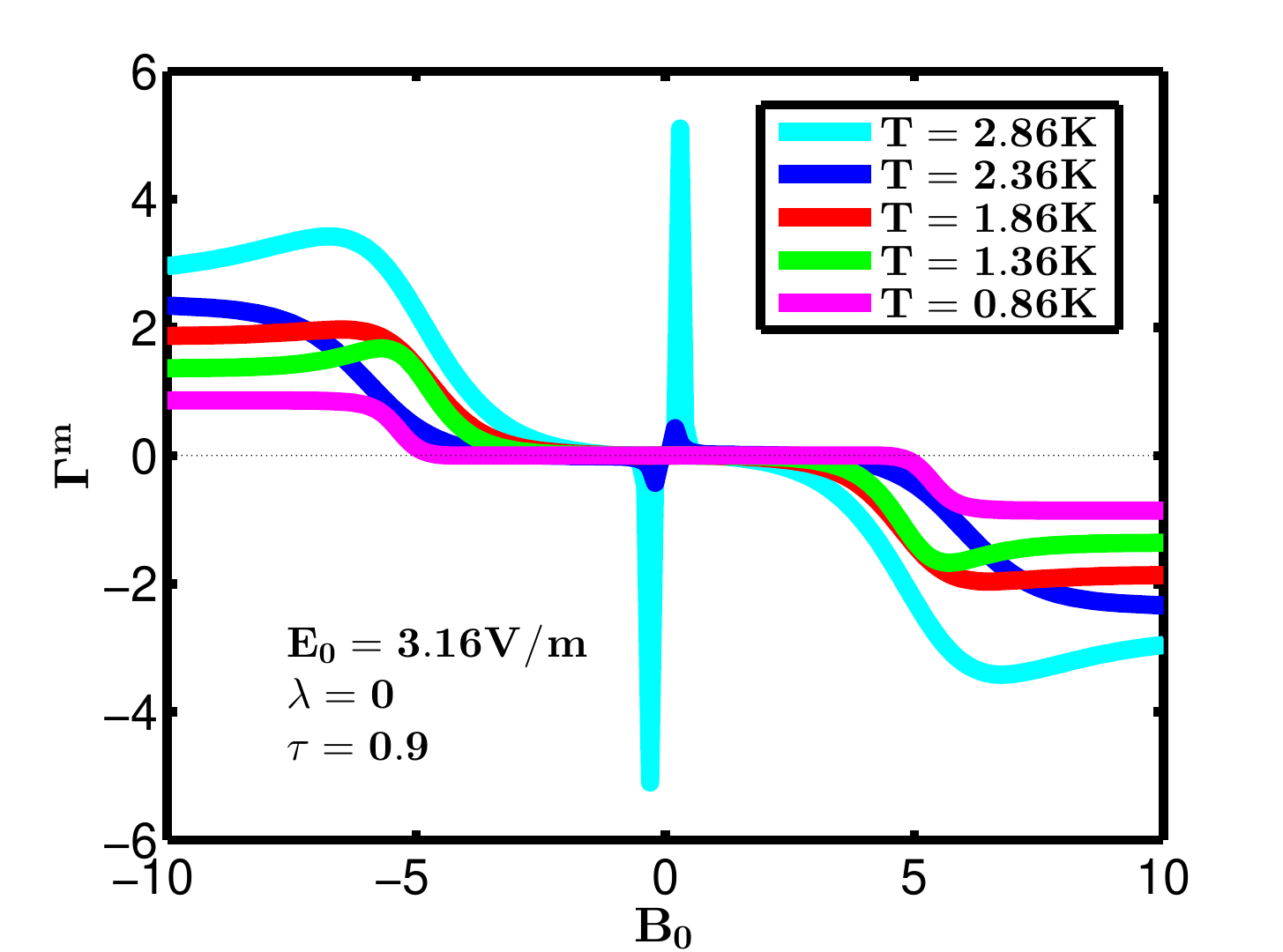}}
\subfigure[]{\includegraphics[width=0.35\textwidth]{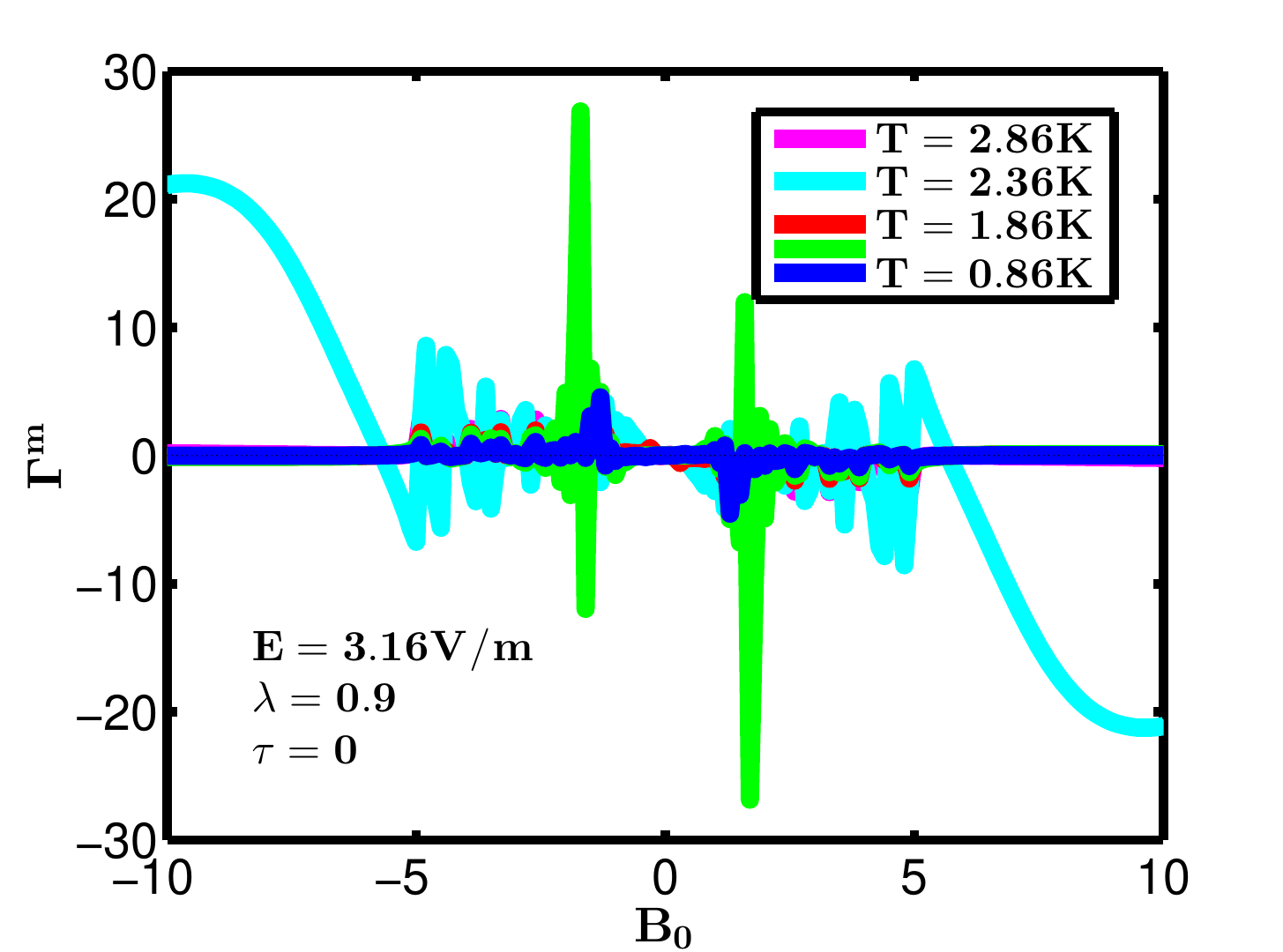}}
}
\centerline{\subfigure[]{\includegraphics[width=0.35\textwidth]{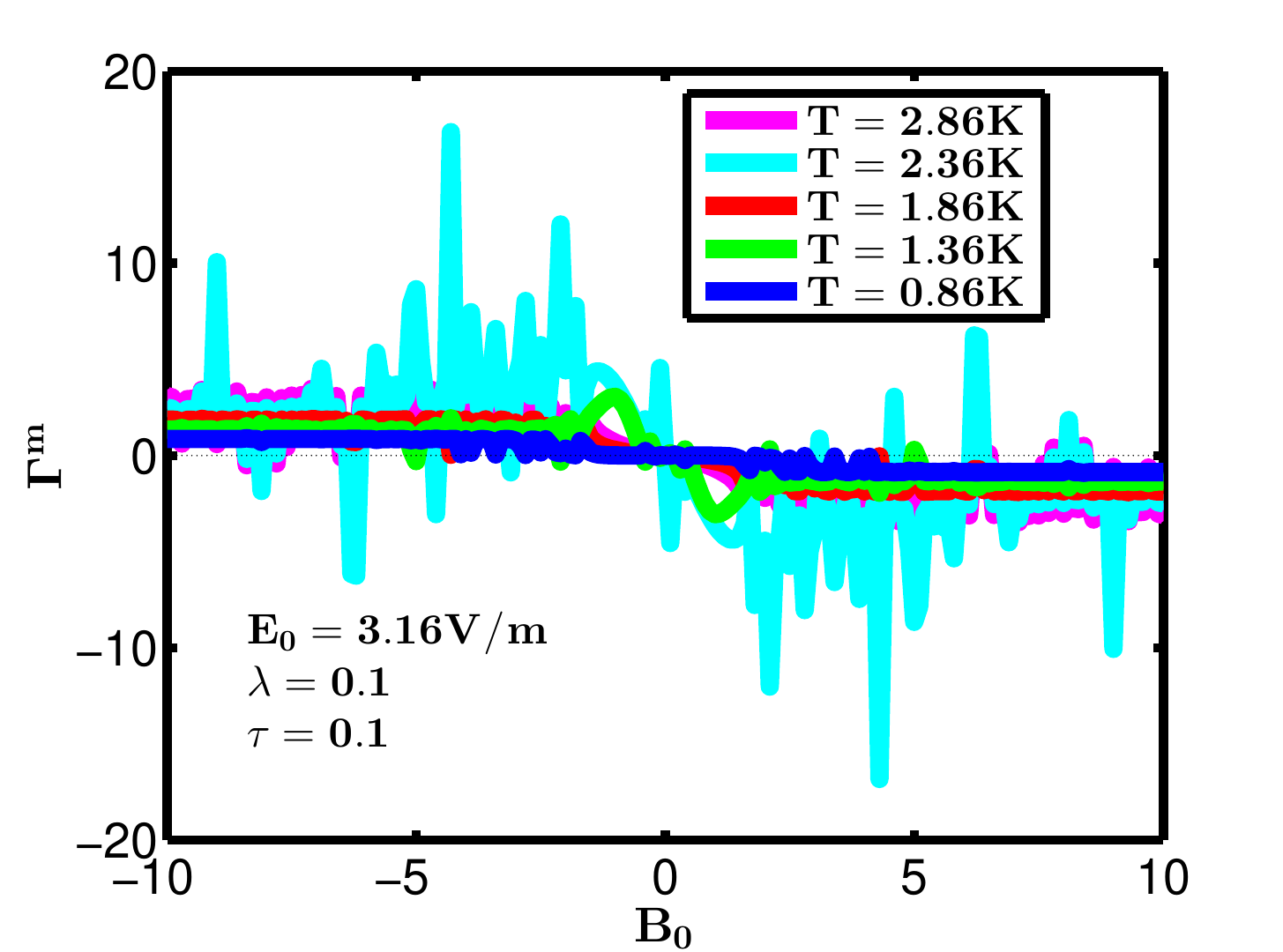}}
\subfigure[]{\includegraphics[width=0.35\textwidth]{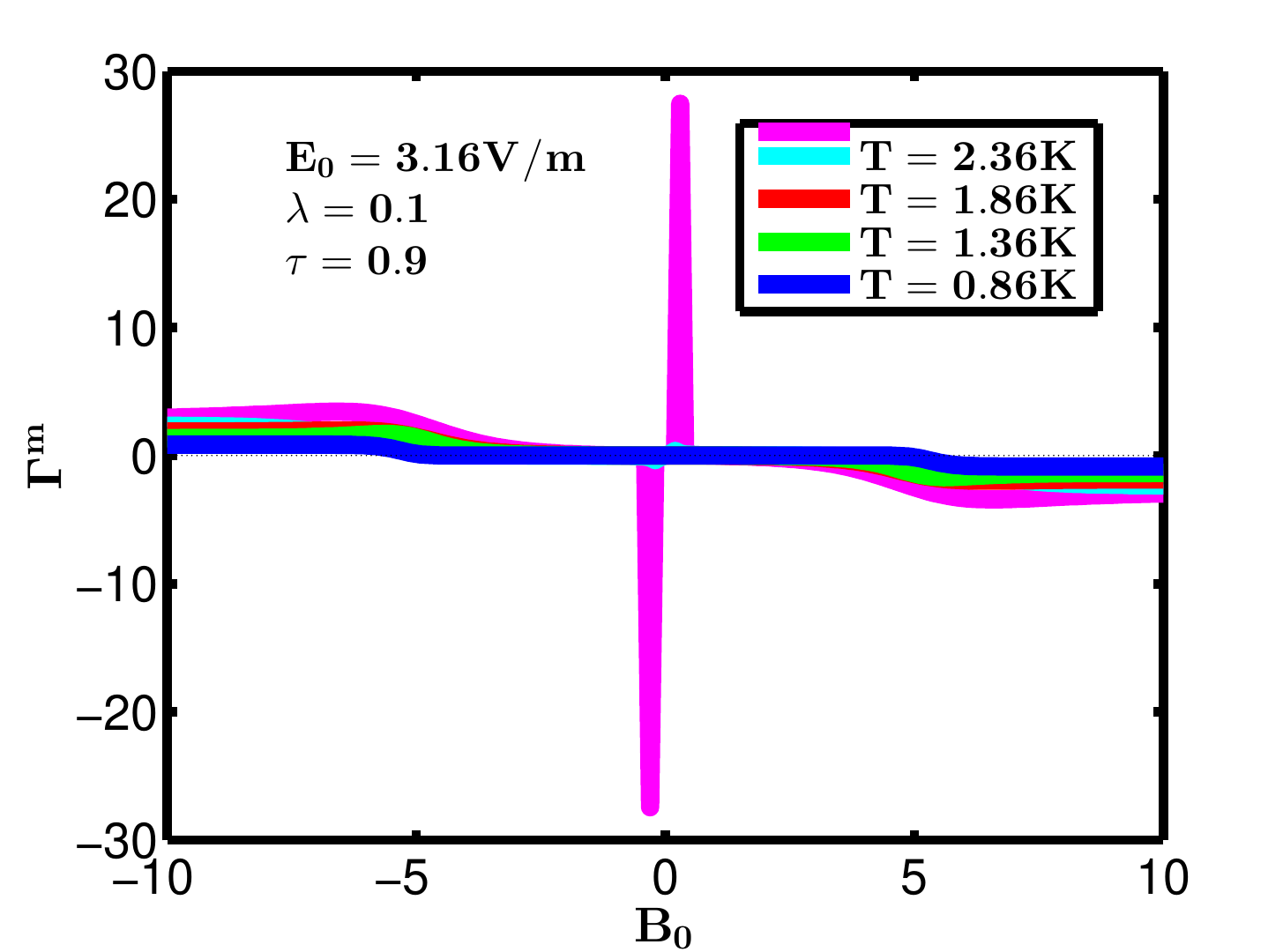}}
\subfigure[]{\includegraphics[width=0.35\textwidth]{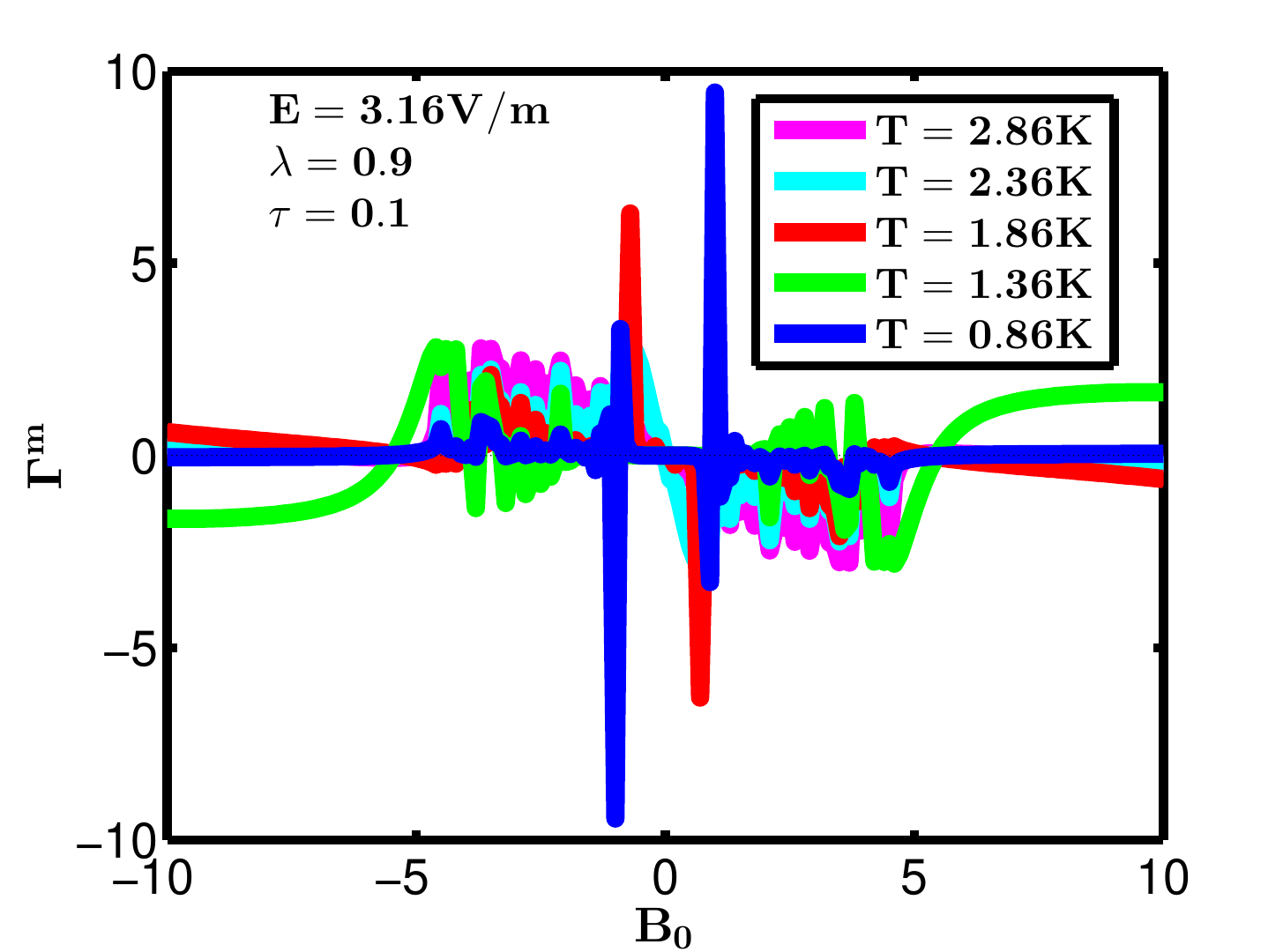}}
}
\caption{magnetic field dependence of the adiabatic magnetic cooling rate by varying the temperature with the following magnetic and electric site-dependent parameters: $\lambda = 0$, $\tau =0.9$ (a); $\lambda = 0.9$, $\tau =0$ (b); $\lambda = 0.1$, $\tau =0.1$ (c); $\lambda = 0.1$, $\tau =0.9$ (d) ; $\lambda = 0.9$, $\tau =0.1$ (e).}
\label{F32}
\end{figure}

\begin{figure}[!htb]
\centerline{
\subfigure[]{\includegraphics[width=0.35\textwidth]{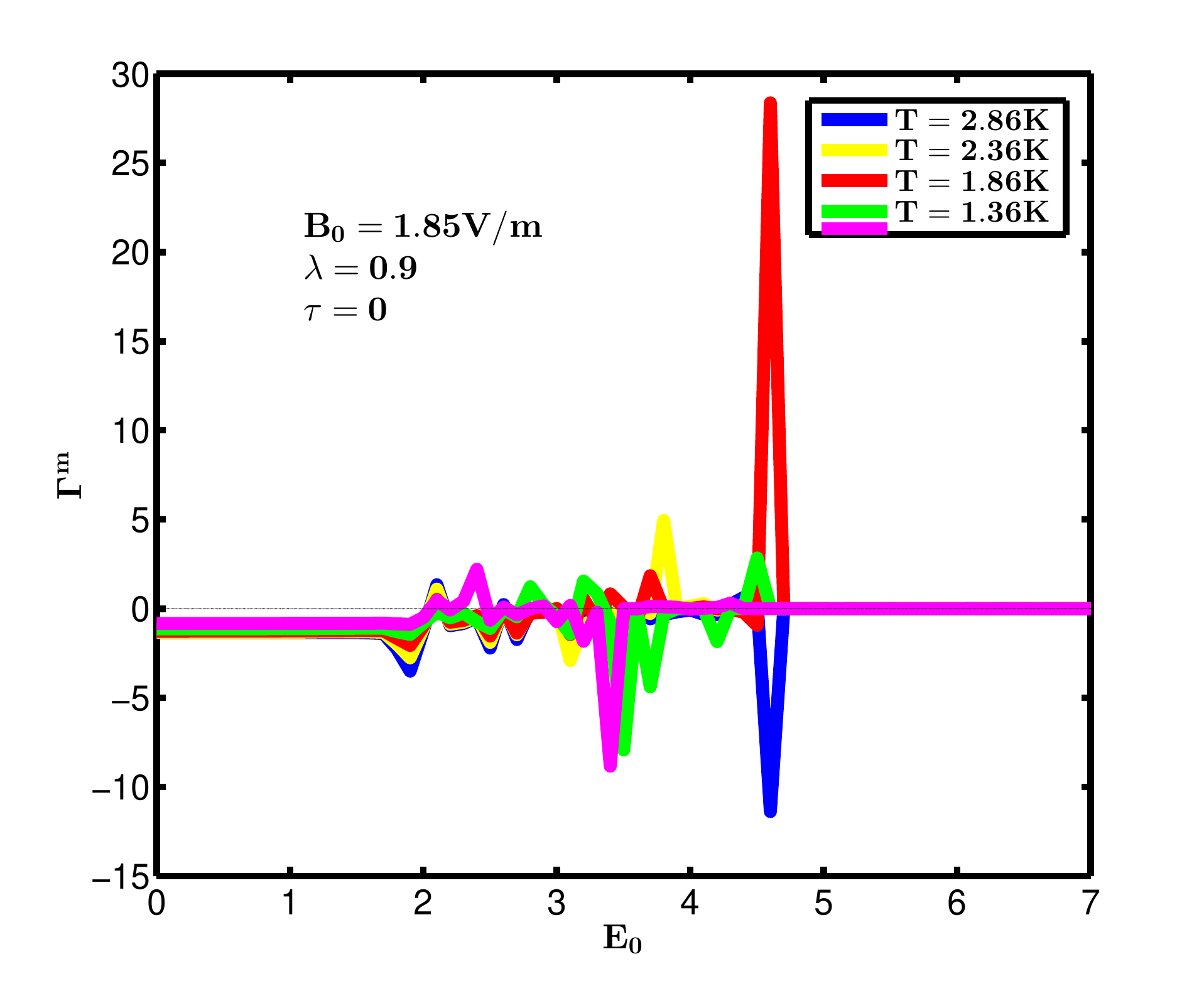}}
\subfigure[]{\includegraphics[width=0.35\textwidth]{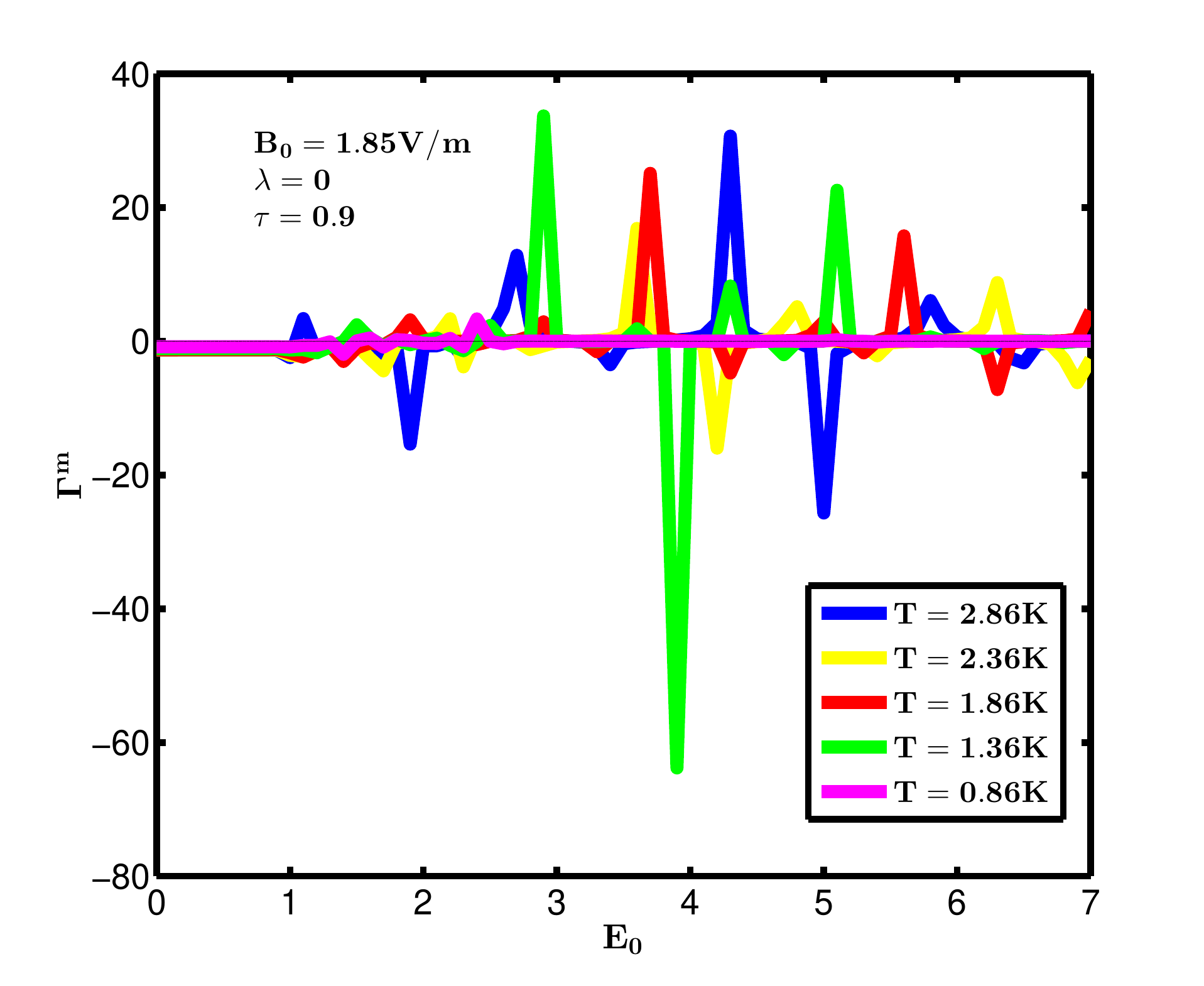}}
}
\centerline{\subfigure[]{\includegraphics[width=0.35\textwidth]{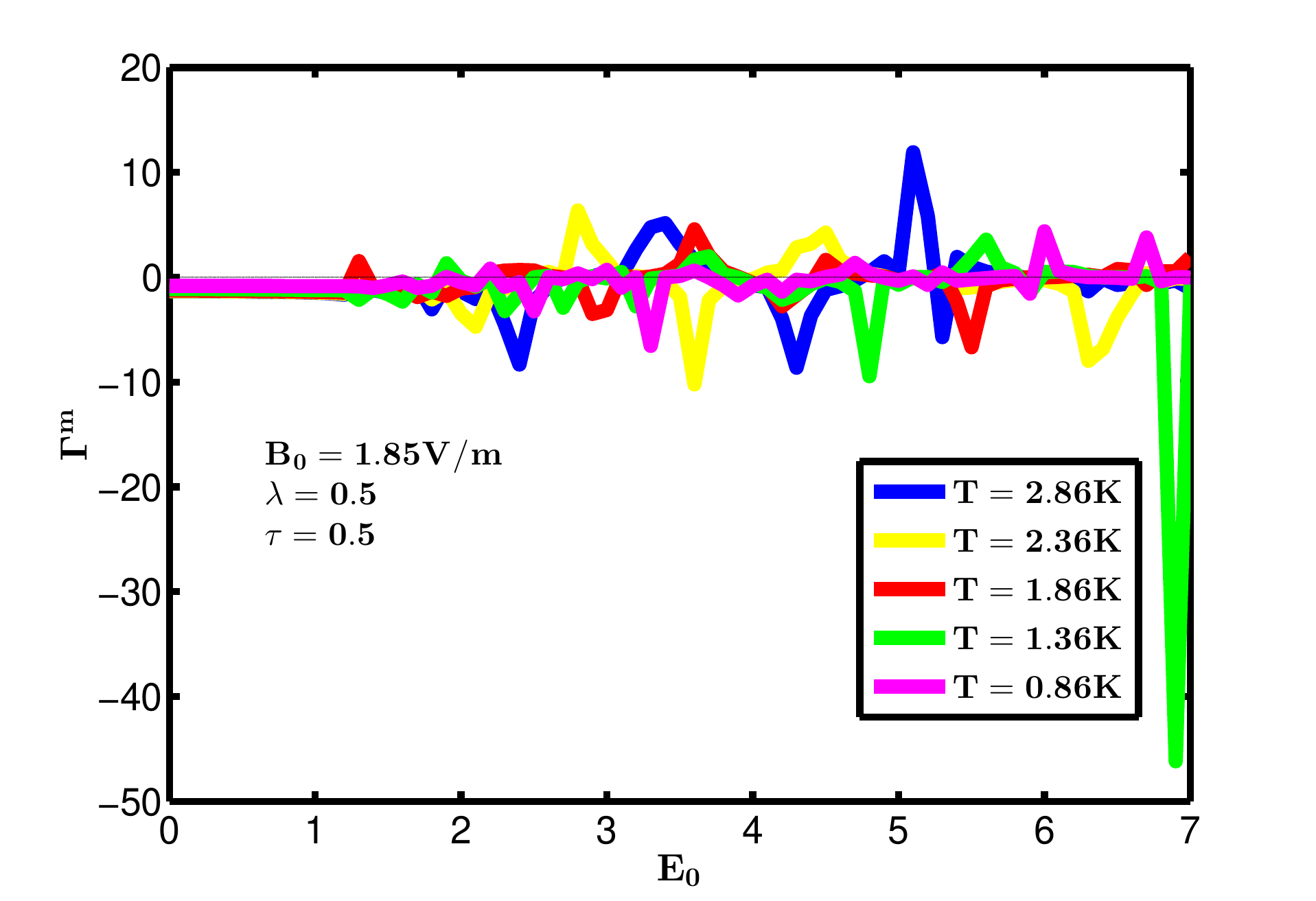}}
\subfigure[]{\includegraphics[width=0.35\textwidth]{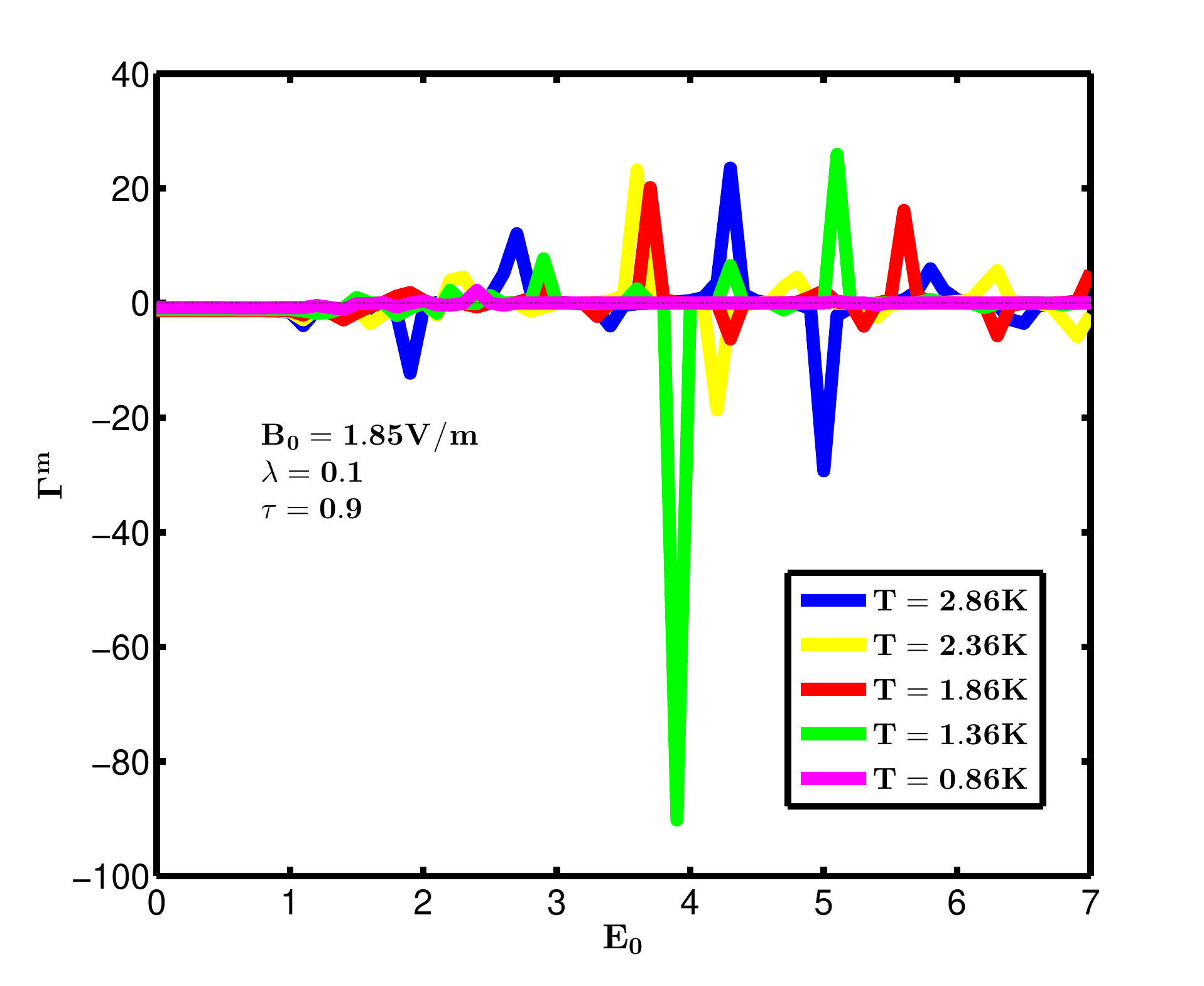}}
\subfigure[]{\includegraphics[width=0.35\textwidth]{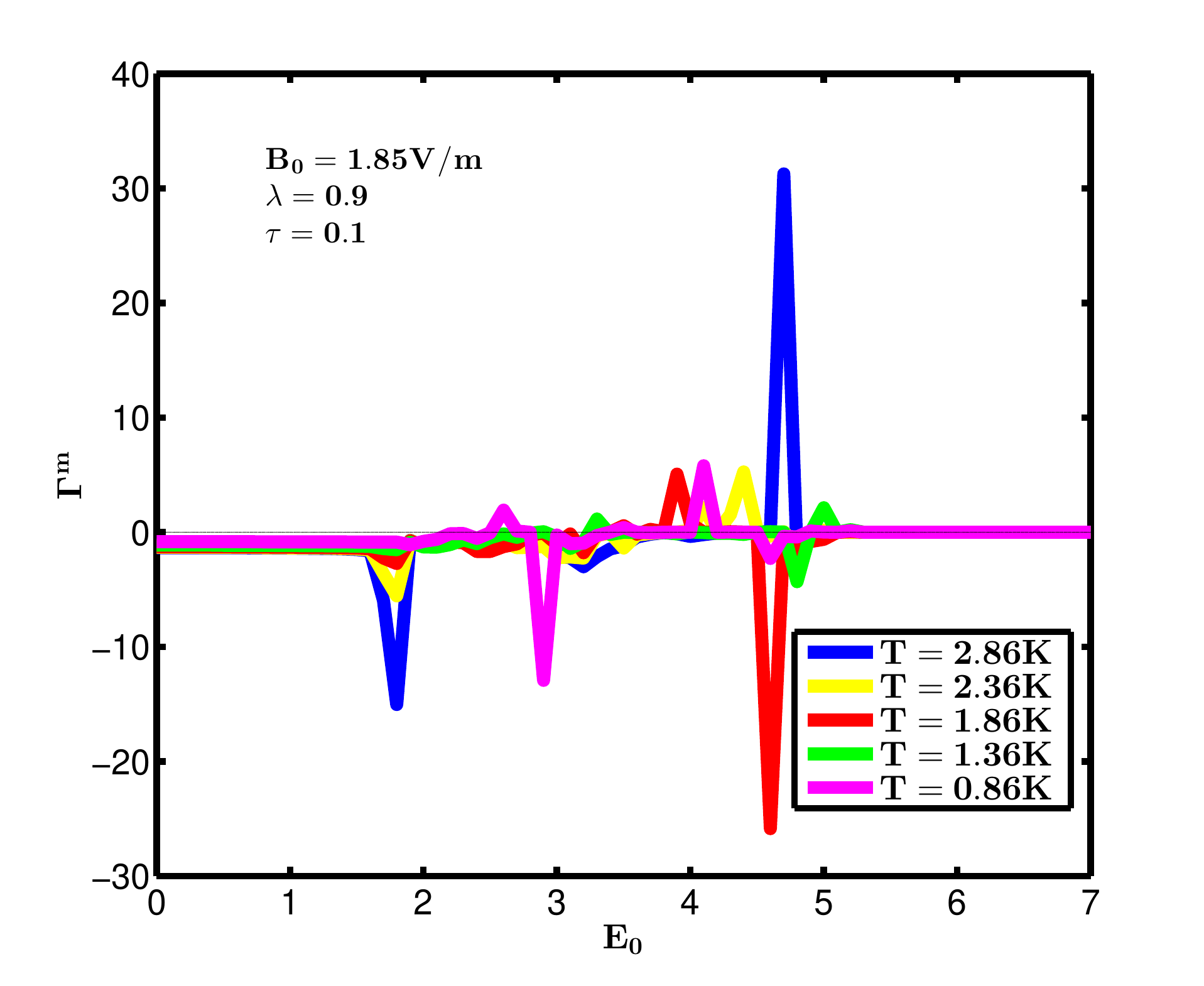}}
}
\caption{electric field dependence of  adiabatic magnetic cooling rate by varying the temperature with the following magnetic and electric site-dependent parameters: $\lambda = 0.9$, $\tau =0$ (a); $\lambda = 0$, $\tau =0.9$ (b); $\lambda = 0.5$, $\tau =0.5$ (c); $\lambda = 0.1$, $\tau =0.9$ (d) ; $\lambda = 0.9$, $\tau =0.1$ (e).}
\label{F33}
\end{figure}

  
Moreover, the magnetic field dependence of the adiabatic magnetic cooling rate is plotted in Fig.~\eqref{F32} for different values of the temperature. It is observed that $\Gamma ^m$ exhibits a peak point near to zero magnetic field followed in some cases by a damped oscillating peaks behavior. Indeed, in the particular case in which the electric site-dependent parameter is equal to $0.9$ the adiabatic magnetic cooling rate does not exhibit any oscillatory part. Furthermore, the electric field response of the adiabatic magnetic cooling rate as plotted in Fig.~\eqref{F33} for different values of temperature shows similar behavior as its temperature dependence (Fig.~\eqref{F31}). However, in contrary to Fig.~\eqref{F31} where the magnetic cooling rate saturates at values different from zero, here it is observed that the magnetic cooling rate always cancels (saturate at zero) after the alternating peaks part. 
\par
\begin{figure}[!htb]
	\centerline{\subfigure[]{\includegraphics[width=0.35\textwidth]{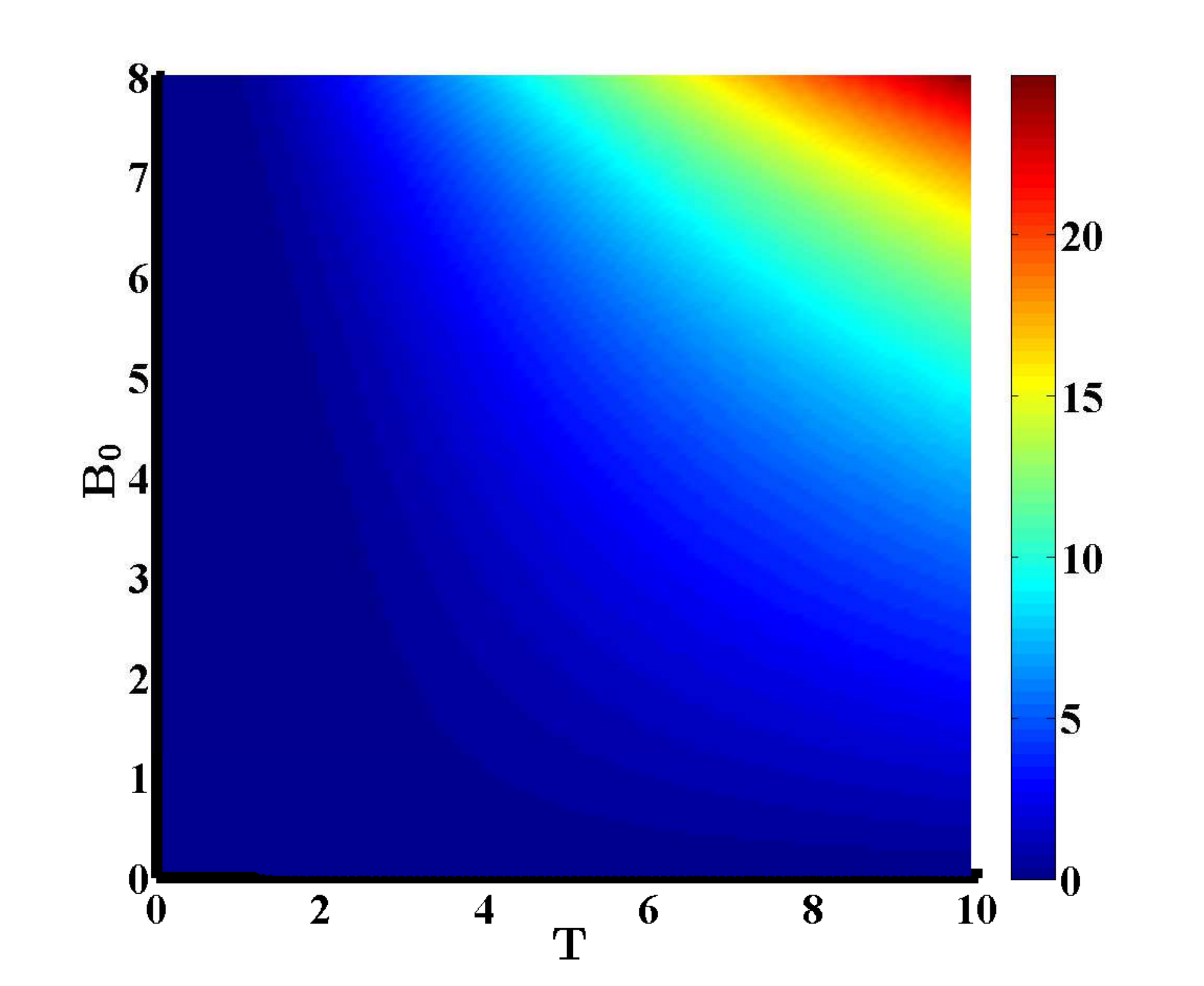}}
		\subfigure[]{\includegraphics[width=0.35\textwidth]{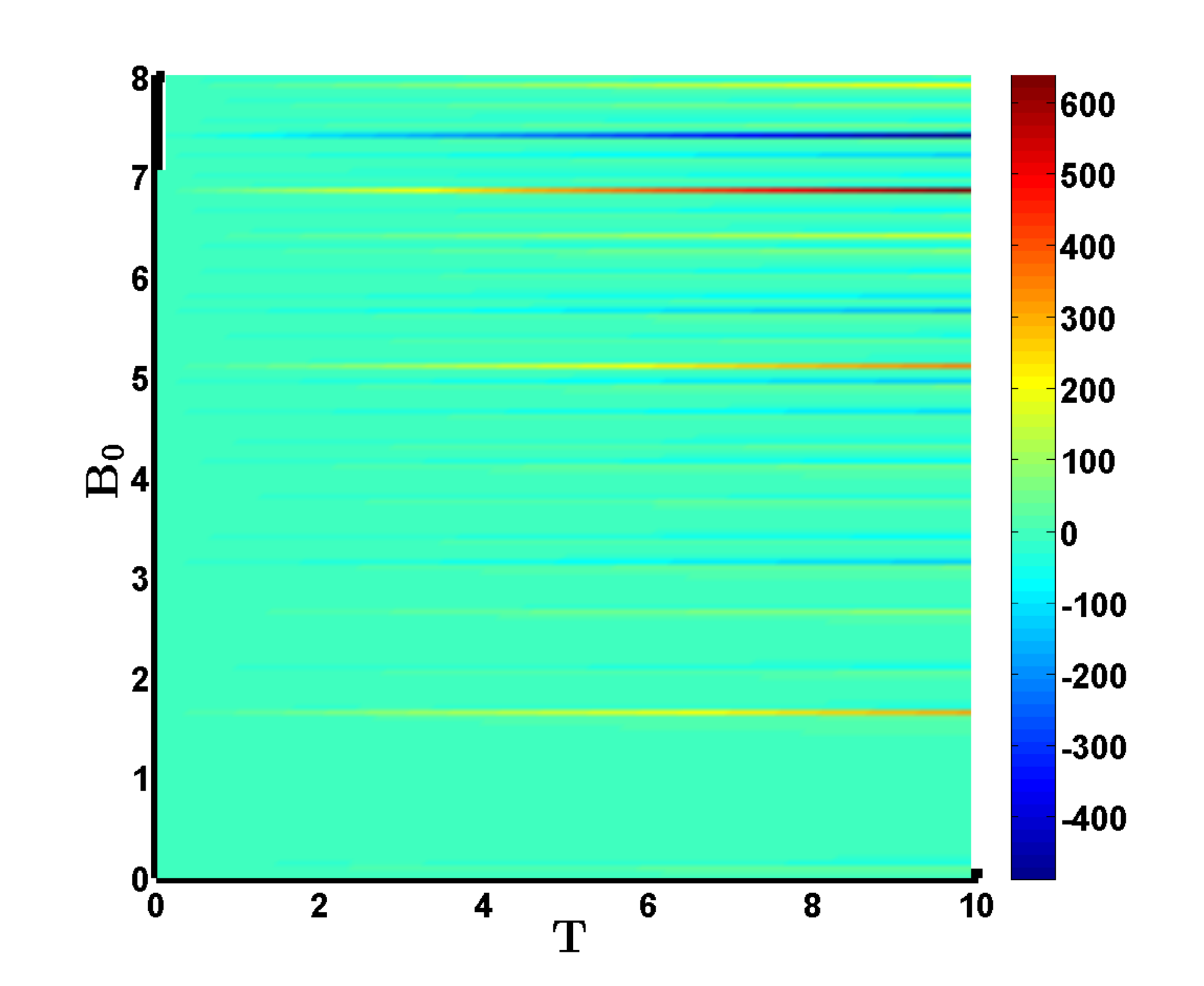}}
		\subfigure[]{\includegraphics[width=0.35\textwidth]{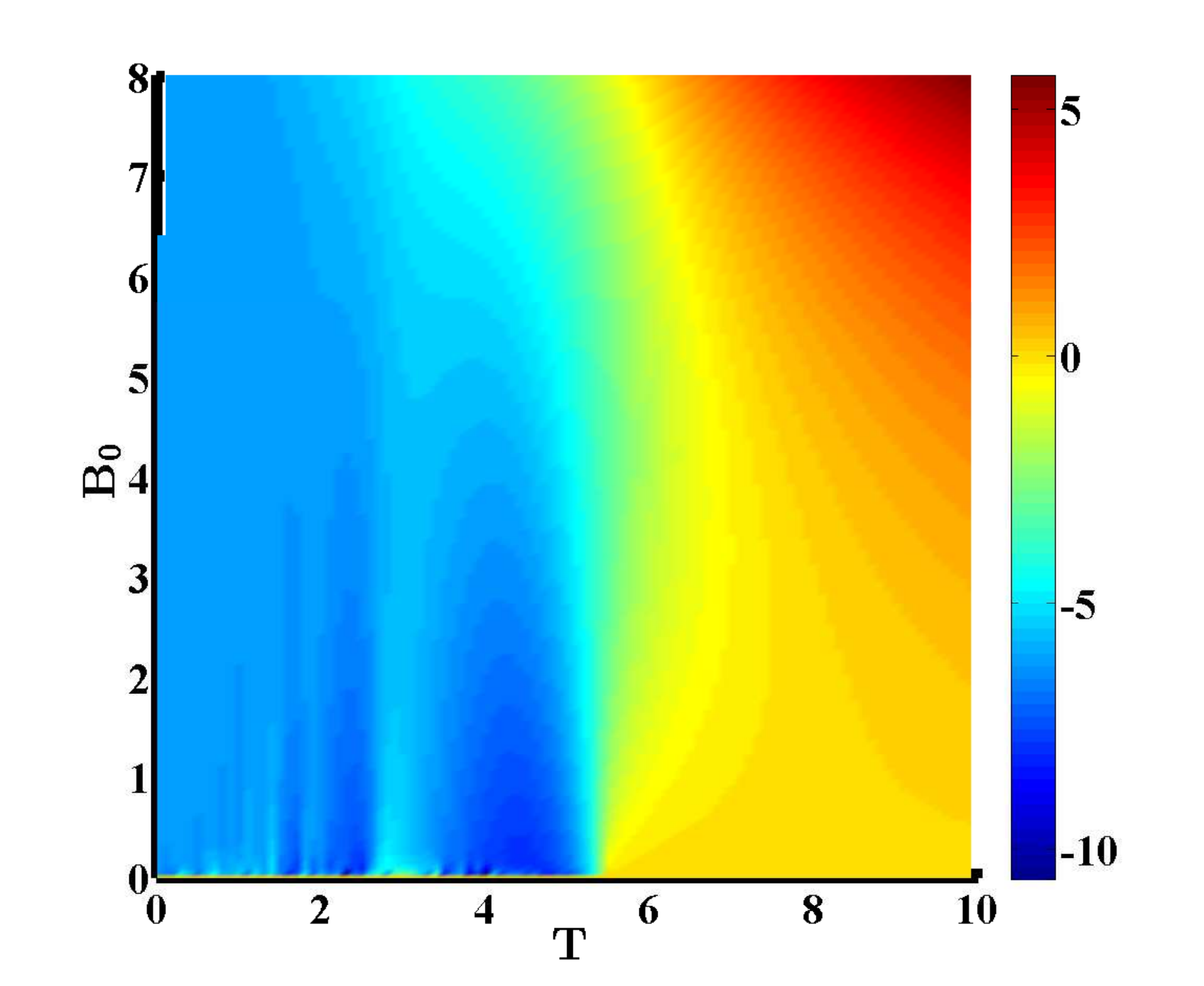}}
	}
	\caption{the surface plot of magnetization against magnetic field and temperature with the following electric field, magnetic and electric site-dependent parameters :(a)$E_0 =0 V/m, \lambda =  0.1$, (b) $ E_0 =3.16 V/m, \lambda = \tau =0$, (c) $E_0 =3.16 V/m, \lambda =0.1, \tau =  0.9$.  }
	\label{FO1}
\end{figure}

 Furthermore, the phase diagram in term of the magnetization obtained within the $B_0 - T$ plane, in  Fig.~\eqref{FO1}, shows three regions (see panel (c)) as function of approximated values of critical magnetic field and temperature: a region where the magnetization varies slightly with the temperature ($T\mathrm{<} 1 K$), an intermediate region where the magnetization oscillates with the temperature ($1 K \mathrm{<} T \mathrm{<} 6 K $) and a region where the magnetization increases both with the magnetic field and the temperature ($T \mathrm{>} 6 K$ ). Note that the critical magnetic field and the critical temperature are tunable by the magnetic and electric site-dependent parameters. Thus, the phase diagram in Fig. ~\eqref{FO1} summarizes the phase transition observed by interpreting the magnetic cooling rate graphs. 

\par
\par
\par
\subsubsection{Adiabatic electric cooling rate}
\begin{figure}[!htb]
\centerline{
\subfigure[]{\includegraphics[width=0.35\textwidth]{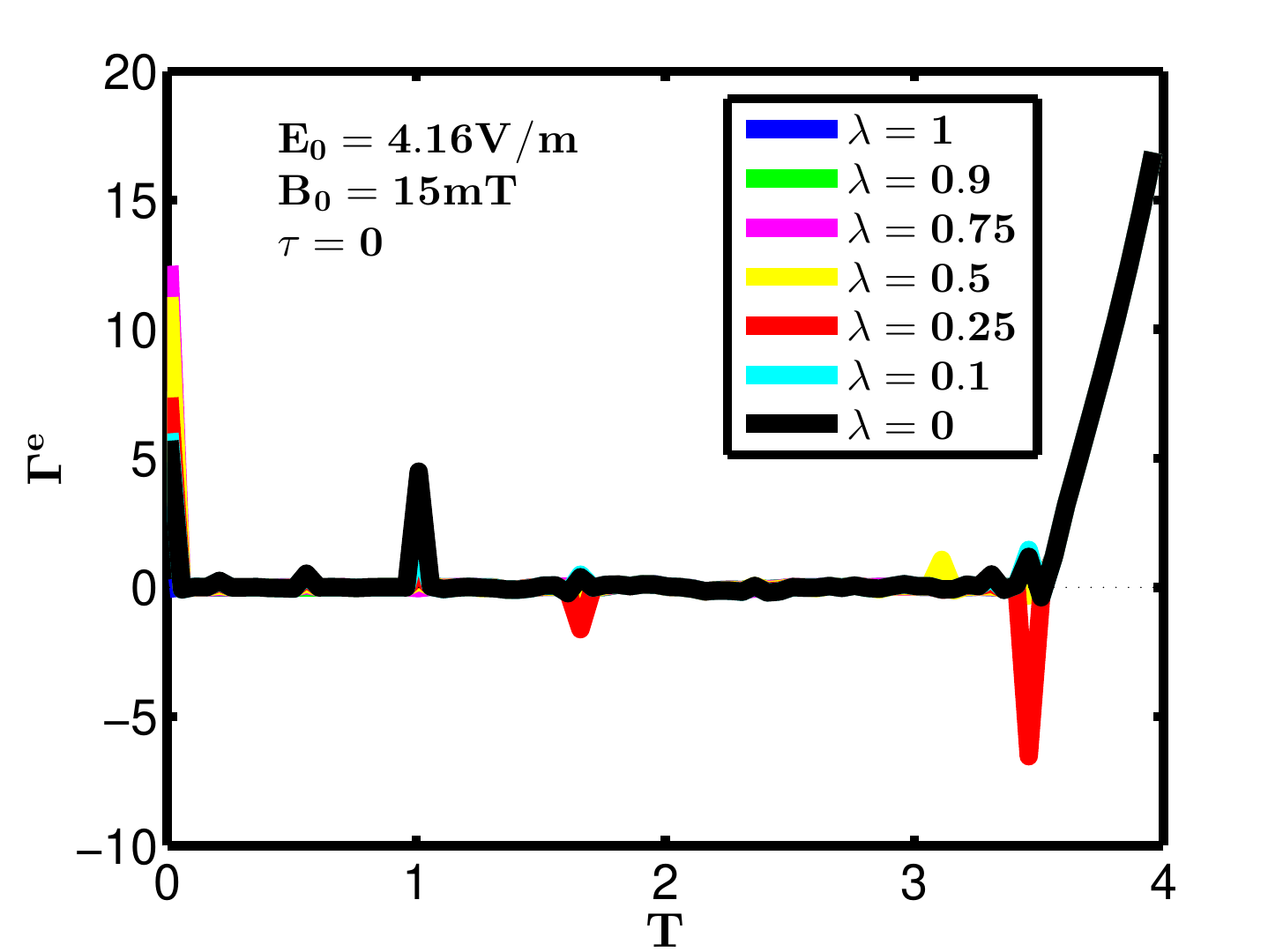}}
\subfigure[]{\includegraphics[width=0.35\textwidth]{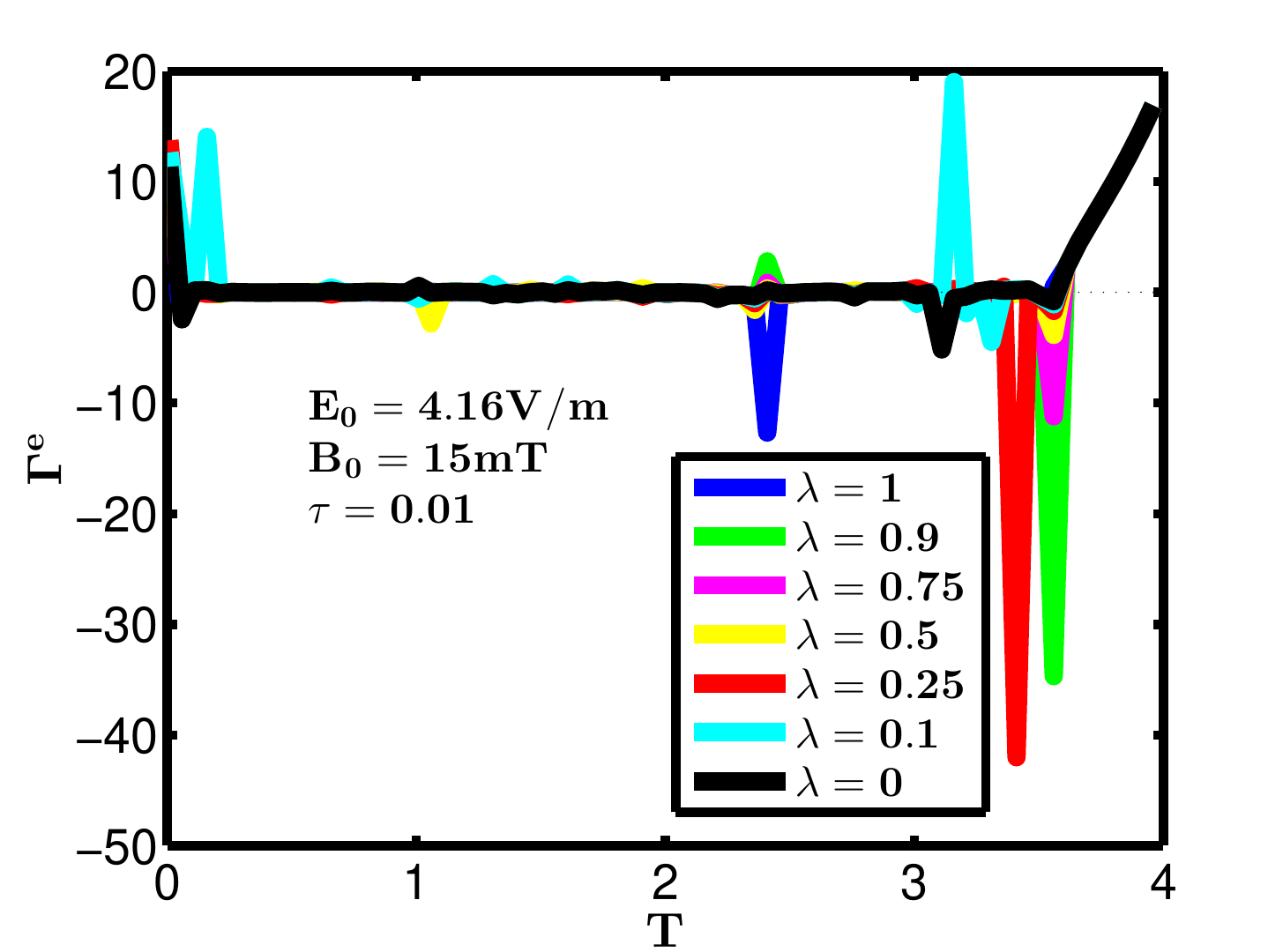}}
\subfigure[]{\includegraphics[width=0.35\textwidth]{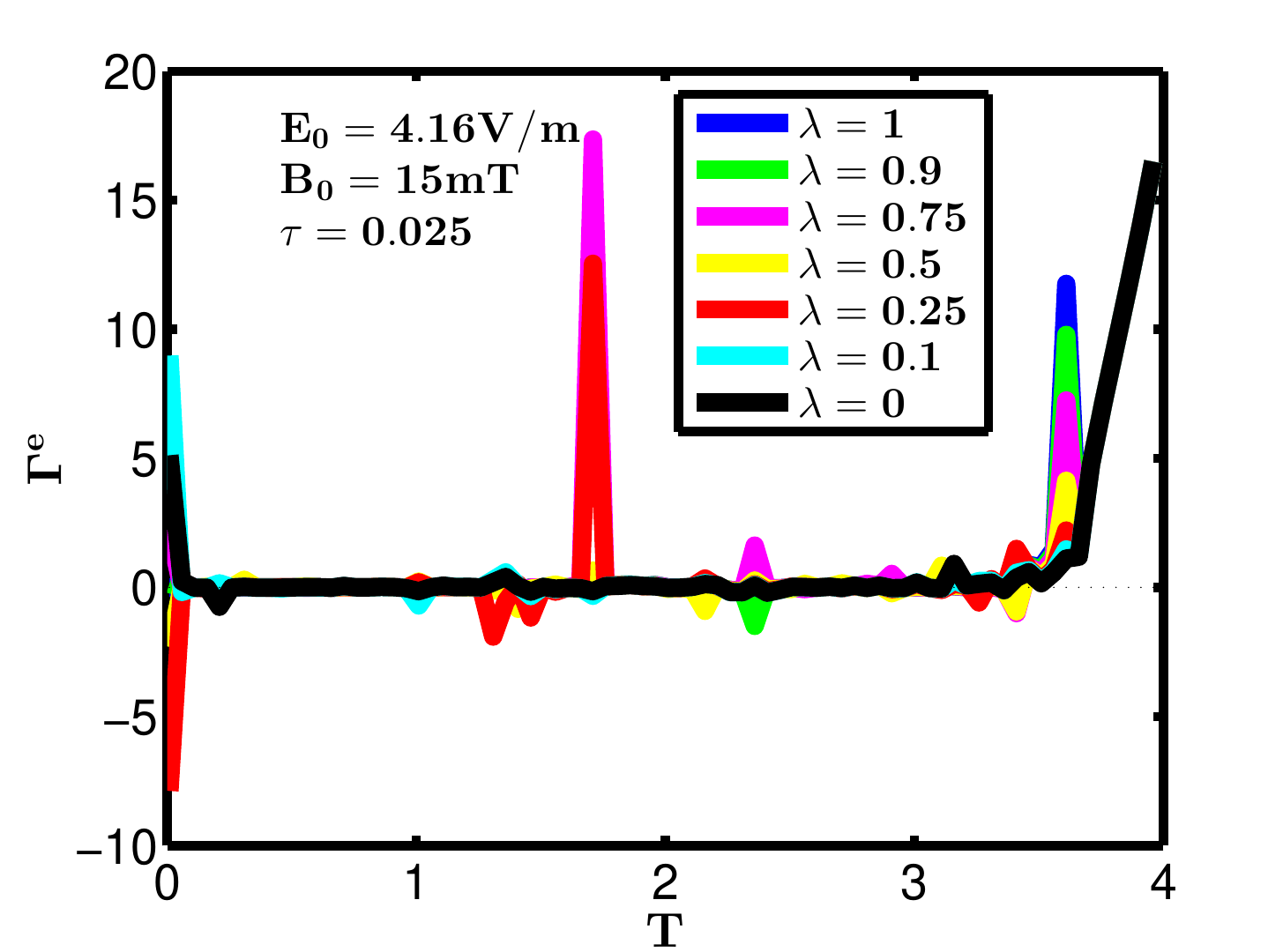}}  
}
\centerline{\subfigure[]{\includegraphics[width=0.35\textwidth]{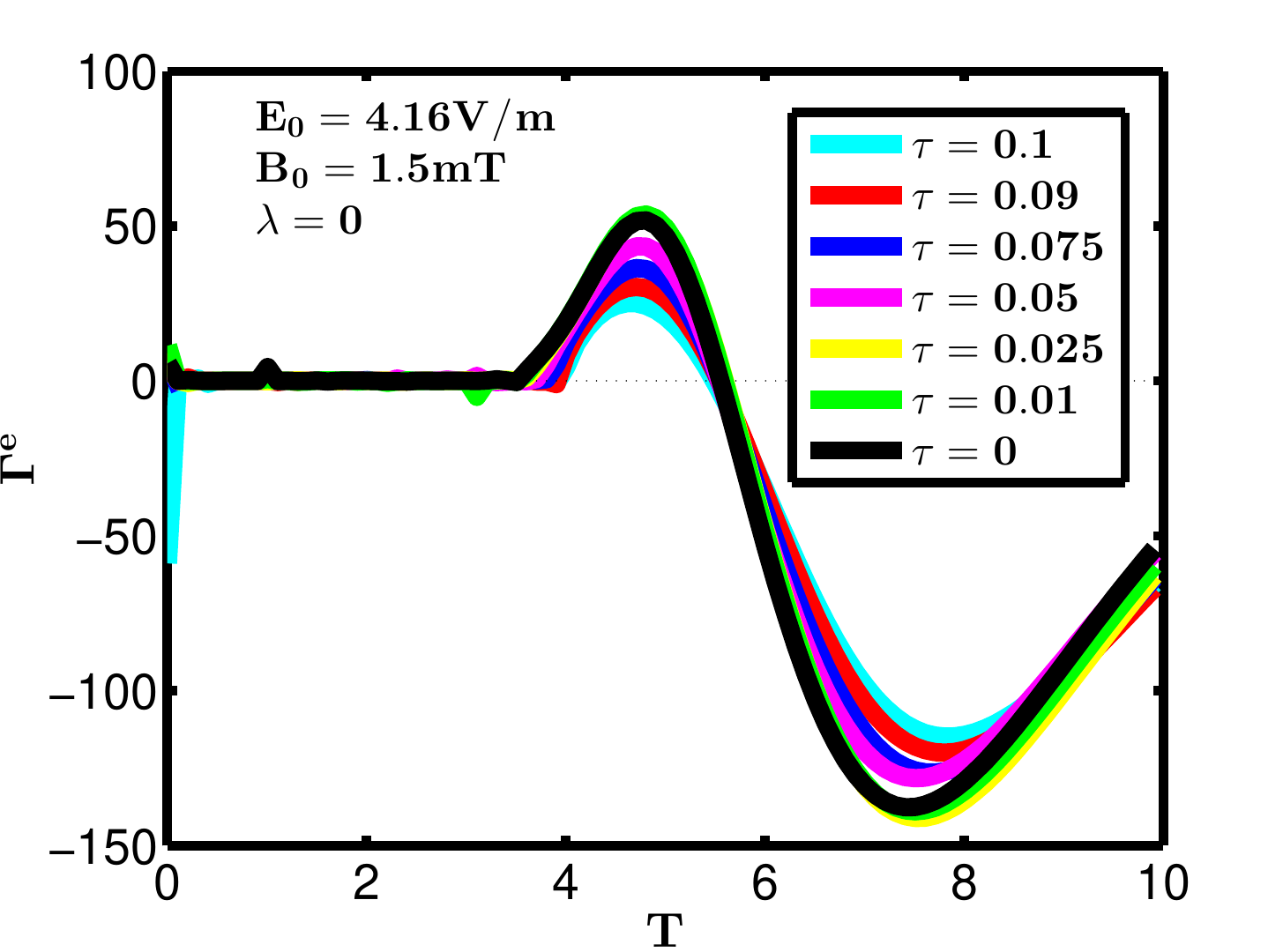}}
\subfigure[]{\includegraphics[width=0.35\textwidth]{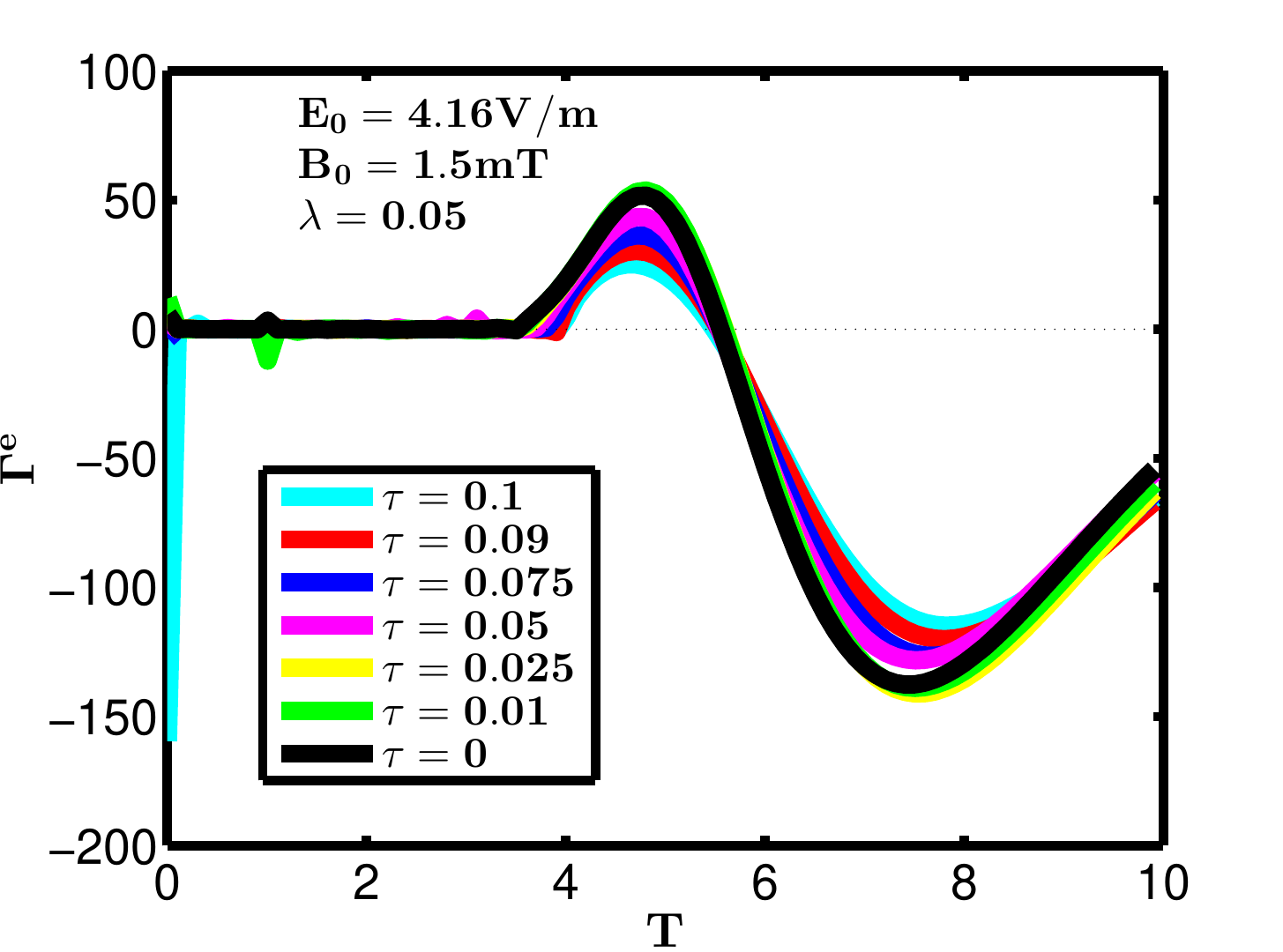}}
\subfigure[]{\includegraphics[width=0.35\textwidth]{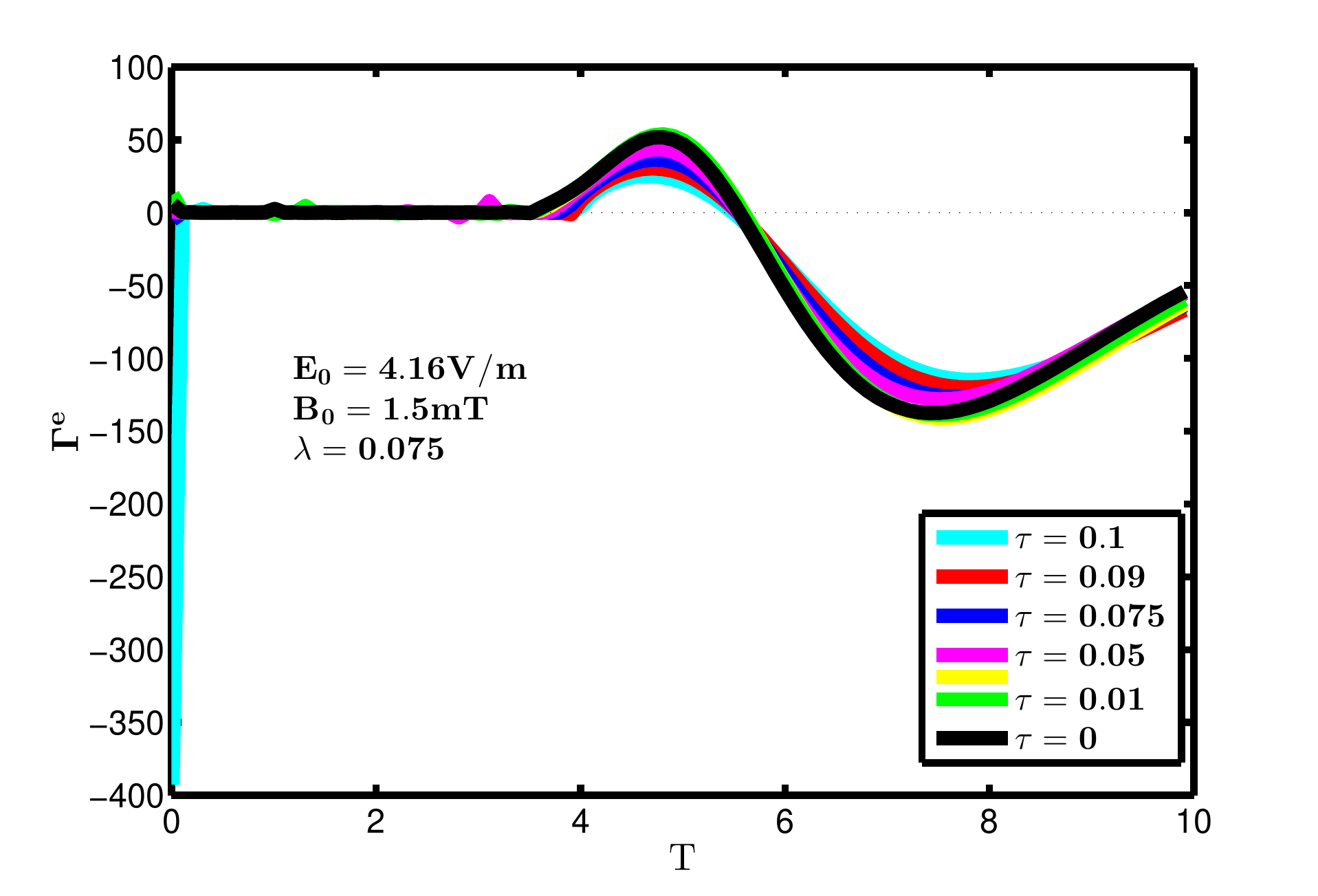}}
}
\caption{the upper panels show the evolution of adiabatic electric cooling rate of the system for different values of the magnetic site-dependent parameters and for three values of the electric site-dependent parameter namely 0 (a), 0.01 (b) and 0.025 (c). In the lower panels we plotted the evolution of adiabatic electric cooling rate for different values of the electric site-dependent parameters and for three values of the magnetic site-dependent parameter namely 0 (d) 0.05 (e) and 0.075 (f). }
\label{F34}
\end{figure}

 In Fig.~\eqref{F34}, the adiabatic electric cooling rate is plotted as a function of temperature and for different values of $\lambda $ and $\tau $. We observed that the site-dependent magnetic field affects the temperature response of the adiabatic electric cooling rate only for temperatures approximately less than $3.7K$, both under uniform and the site-dependent electric field. The evidence of that influence is demonstrated by the presence of the peak points at the different or same temperature for different values of the magnetic site-dependent parameter. When $T \mathrm{>} 3.7 K$, both the site dependent magnetic and electric parameters have no more effect on the adiabatic electric cooling rate.  However, the site-dependent electric field affects the temperature dependence of the electric cooling rate from $T \mathrm{>} 3.5 K$ which is the temperature from which the adiabatic electric cooling rate exhibits alternating negative and positive sign maxima (down panels). 

\par
\begin{figure}[!htb]
\centerline{
\subfigure[]{\includegraphics[width=0.35\textwidth]{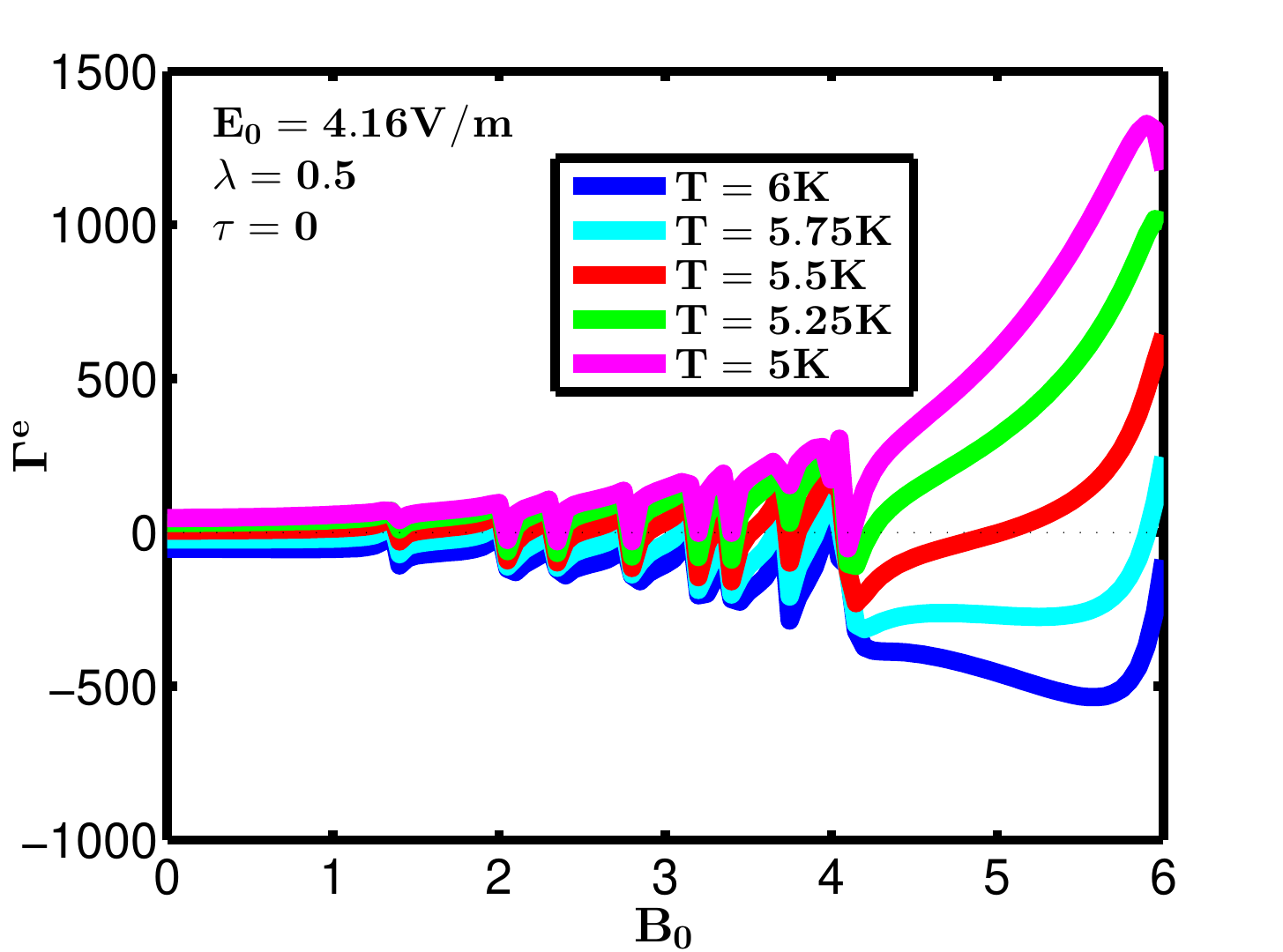}}
\subfigure[]{\includegraphics[width=0.35\textwidth]{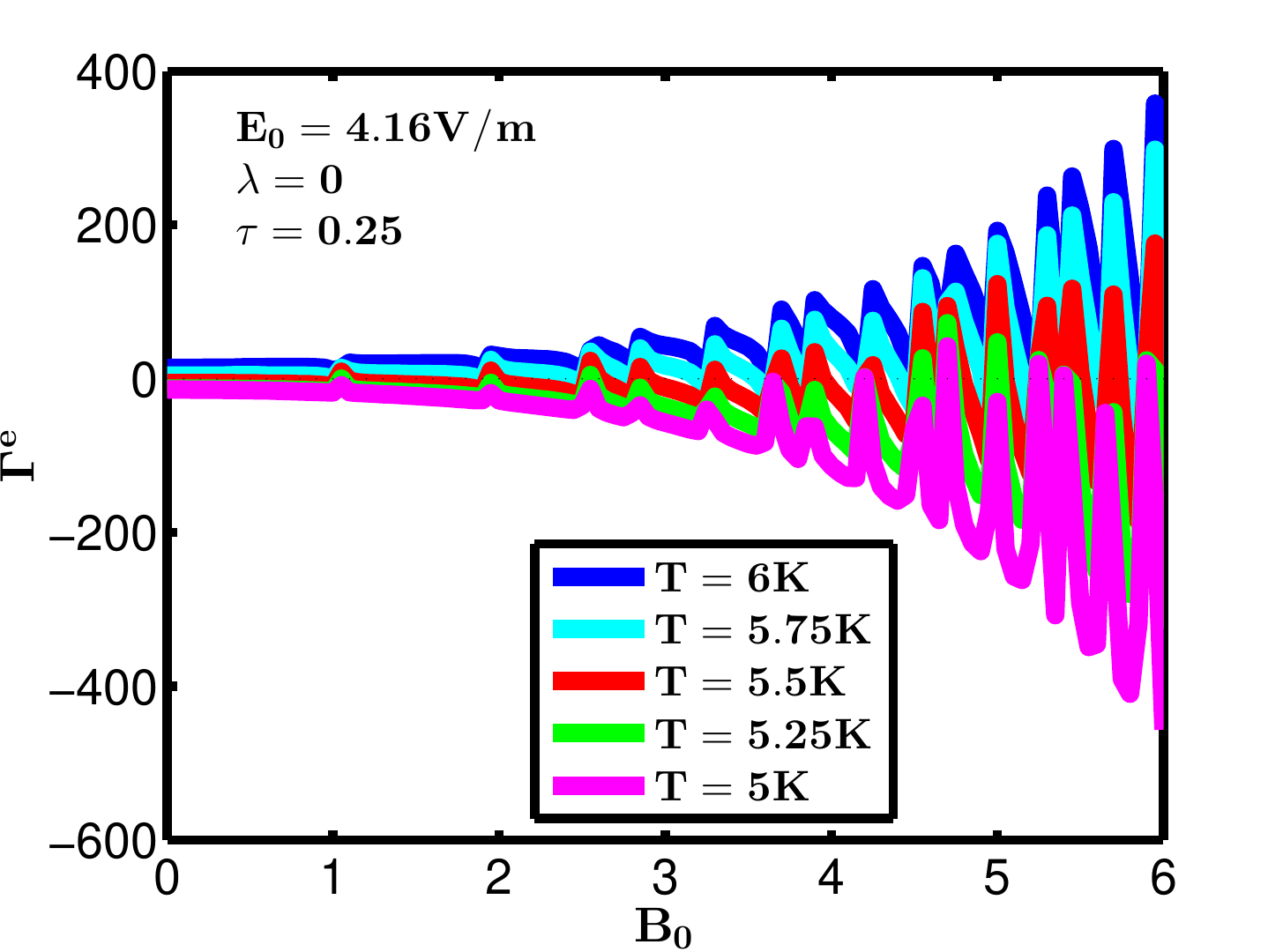}}
}
\centerline{\subfigure[]{\includegraphics[width=0.35\textwidth]{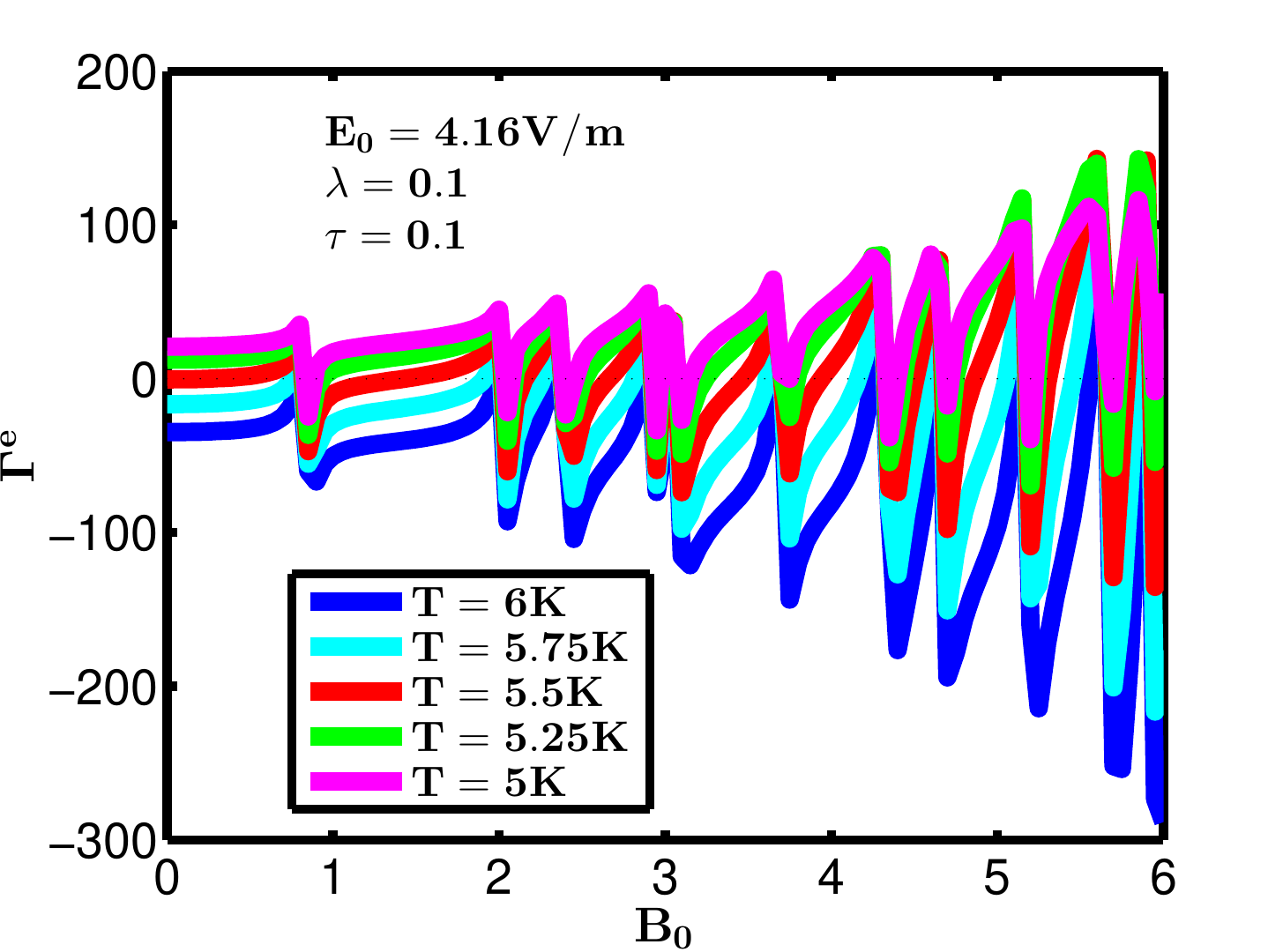}}
\subfigure[]{\includegraphics[width=0.35\textwidth]{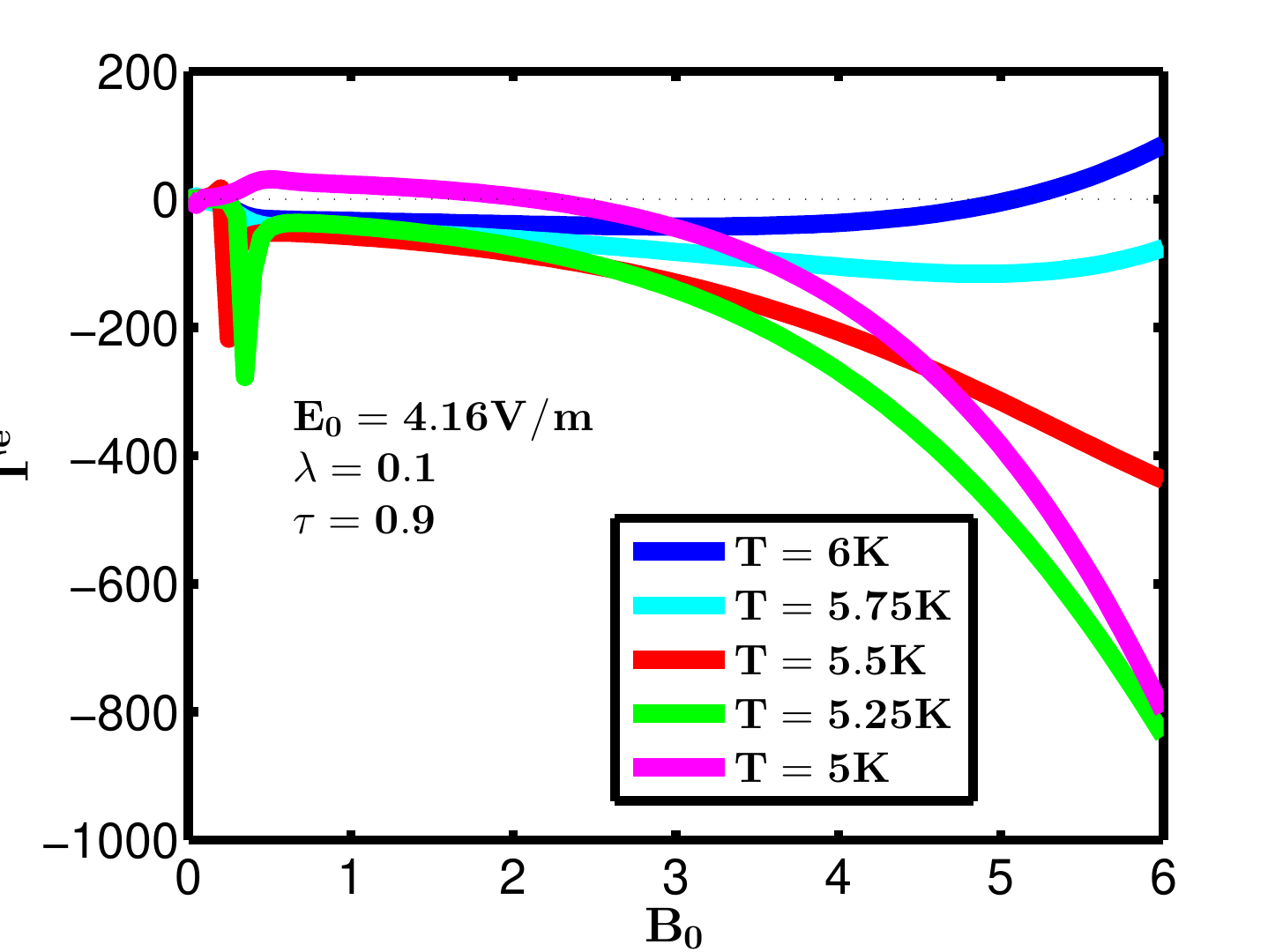}}
\subfigure[]{\includegraphics[width=0.35\textwidth]{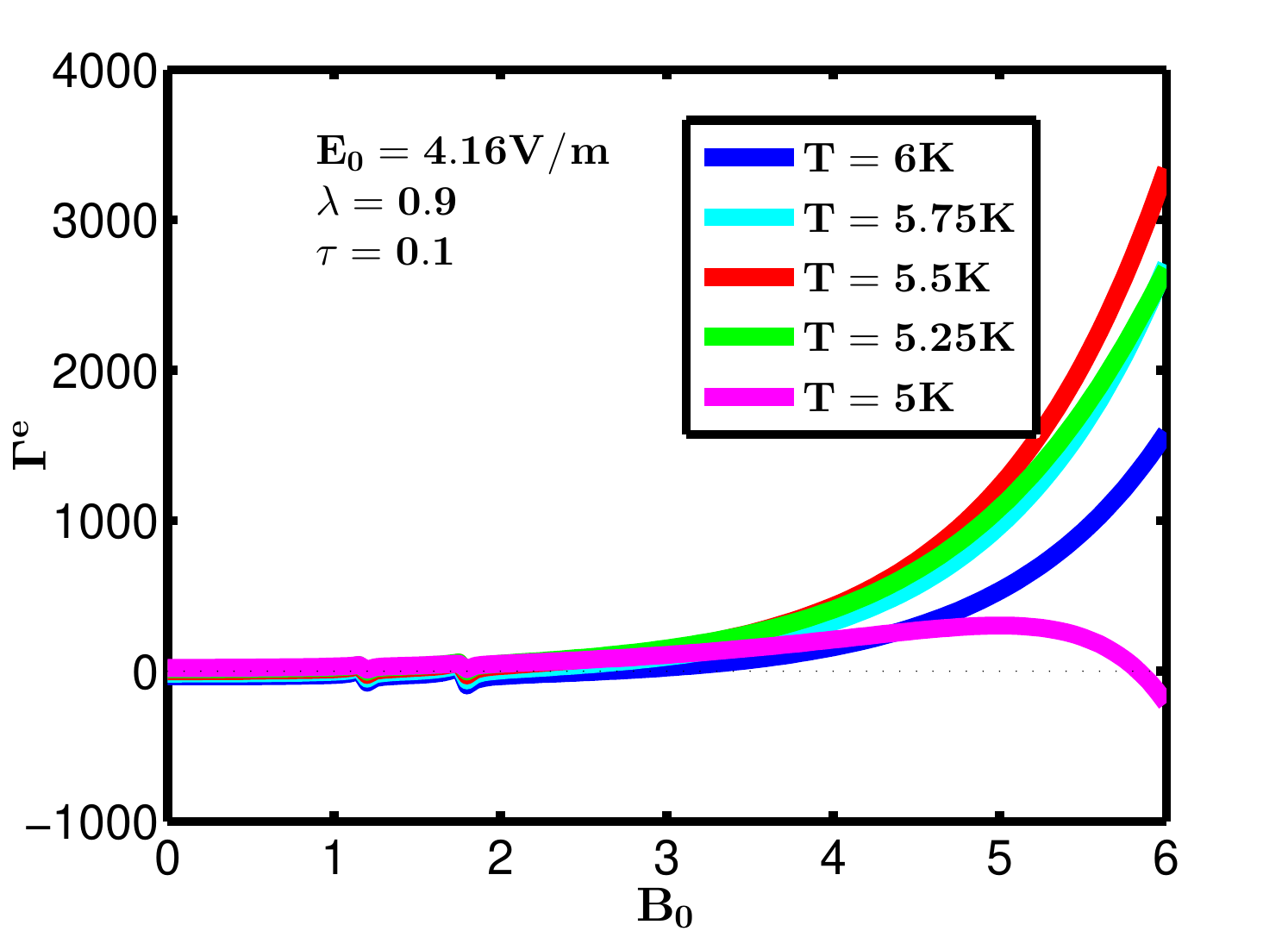}}
}
\caption{magnetic field dependence of the adiabatic electric cooling rate by varying the temperature with the following magnetic and electric site-dependent parameters: $\lambda = 0.5$, $\tau =0$ (a); $\lambda = 0$, $\tau =0.25$ (b); $\lambda = 0.1$, $\tau =0.1$ (c); $\lambda = 0.1$, $\tau =0.9$ (d) ; $\lambda = 0.9$, $\tau =0.1$ (e).}
\label{F35}
\end{figure}

 The plots in Fig.~\eqref{F35} show the magnetic dependence of the adiabatic electric cooling rate. It observed that when the electric field is uniform ($\tau =0$) and the magnetic field site-dependent ($\lambda \mathrm{\neq} 0$), the adiabatic electric cooling rate can be increased by decreasing the temperature of the system. The situation is completely reversed when the electric field becomes site-dependent ($\tau \mathrm{\neq} 0$) and the magnetic field uniform ($\lambda =0$). Overall, it is observed that the adiabatic electric cooling rate can increase or decrease linearly or nonlinearly with the magnetic field, depending on the temperature of the system. Indeed, when the magnetic and electric site-dependent parameters are combined as follow $\lambda =0.1$, $\tau =0.9$ and vice-versa, the nonlinear behavior of the electric cooling rate is strongly attenuated (panels (d) and (e)). 

\par
\begin{figure}[!htb]
\centerline{
\subfigure[]{\includegraphics[width=0.35\textwidth]{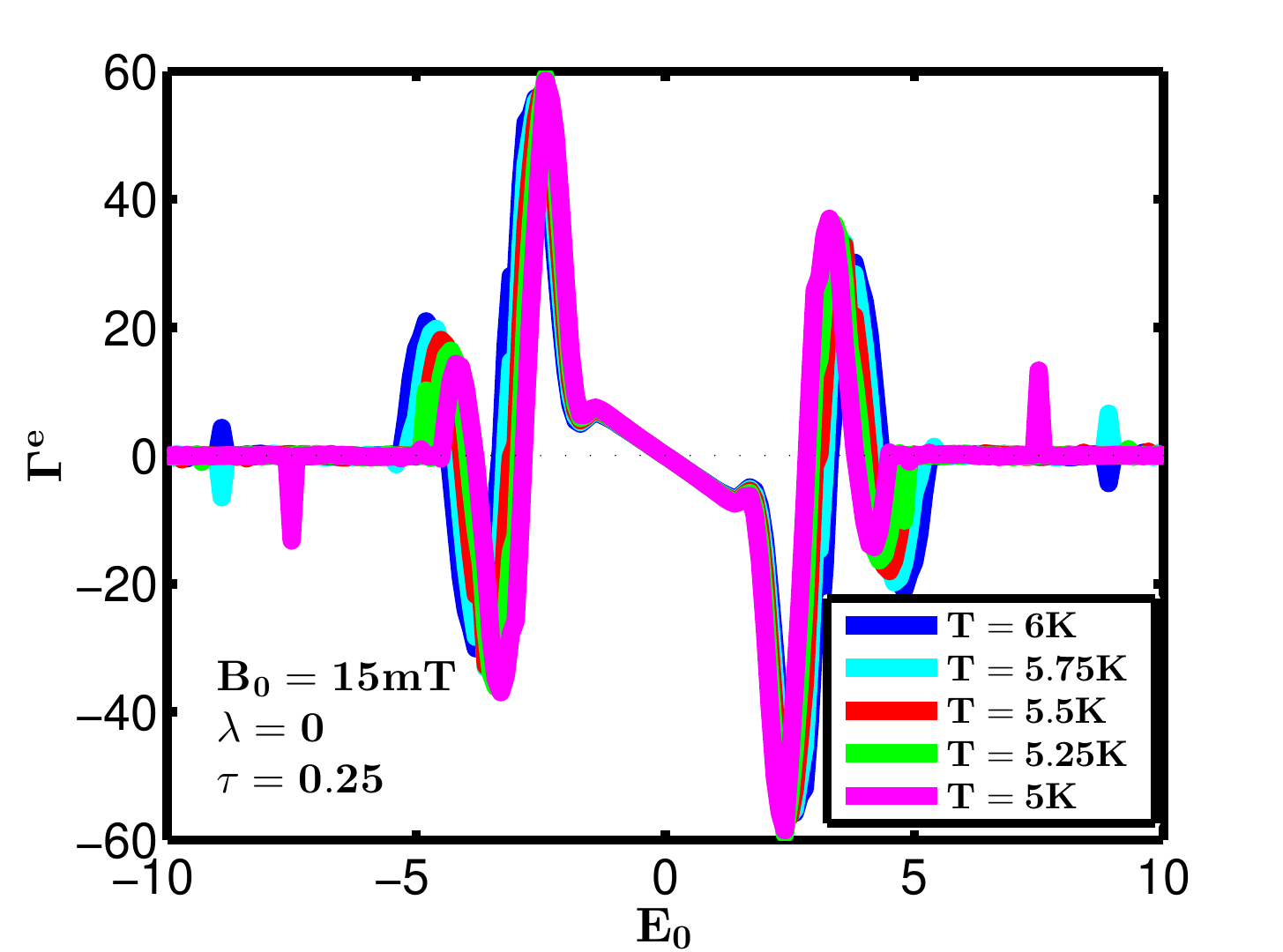}}
\subfigure[]{\includegraphics[width=0.35\textwidth]{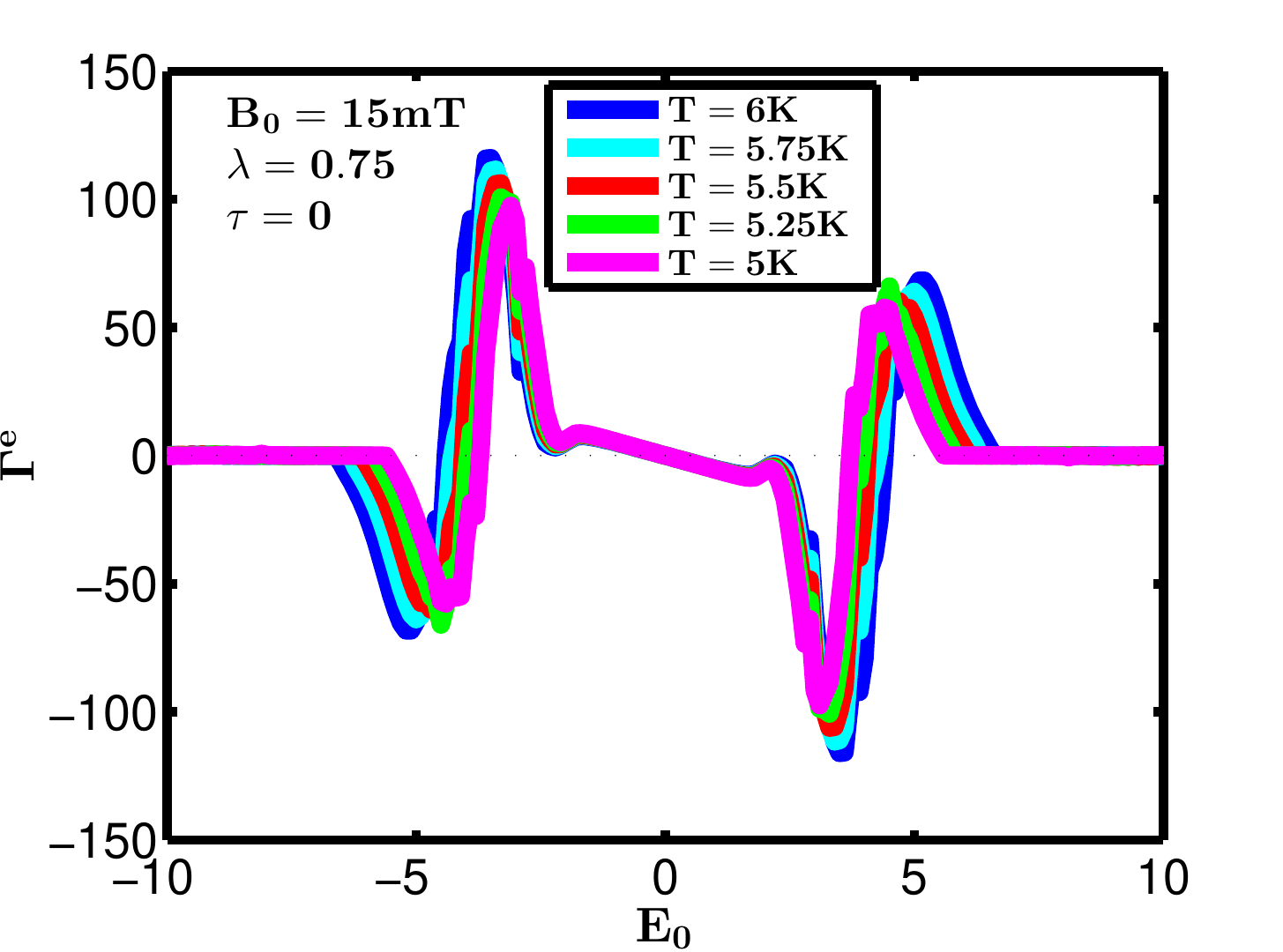}}
}
\centerline{\subfigure[]{\includegraphics[width=0.35\textwidth]{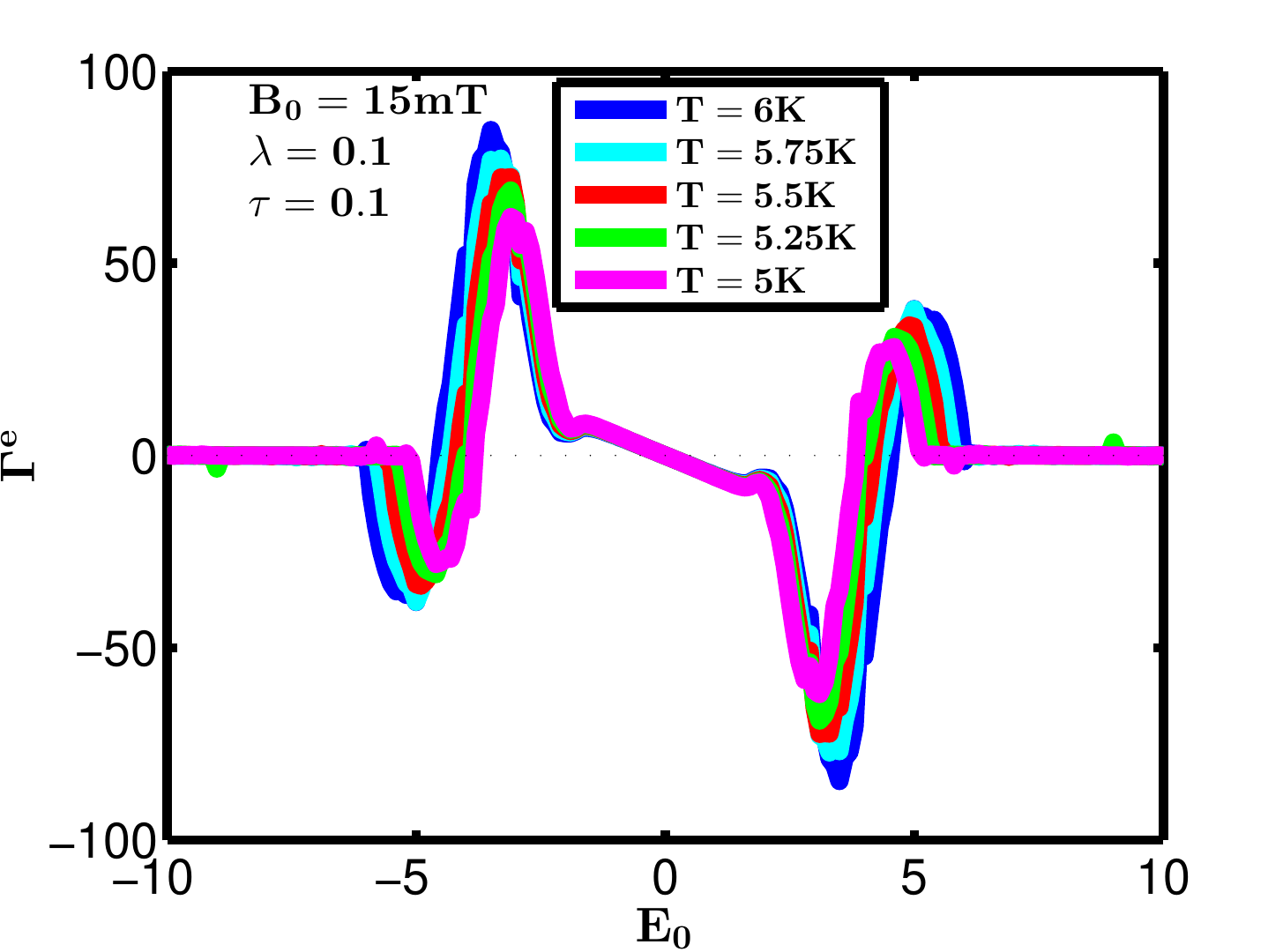}}
\subfigure[]{\includegraphics[width=0.35\textwidth]{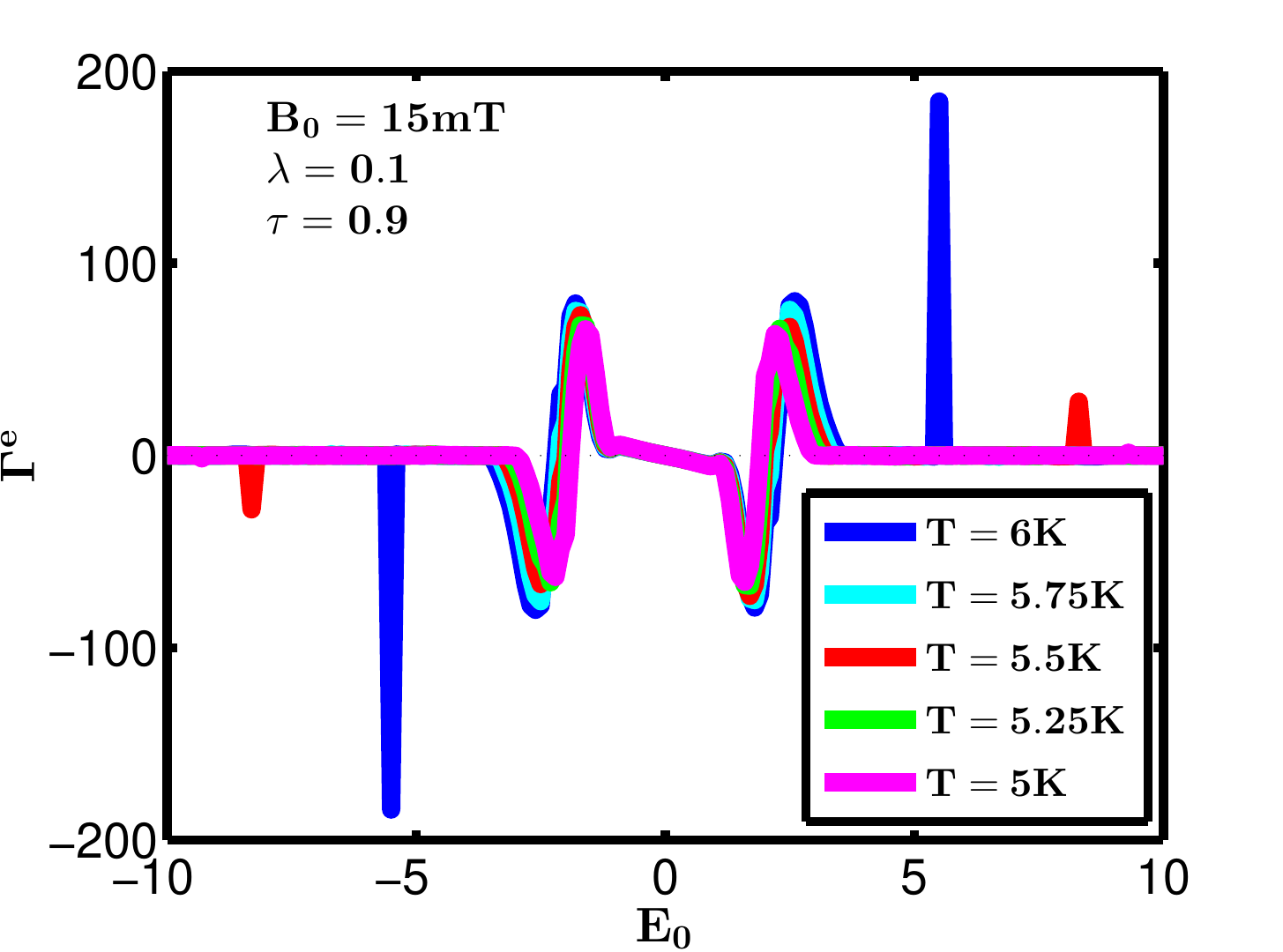}}
\subfigure[]{\includegraphics[width=0.35\textwidth]{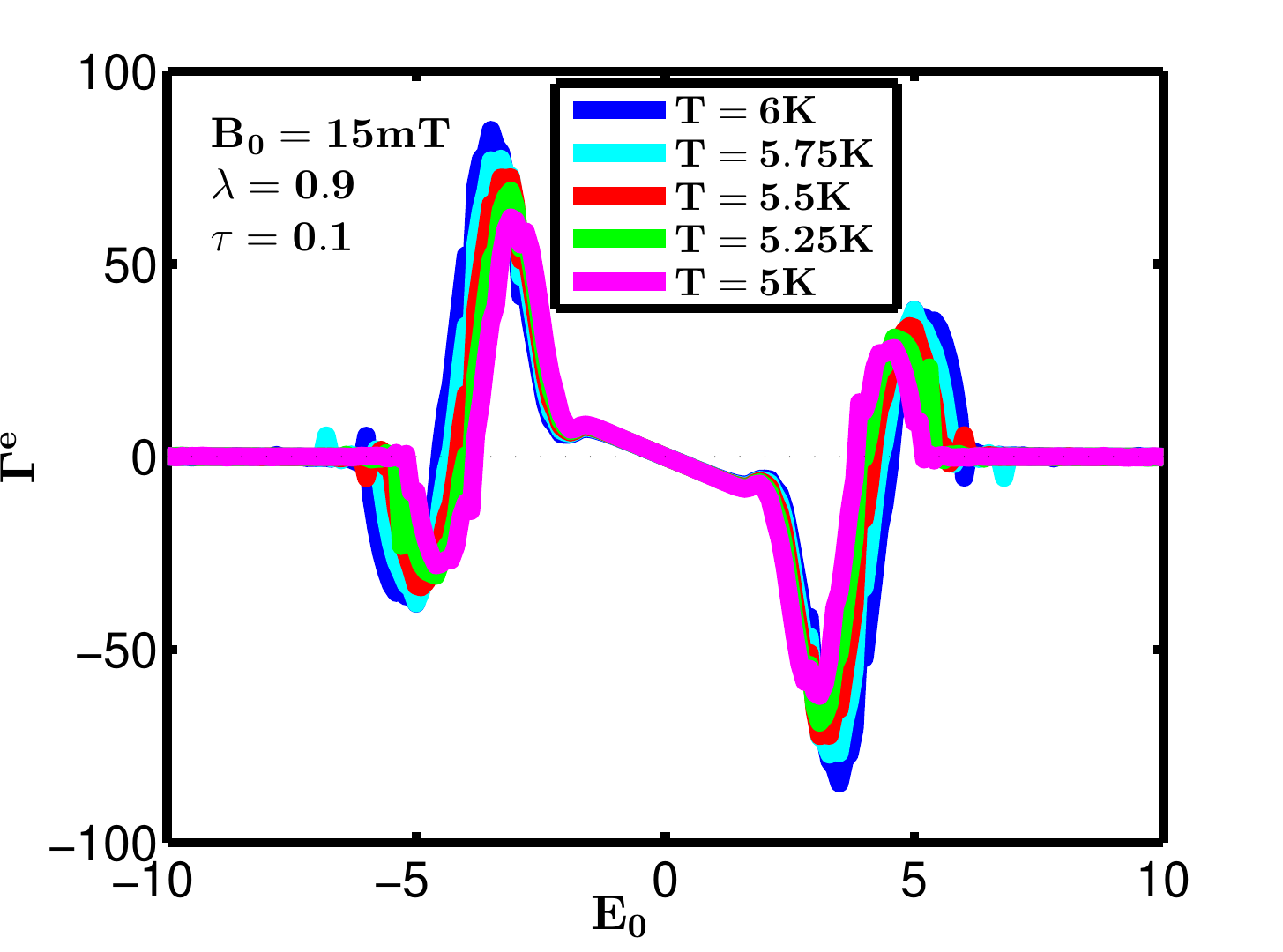}}
}
\caption{electric field dependence of  adiabatic electric cooling rate by varying the temperature with the following magnetic and electric site-dependent parameters: $\lambda = 0$, $\tau =0.25$ (a); $\lambda = 0.75$, $\tau =0$ (b); $\lambda = 0.1$, $\tau =0.1$ (c); $\lambda = 0.1$, $\tau =0.9$ (d) ; $\lambda = 0.9$, $\tau =0.1$ (e).}
\label{F36}
\end{figure}

Moreover, the curves in Fig.~\eqref{F36} highlighting the electric dependence of the adiabatic electric cooling rate show that it displays alternating maxima at weak electric field ($E_0 \mathrm{<} 6 V/m$), and finally freezes at zero with the presence of peak points for certain values of temperature. Overall, the results obtained here show that the adiabatic electric cooling rate displays the characteristic behavior of the electrocaloric effect as observed in experimental results \cite{12,14}. Notice that the coexistence of positive and negative values of the electric cooling indicates the heating (positive electrocaloric effect) and cooling (negative electrocaloric effect) of the system, respectively. 
\par
\begin{figure}[!htb]
	\centerline{\subfigure[]{\includegraphics[width=0.35\textwidth]{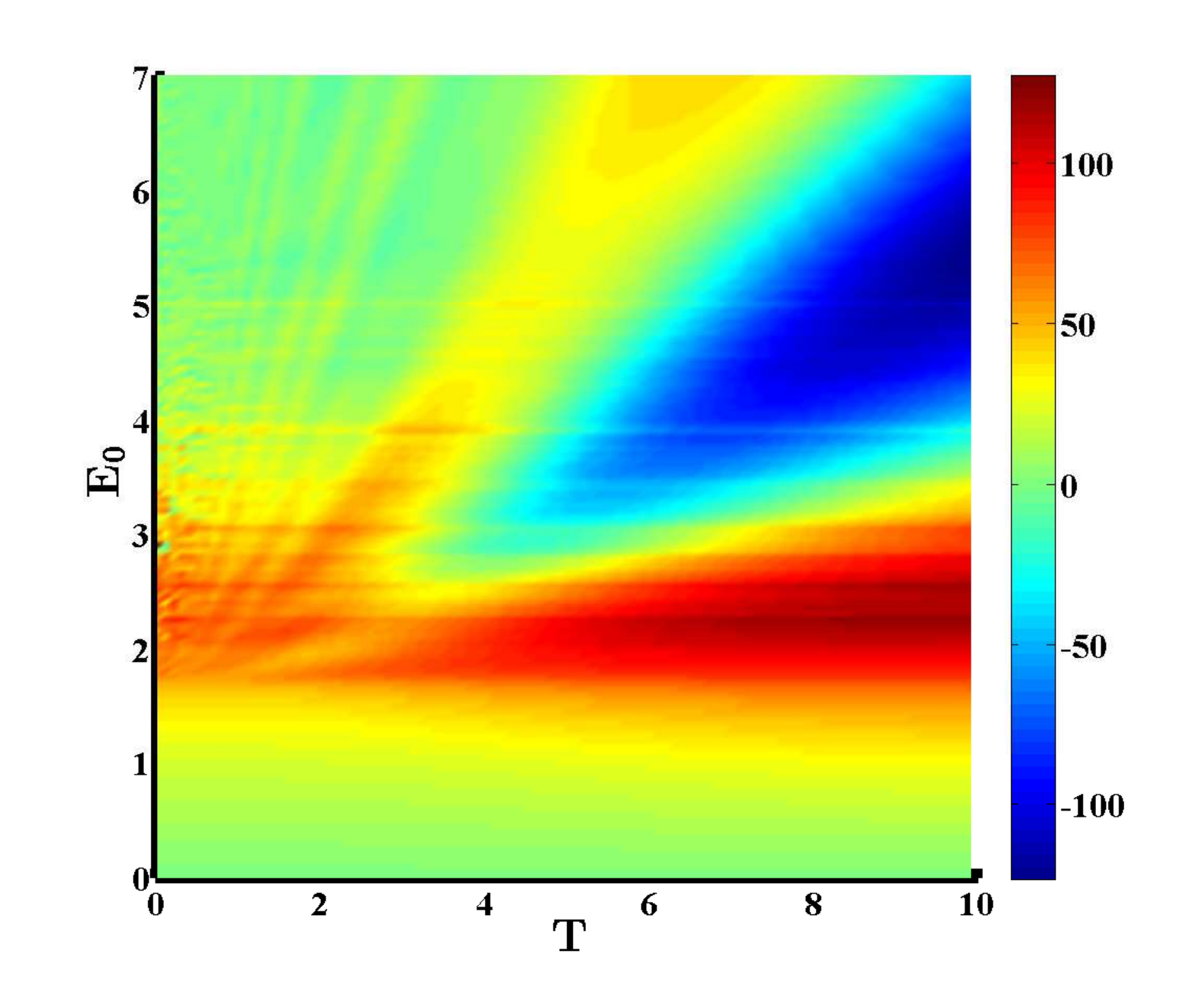}}
		\subfigure[]{\includegraphics[width=0.35\textwidth]{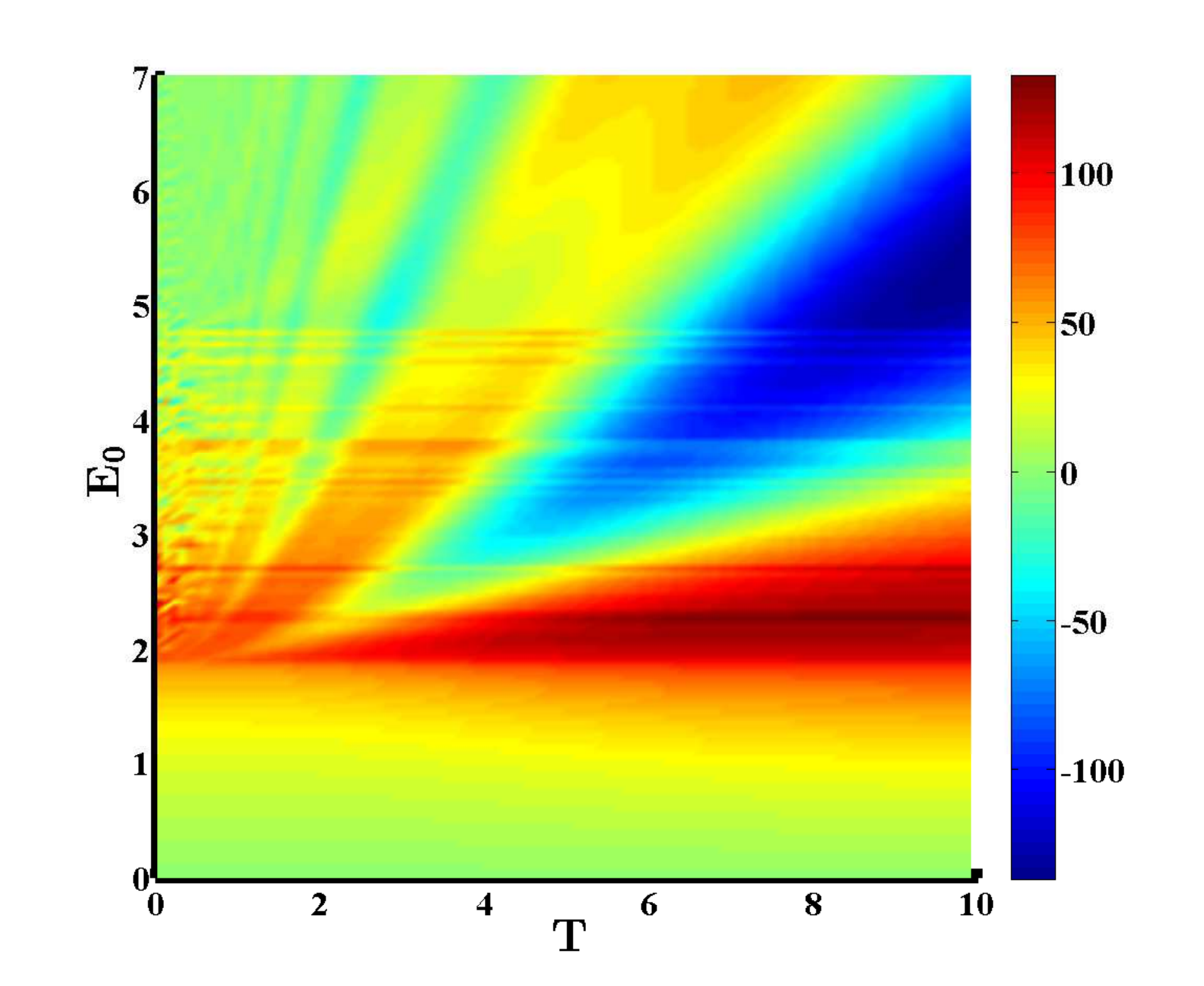}}
		\subfigure[]{\includegraphics[width=0.35\textwidth]{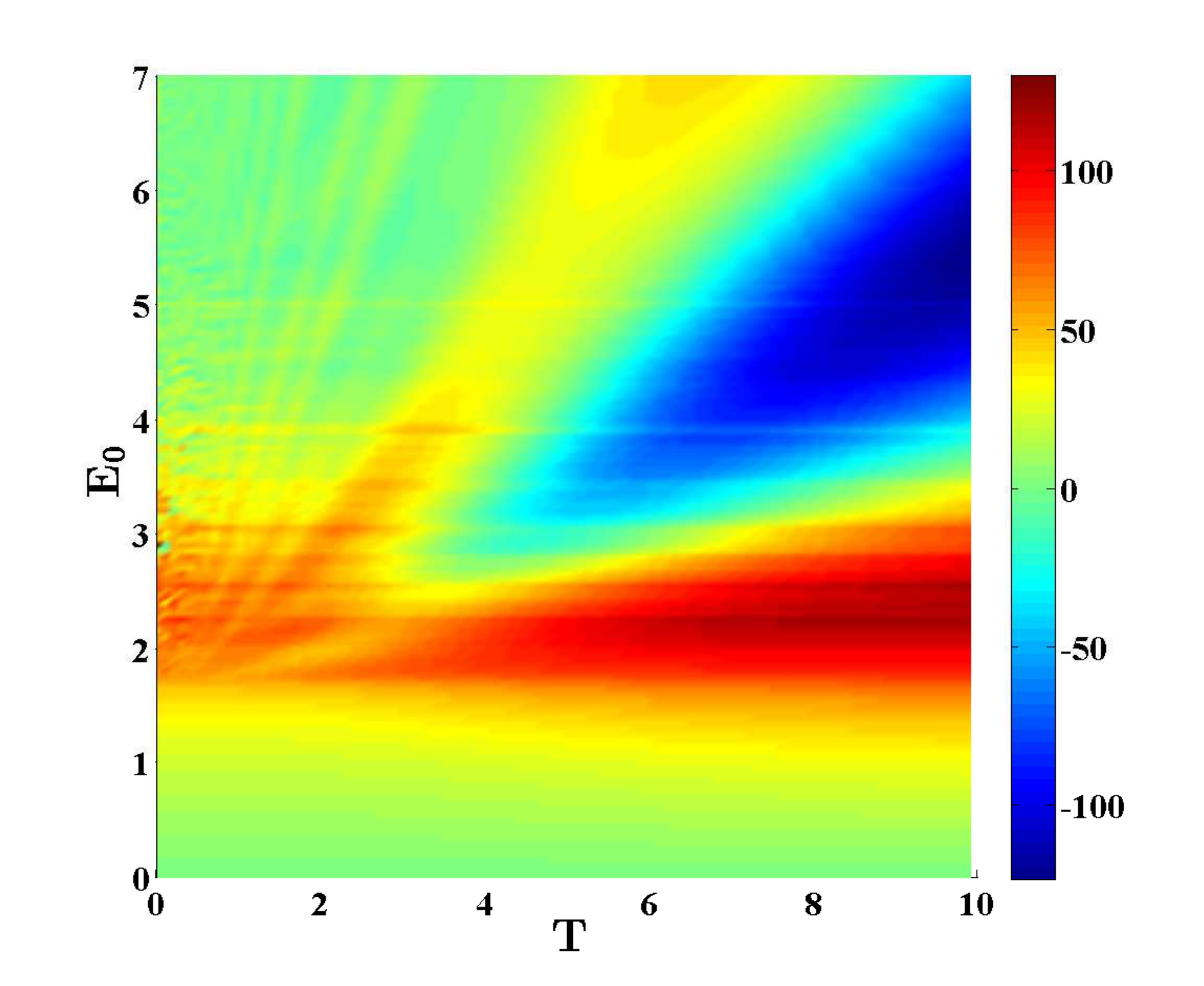}}
	}
	\caption{the surface plot of electric polarization against electric field and temperature with the following magnetic field, magnetic and electric site-dependent parameters :(a)$B_0 =0 T, \tau =  0.1$, (b) $ B_0 =1.5 mT, \lambda = \tau =0$, (c) $B_0 =1.5 mT, \lambda =0.05, \tau =   0.1$.  }
	\label{FO2}
\end{figure}

 
Furthermore, the phase diagram in term of the electric polarization obtained within the $E_0 - T$ plane, in  Fig.~\eqref{FO2}, shows three regions as function of approximated values of critical electric field and temperature: a region where the electric polarization varies slightly with the electric field ($E_0 \mathrm{<} 1 V/m$), a region where the electric polarization increases with the electric field but oscillates with the temperature ($1 V/m \mathrm{<} E_0 \mathrm{<} 2.1 V/m $) and a region where the electric polarization oscillates both with the electric field and the temperature ($E_0 \mathrm{>} 2.1 V/m$ ). Note that the critical electric field and the critical temperature are tunable by the magnetic and electric site-dependent parameters. 

\par
\par
\par
\subsection{Adiabatic magnetoelectric cooling rate}
\begin{figure}[!htb]
\centerline{
\subfigure[]{\includegraphics[width=0.35\textwidth]{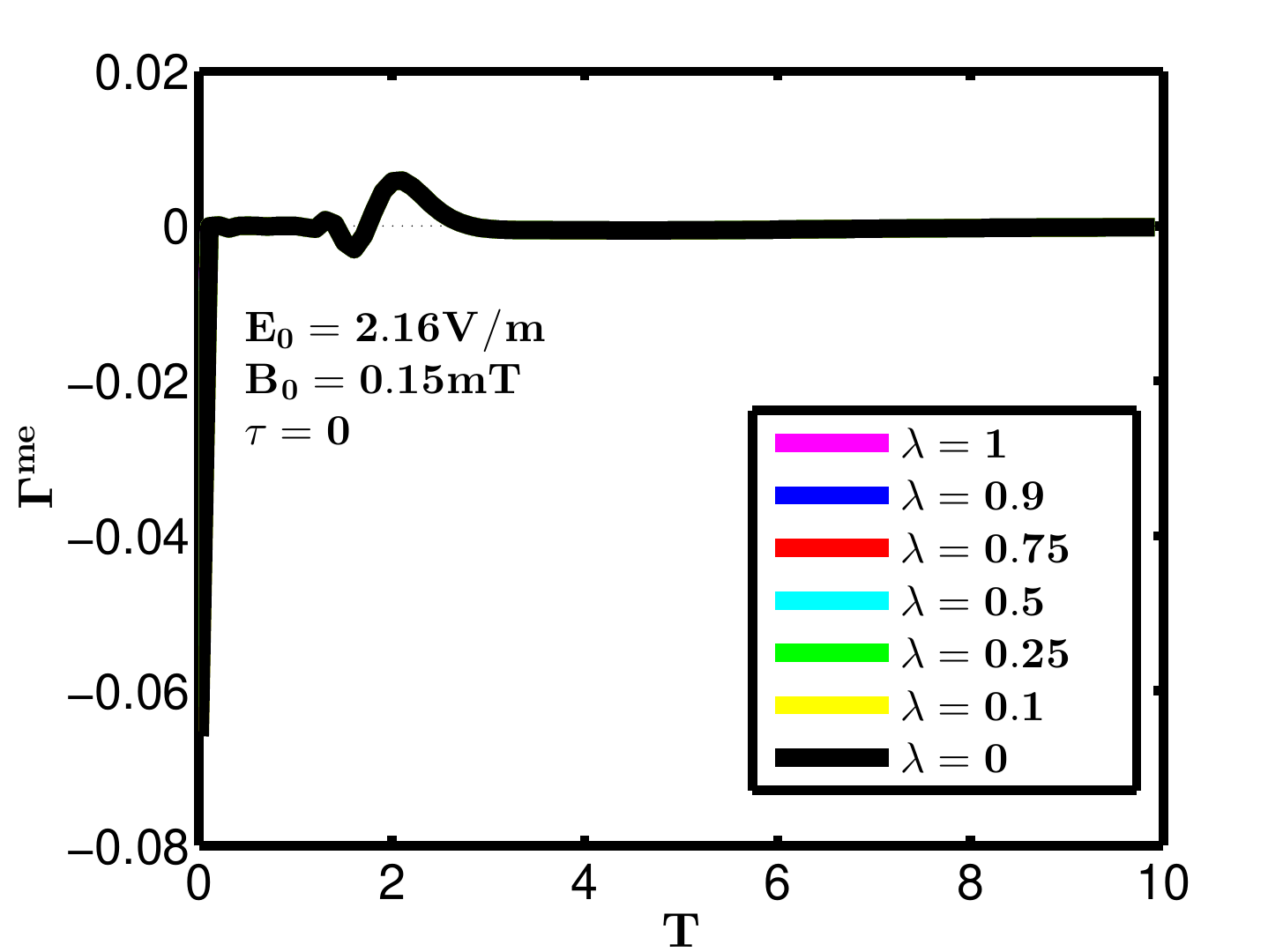}}
\subfigure[]{\includegraphics[width=0.35\textwidth]{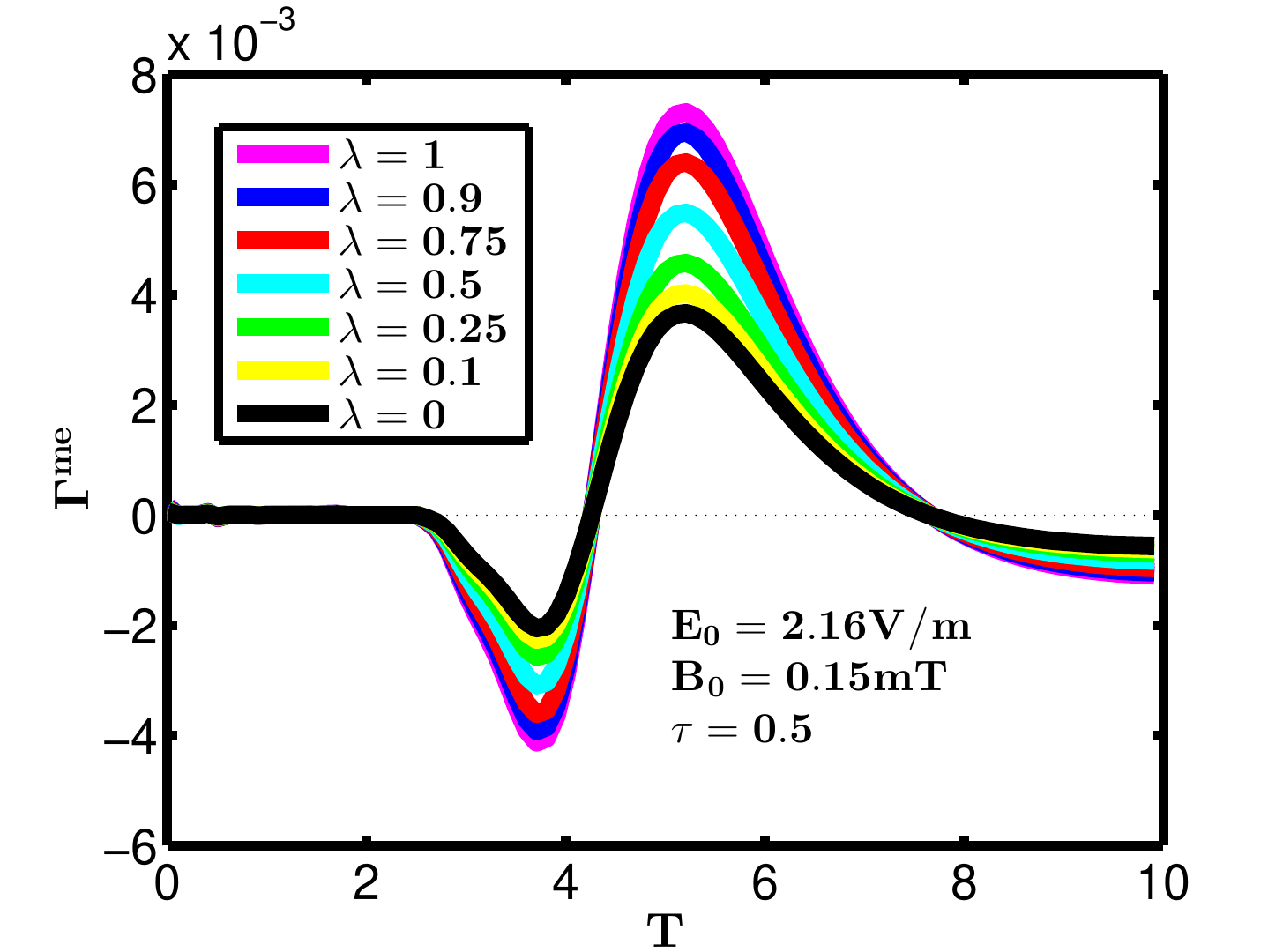}}
\subfigure[]{\includegraphics[width=0.35\textwidth]{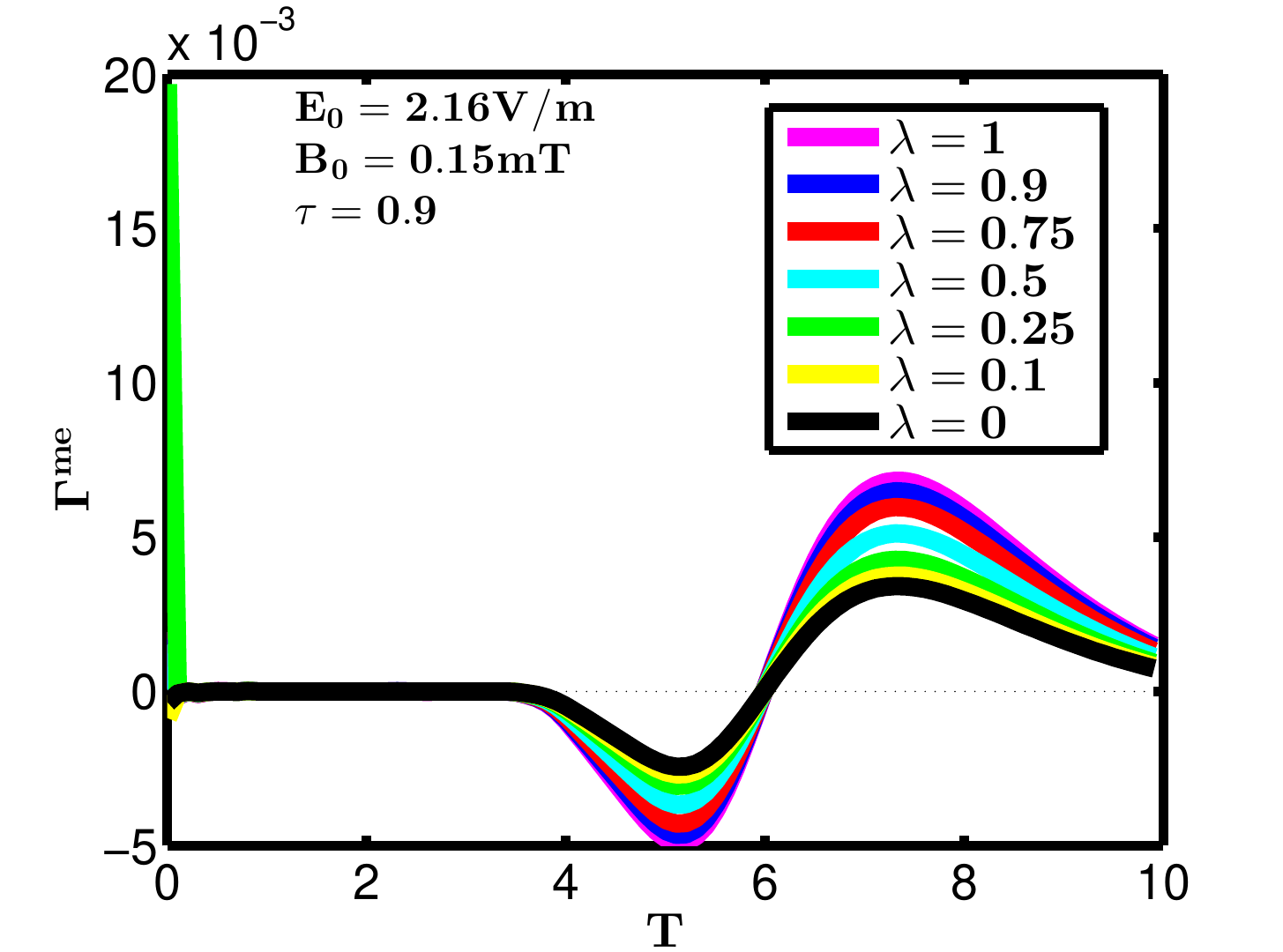}}  
}
\centerline{\subfigure[]{\includegraphics[width=0.35\textwidth]{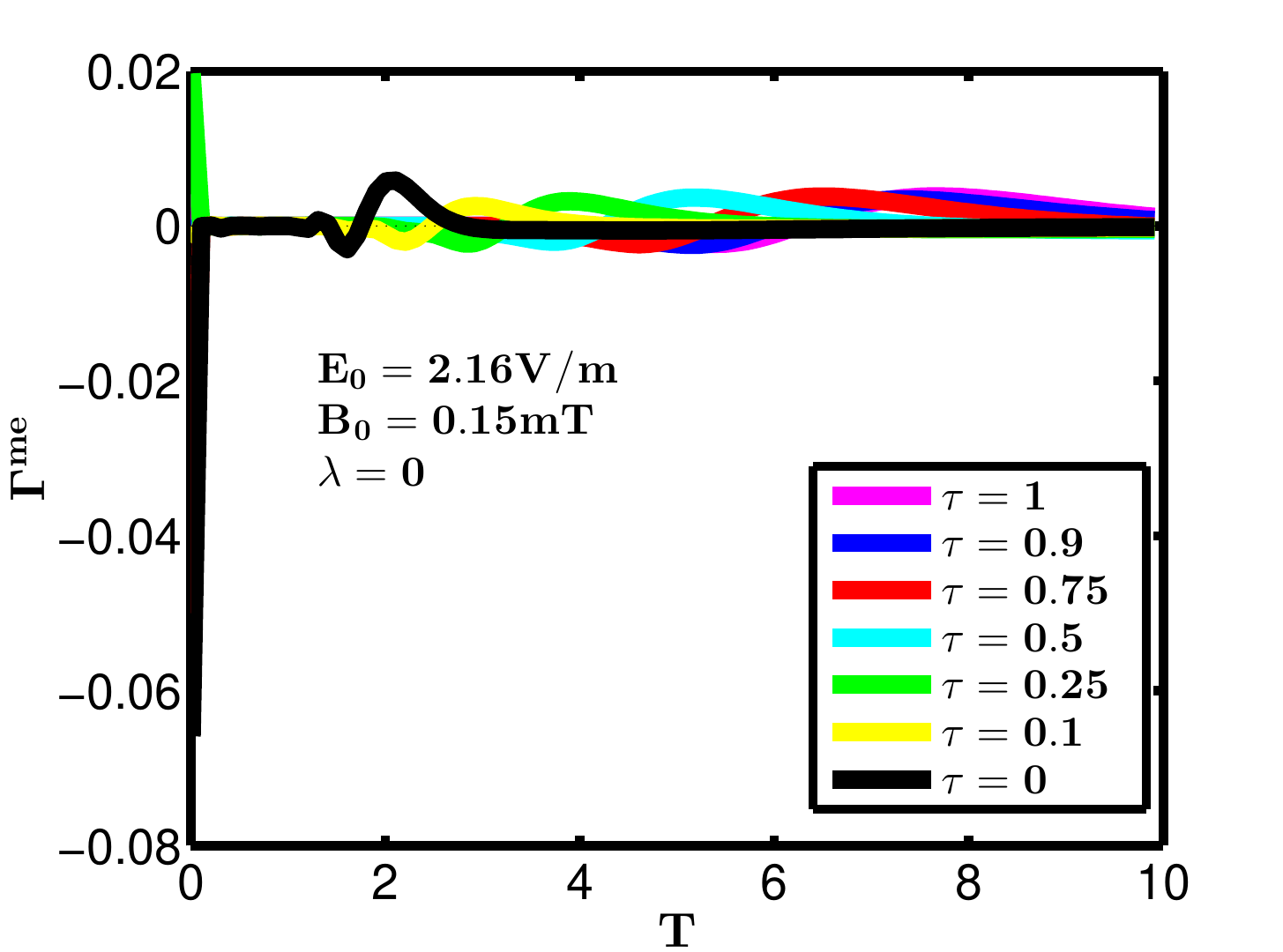}}
\subfigure[]{\includegraphics[width=0.35\textwidth]{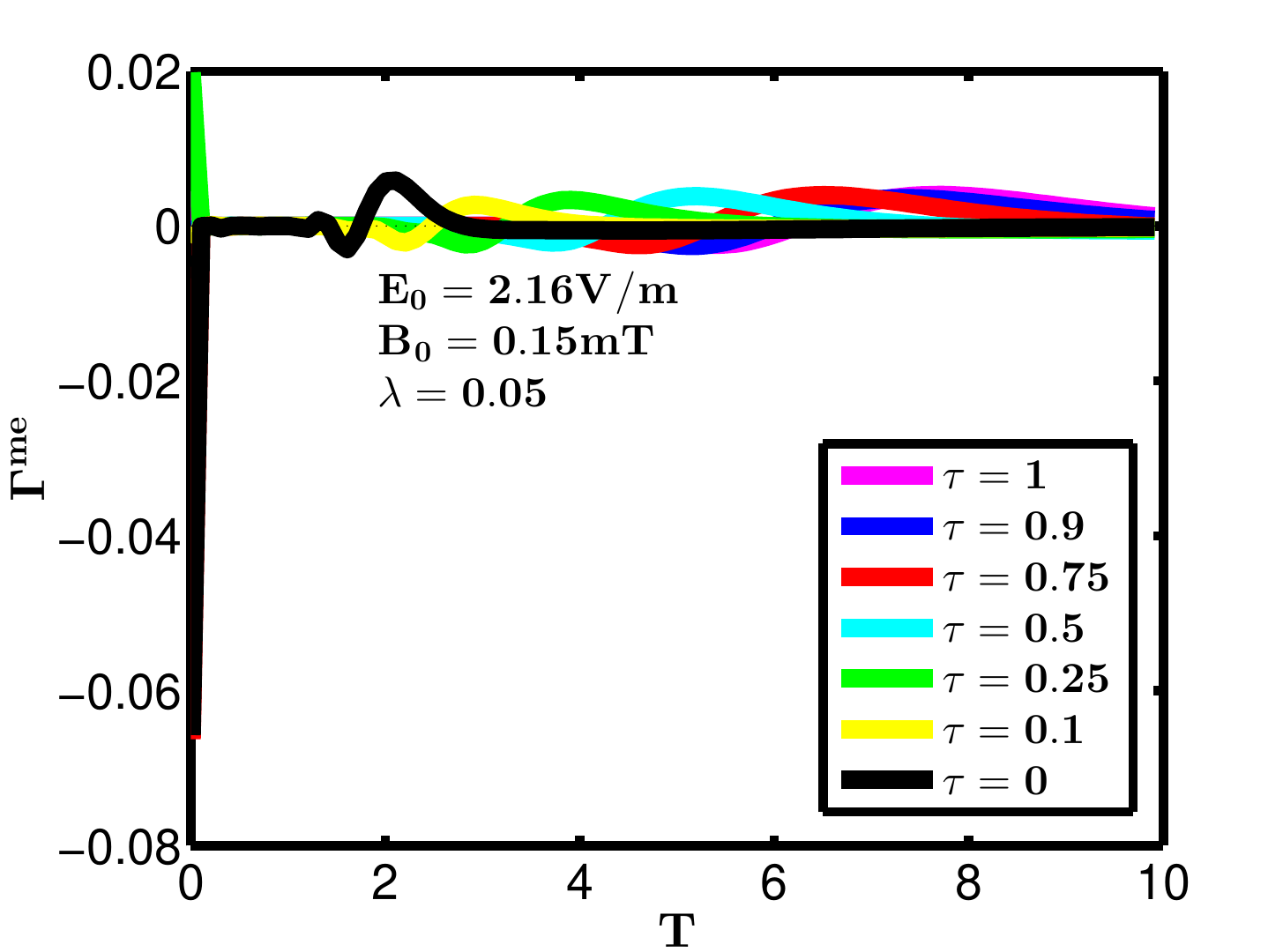}}
\subfigure[]{\includegraphics[width=0.35\textwidth]{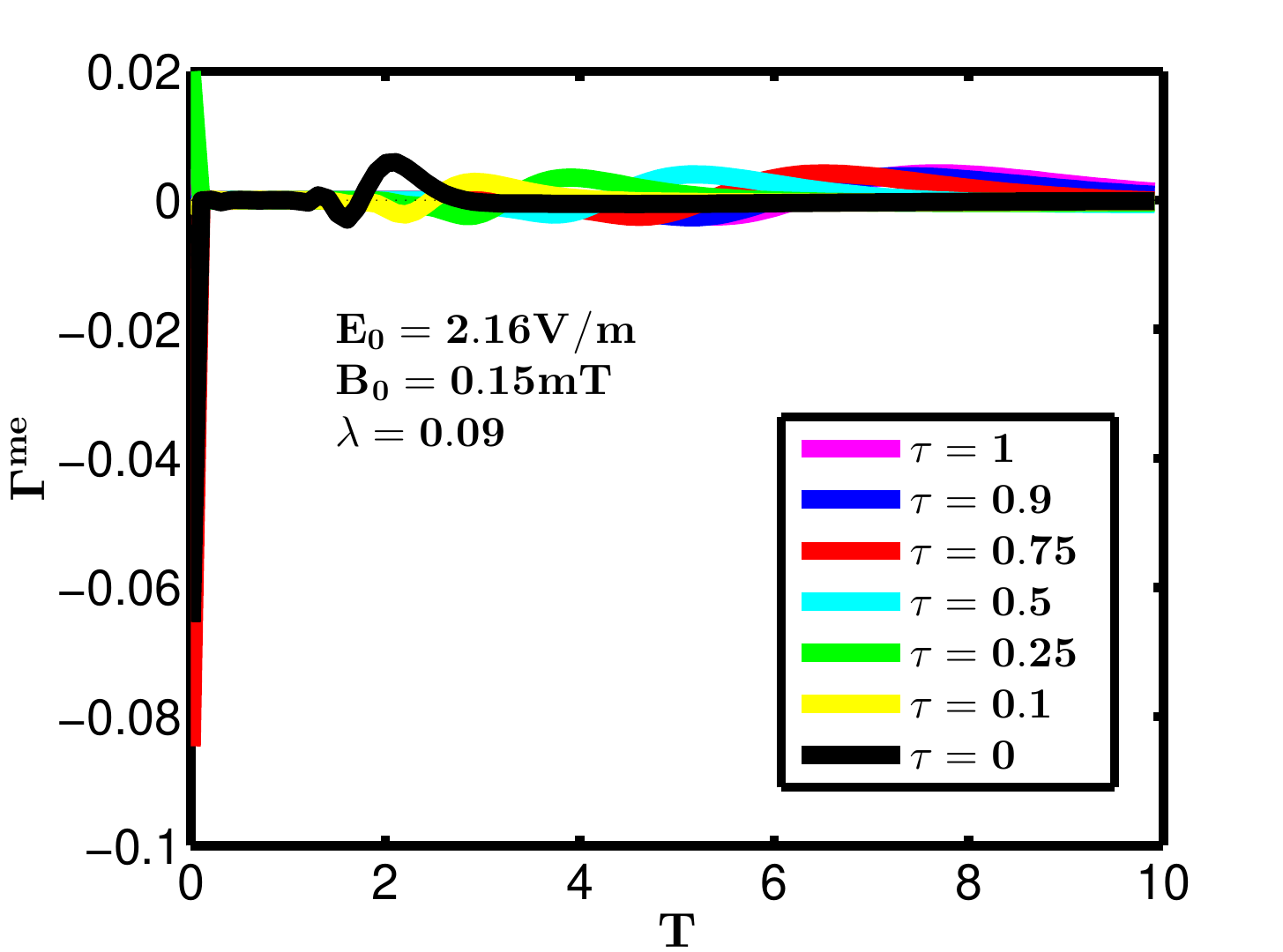}}
}
\caption{the upper panels show the evolution of adiabatic magnetoelectric cooling rate of the system for different values of the magnetic site-dependent parameters and for three values of the electric site-dependent parameter namely 0 (a), 0.5 (b) and 0.9 (c). In the lower panels we plotted the evolution of adiabatic magnetoelectric cooling rate for different values of the electric site-dependent parameters and for three values of the magnetic site-dependent parameter namely 0 (d) 0.05 (e) and 0.09 (f). }
\label{F41}
\end{figure}

 Fig.~\eqref{F41} clearly shows in panel (a) that the site-dependent magnetic field doesn't affect the temperature response of the magnetoelectric cooling rate under the uniform electric field ($\tau =0$). However, by varying the magnetic site-dependent parameter under the site-dependent electric field, alternating negative and positive maxima appear from $T \mathrm{>} 1.9 K$. This thus shows the influence of both site-dependent fields (panels b and c). Note that such a behavior of the magnetoelectric cooling indicates the magneto-electrocaloric effect. In addition, the maximum value of the magneto-electric cooling increase with the magnetic site-dependent parameter. Thus, the site-dependent magnetic field enhance the magneto-electrocaloric effect.  Moreover, by varying the electric site-dependent parameter it is observed that the temperature dependence of the magnetoelectric cooling rate exhibits an oscillating behavior both when the system is suggested to a uniform magnetic field ($\lambda =0$) and site-dependent magnetic field ($\lambda \mathrm{\neq} 0$) (lower panels). Indeed, contrary to the case of uniform electric field where the site-dependent magnetic field has no effect on the evolution of the magnetoelectric cooling rate, it is observed that the magnetoelectric cooling rate is affected by the site-dependent electric field when the magnetic field is uniform.  
\par
\begin{figure}[!htb]
\centerline{
\subfigure[]{\includegraphics[width=0.35\textwidth]{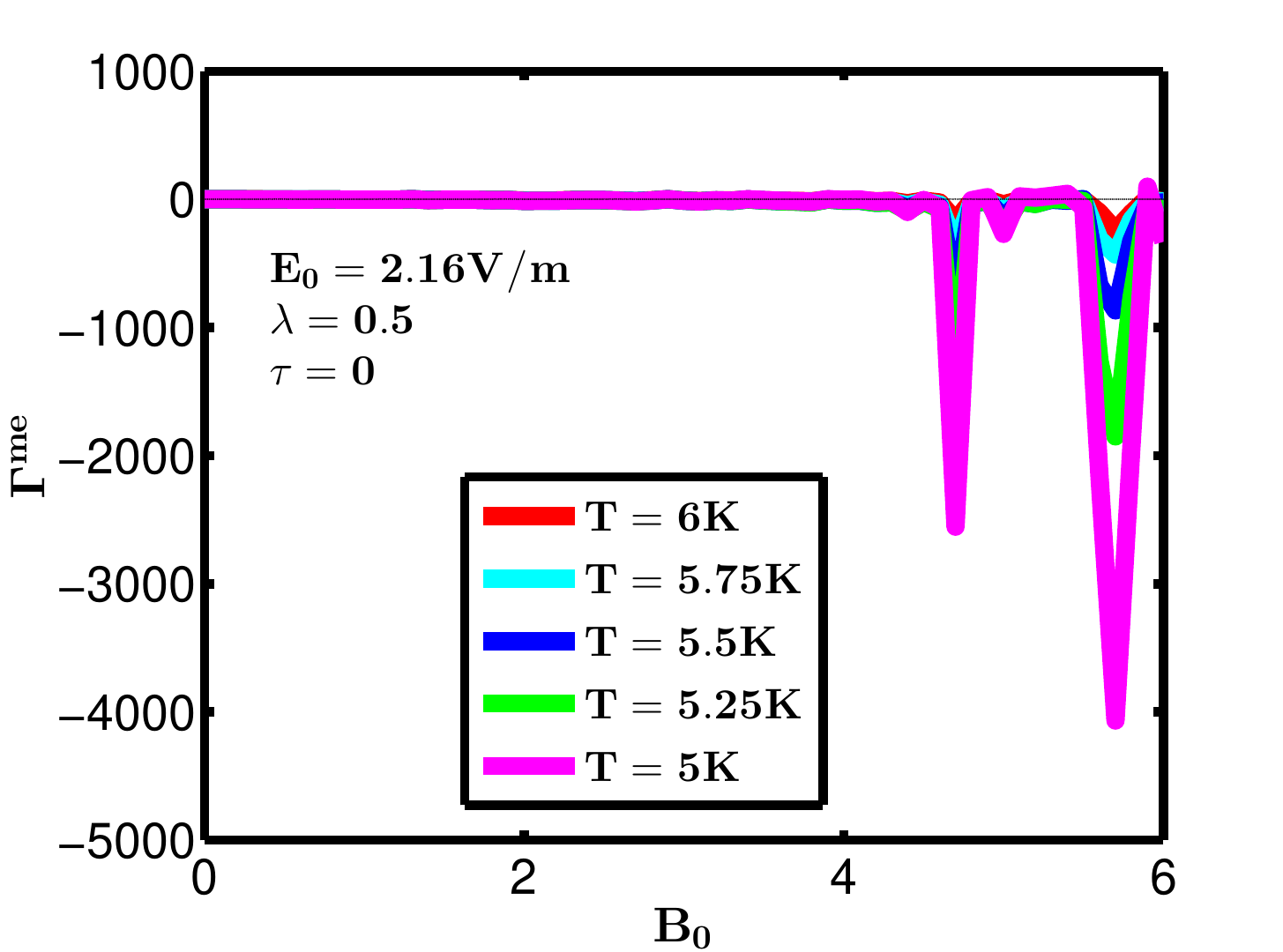}}
\subfigure[]{\includegraphics[width=0.35\textwidth]{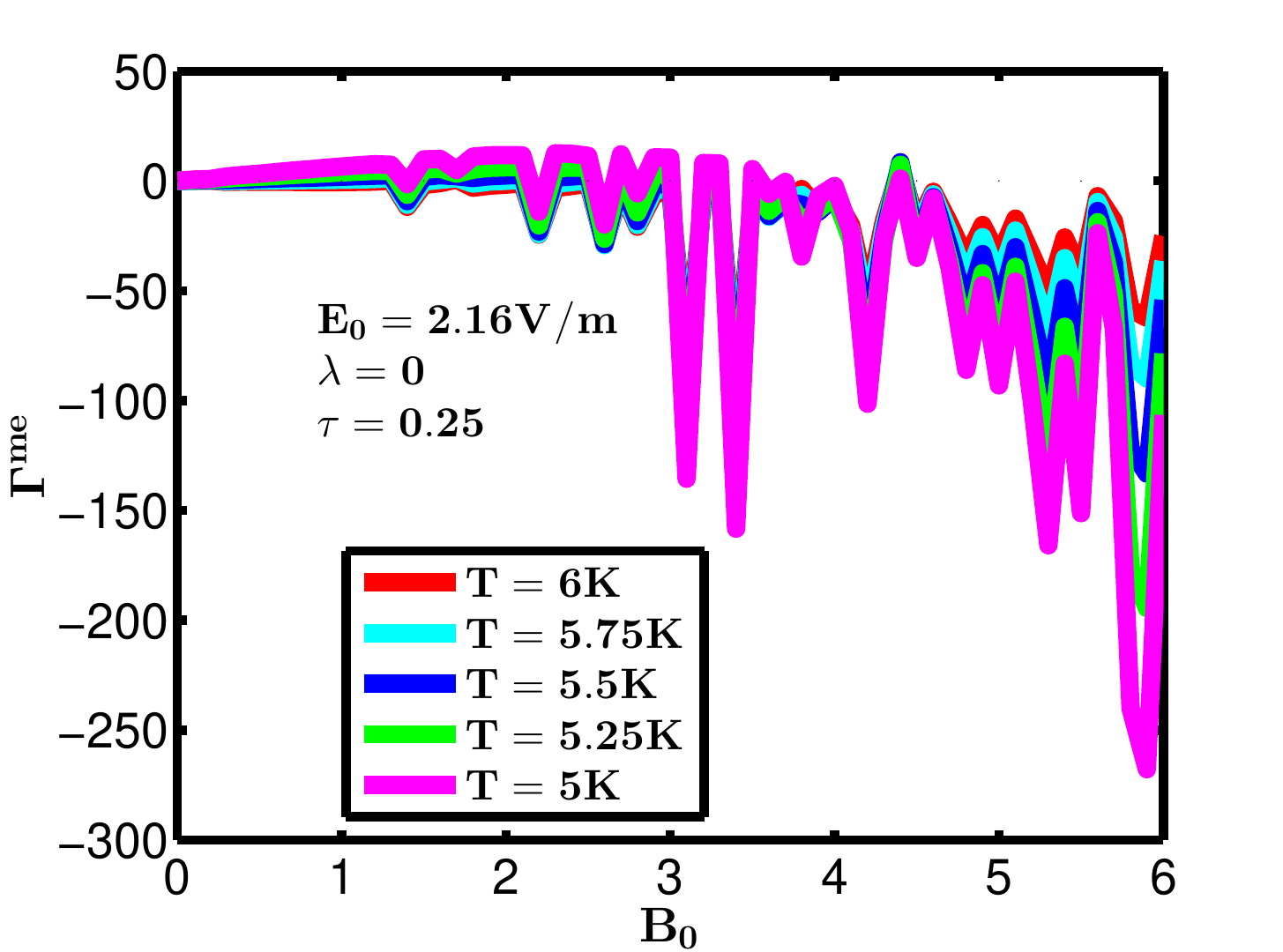}}
}
\centerline{\subfigure[]{\includegraphics[width=0.35\textwidth]{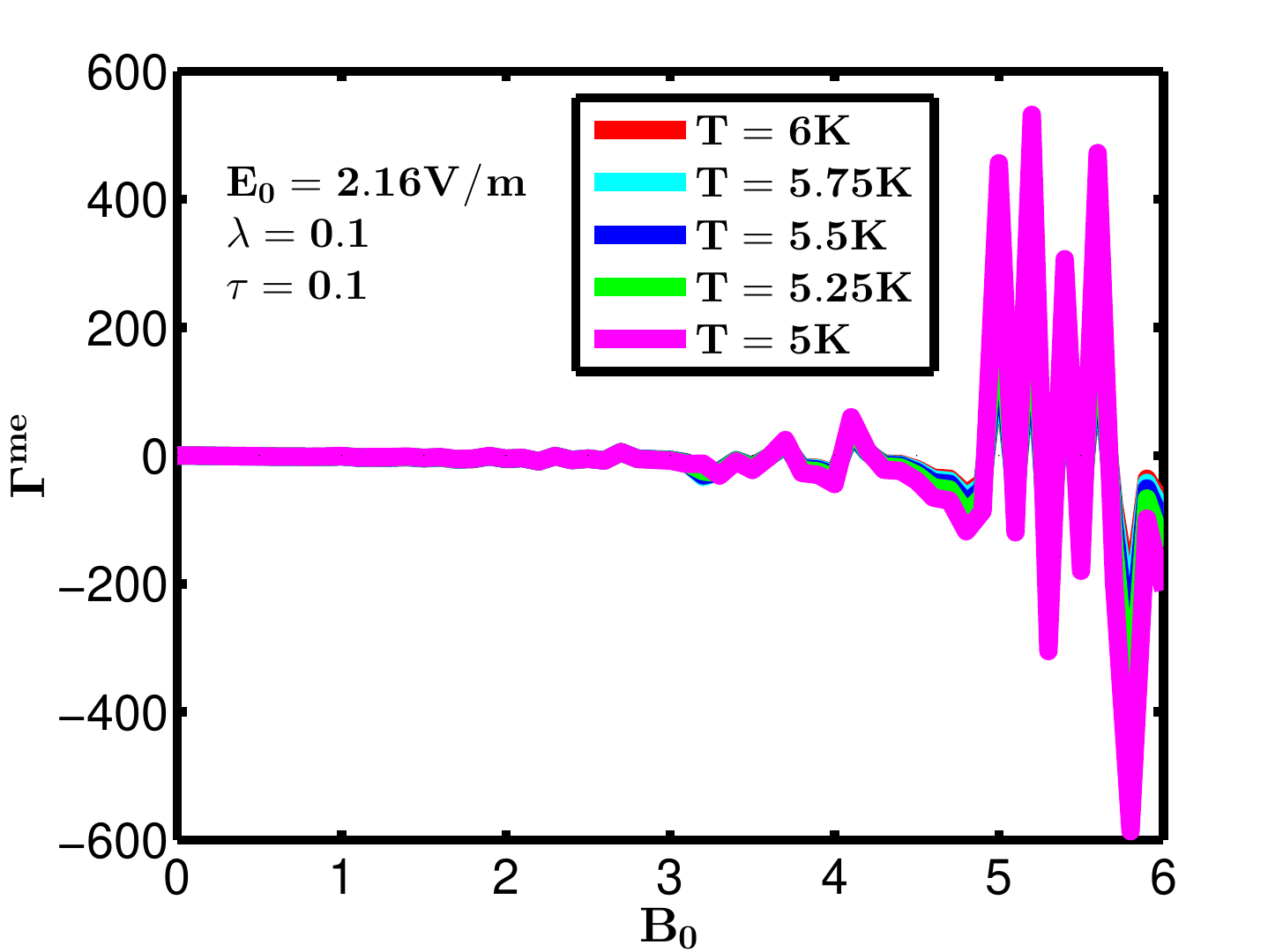}}
\subfigure[]{\includegraphics[width=0.35\textwidth]{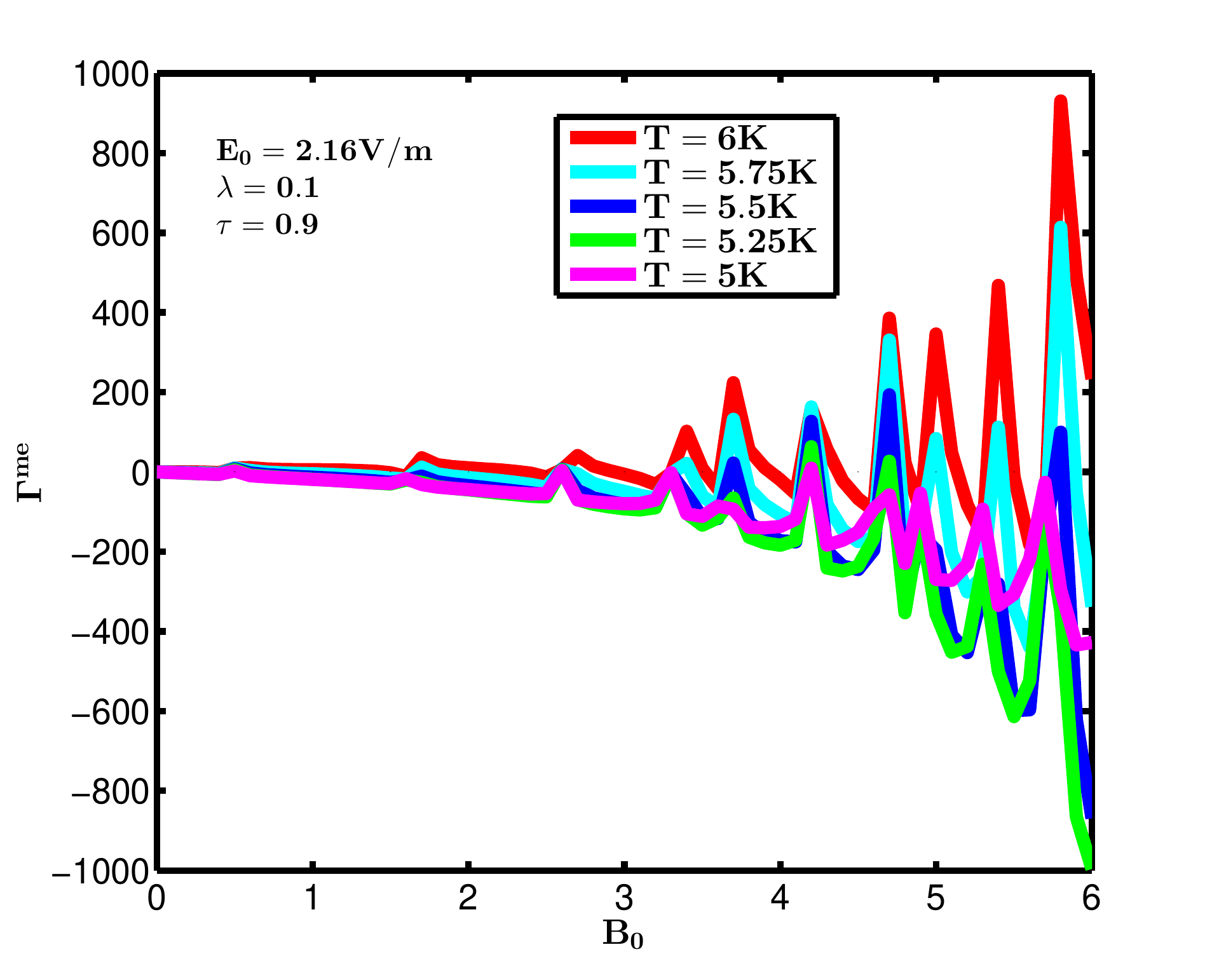}}
\subfigure[]{\includegraphics[width=0.35\textwidth]{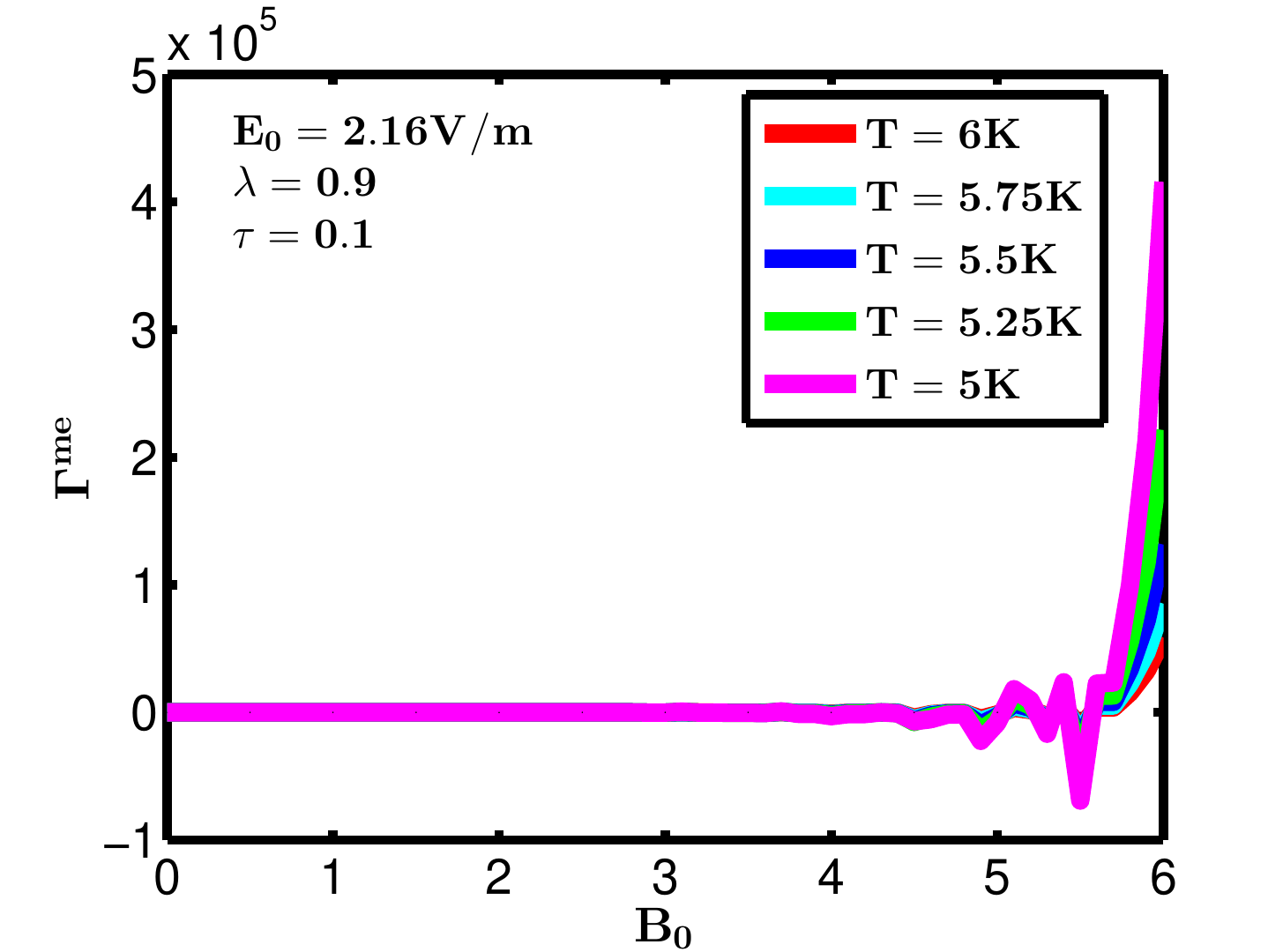}}
}
\caption{magnetic field dependence of the adiabatic magnetoelectric cooling rate by varying the temperature with the following magnetic and electric site-dependent parameters: $\lambda = 0.5$, $\tau =0$ (a); $\lambda = 0$, $\tau =0.25$ (b); $\lambda = 0.1$, $\tau =0.1$ (c); $\lambda = 0.1$, $\tau =0.9$ (d) ; $\lambda = 0.9$, $\tau =0.1$ (e).}
\label{F42}
\end{figure}

Furthermore, the magnetic field dependence of the magnetoelectric cooling rate is depicted in Fig.~\eqref{F42}. When $\lambda =0$ and $\tau \mathrm{\neq} 0$ and vice-versa, it is observed an appearance of the negative peak points from certain values of the magnetic field (up panels). However, under the influence of both the site-dependent magnetic and electric fields (that is $\tau \mathrm{\neq} 0$ and $\lambda \mathrm{\neq} 0$), the curves display almost the same qualitative behavior as the magnetic response of the electric cooling rate.

\par
\begin{figure}[!htb]
\centerline{
\subfigure[]{\includegraphics[width=0.35\textwidth]{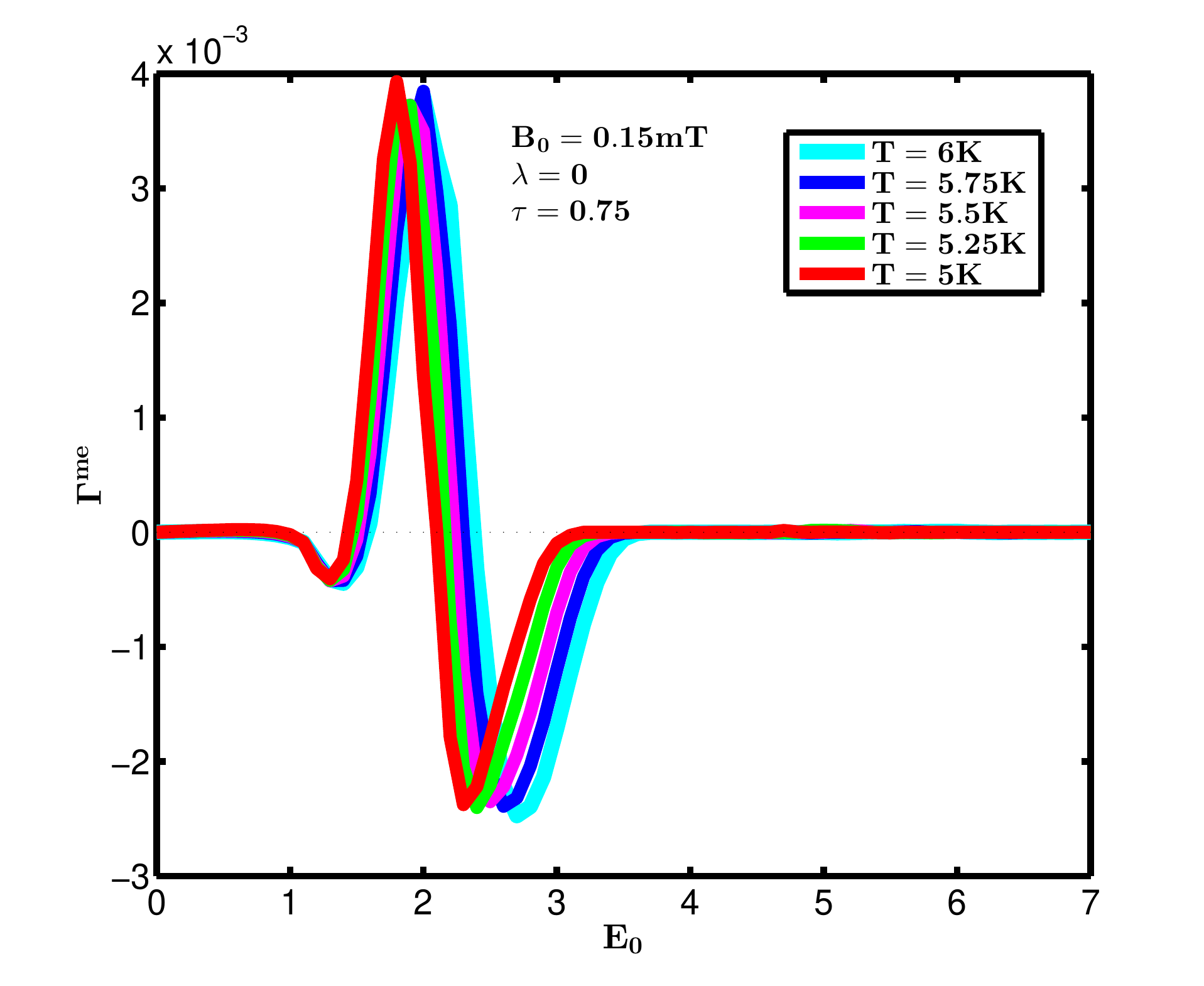}}
\subfigure[]{\includegraphics[width=0.35\textwidth]{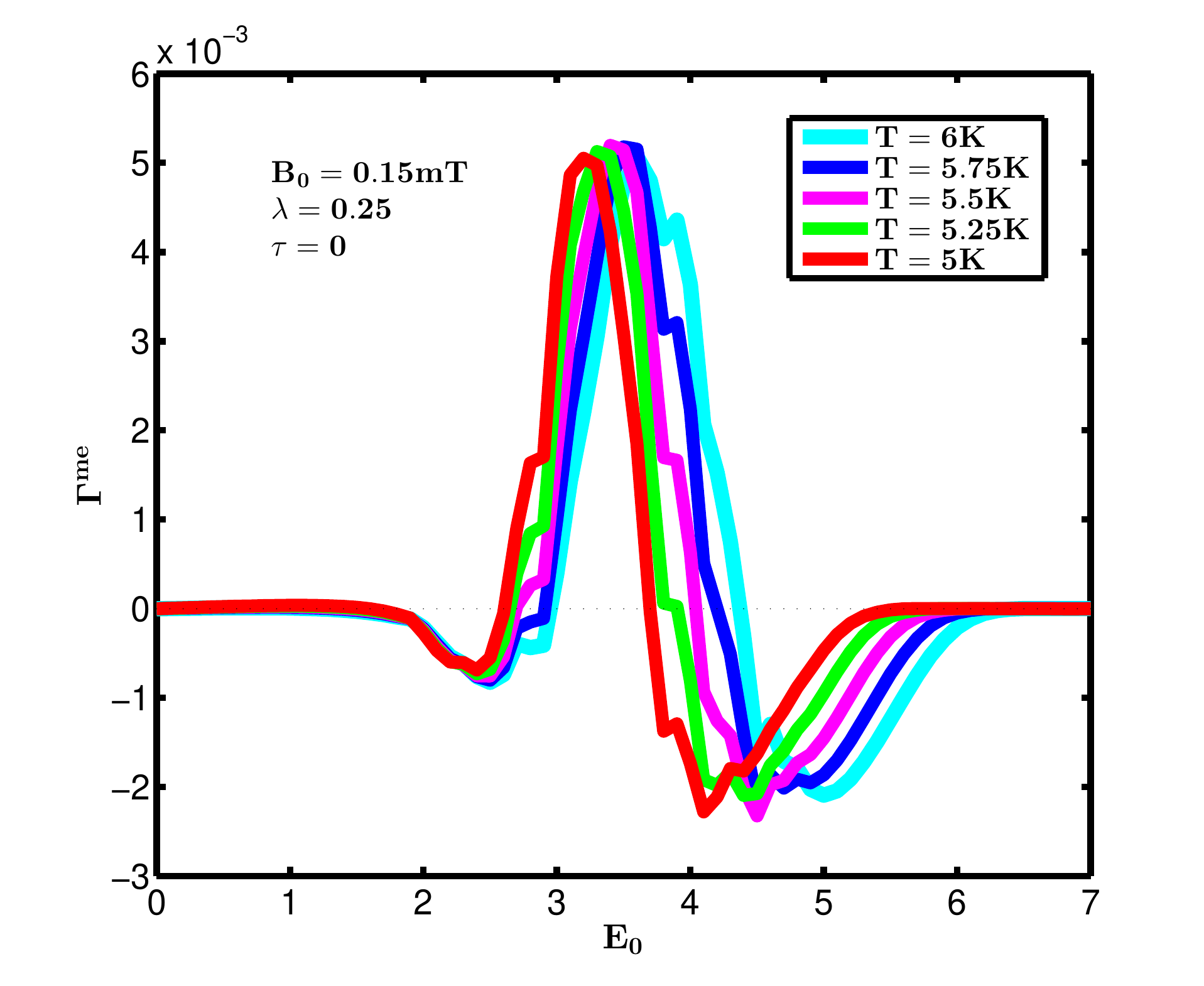}}
}
\centerline{\subfigure[]{\includegraphics[width=0.35\textwidth]{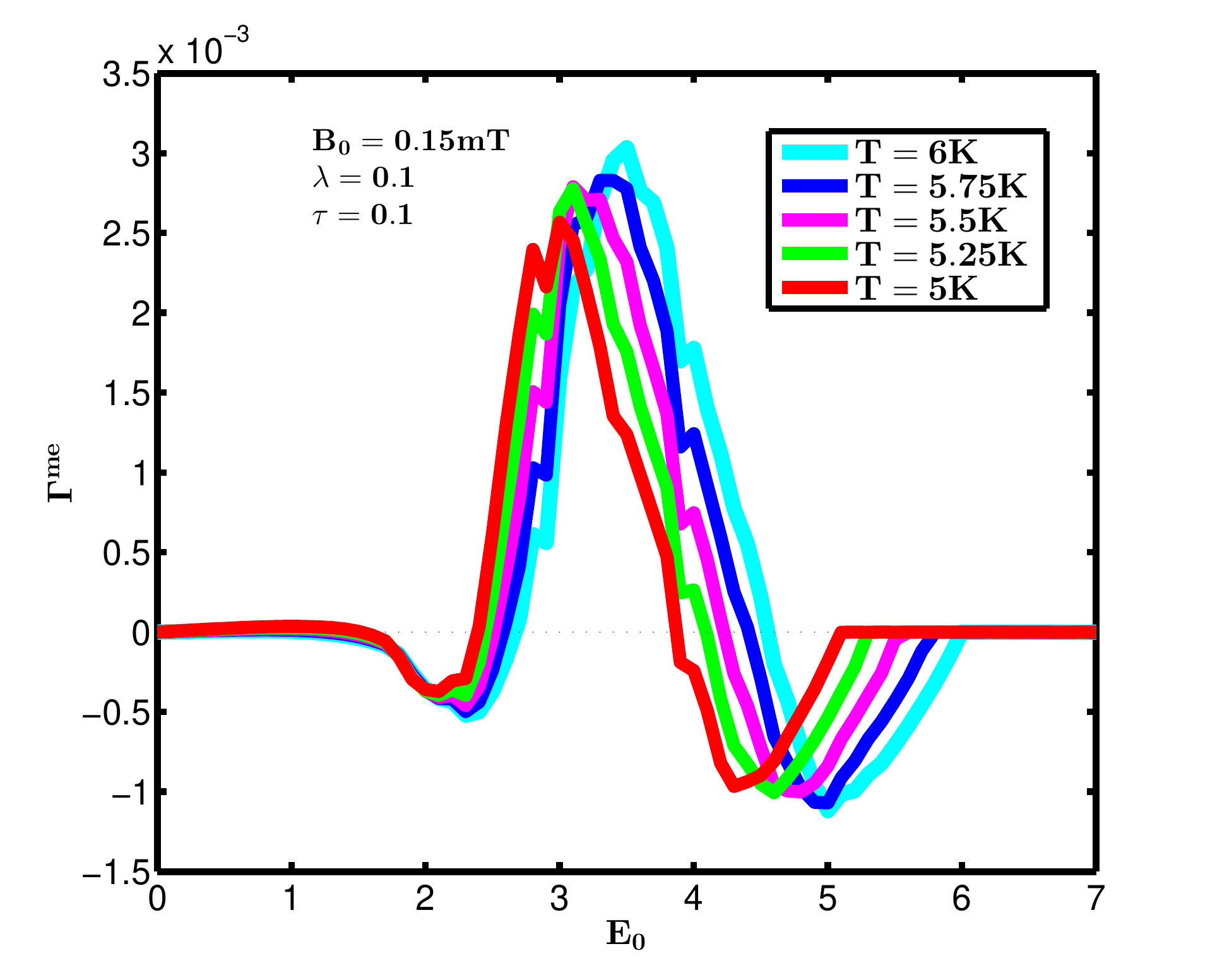}}
\subfigure[]{\includegraphics[width=0.35\textwidth]{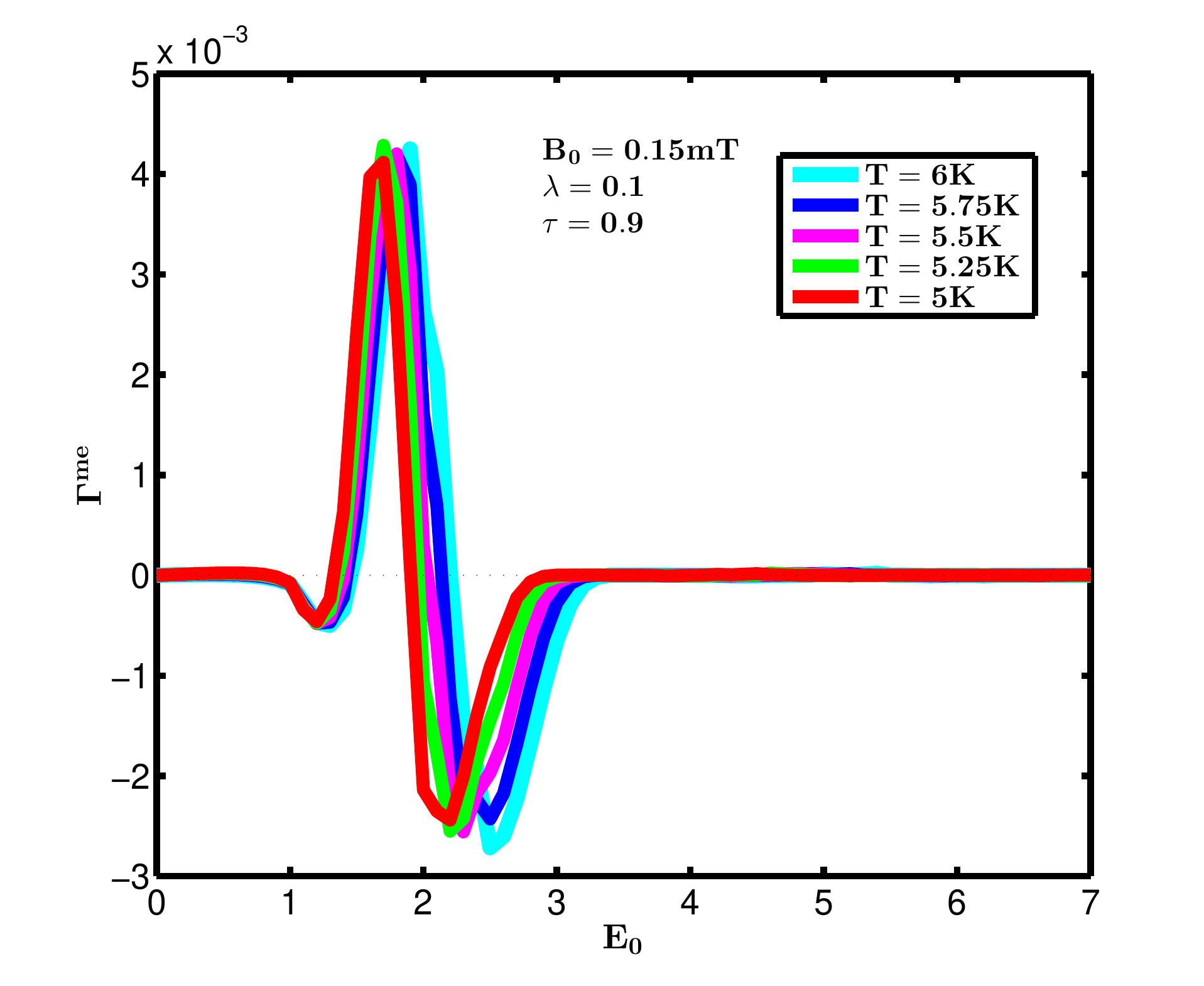}}
\subfigure[]{\includegraphics[width=0.35\textwidth]{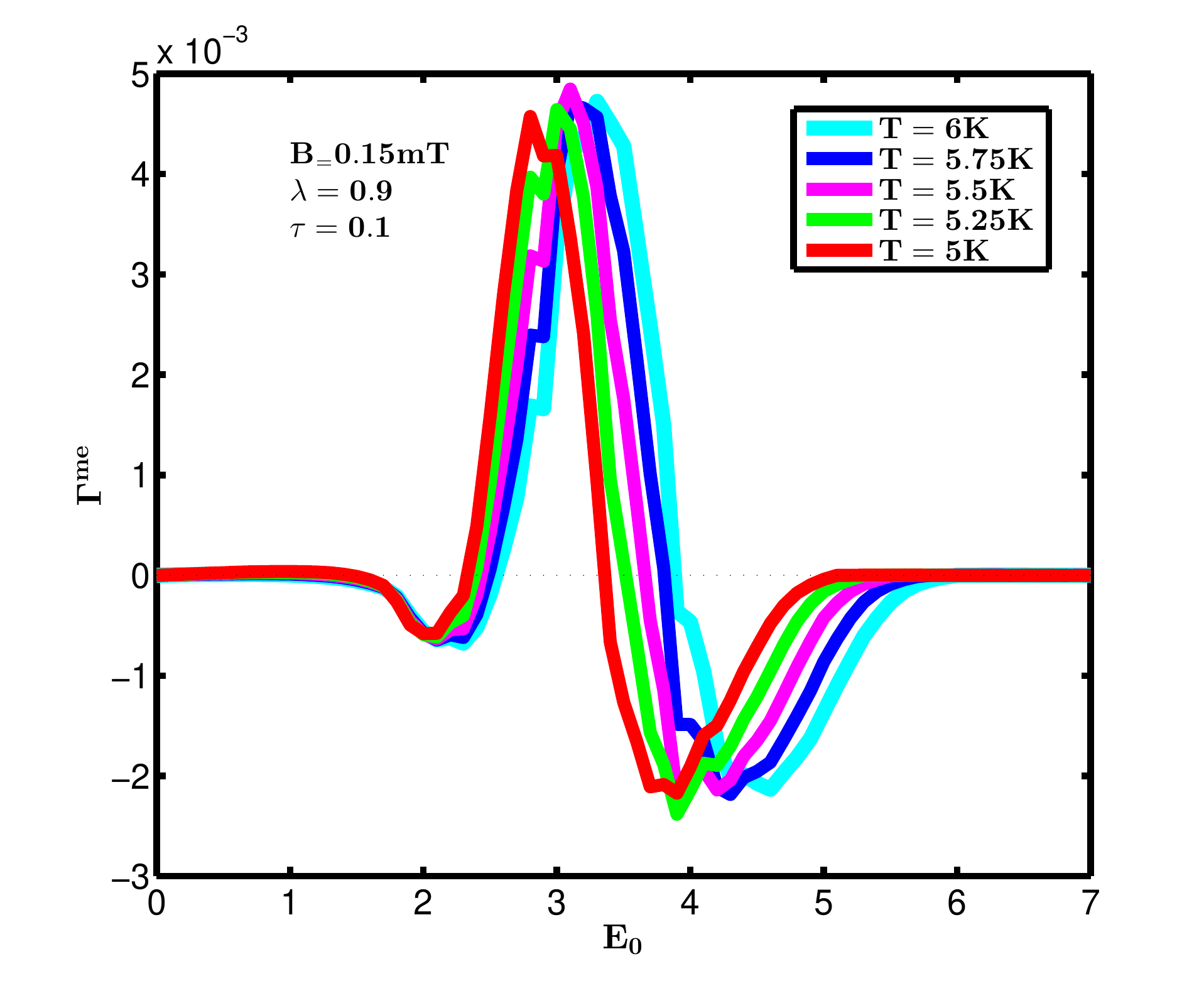}}
}
\caption{electric field dependence of  adiabatic magnetoelectric cooling rate by varying the temperature with the following magnetic and electric site-dependent parameters: $\lambda = 0$, $\tau =0.75$ (a); $\lambda = 0.25$, $\tau =0$ (b); $\lambda = 0.1$, $\tau =0.1$ (c); $\lambda = 0.1$, $\tau =0.9$ (d) ; $\lambda = 0.9$, $\tau =0.1$ (e).}
\label{F43}
\end{figure}

The electric field response of the magnetoelectric cooling rate as depicted in Fig.~\eqref{F43} shows that it displays three successive maxima with a positive middle one as observed in  ref.\cite{52}. Besides, the curve is smooth in the case of the site-dependent electric field than in the case of site-dependent magnetic field and these peaks appear between $1 V/m$ and $6 V/m$. Note that by analogy to the magnetic and electric cooling rates, the negative magnetoelectric cooling rate demonstrates the cooling of the system ( inverse magneto-electro caloric effect) whereas the positive magnetoelectric cooling rate signals the heating of the system (magneto-electro caloric effect).
\par
\begin{figure}[!htb]
	\centerline{\subfigure[]{\includegraphics[width=0.35\textwidth]{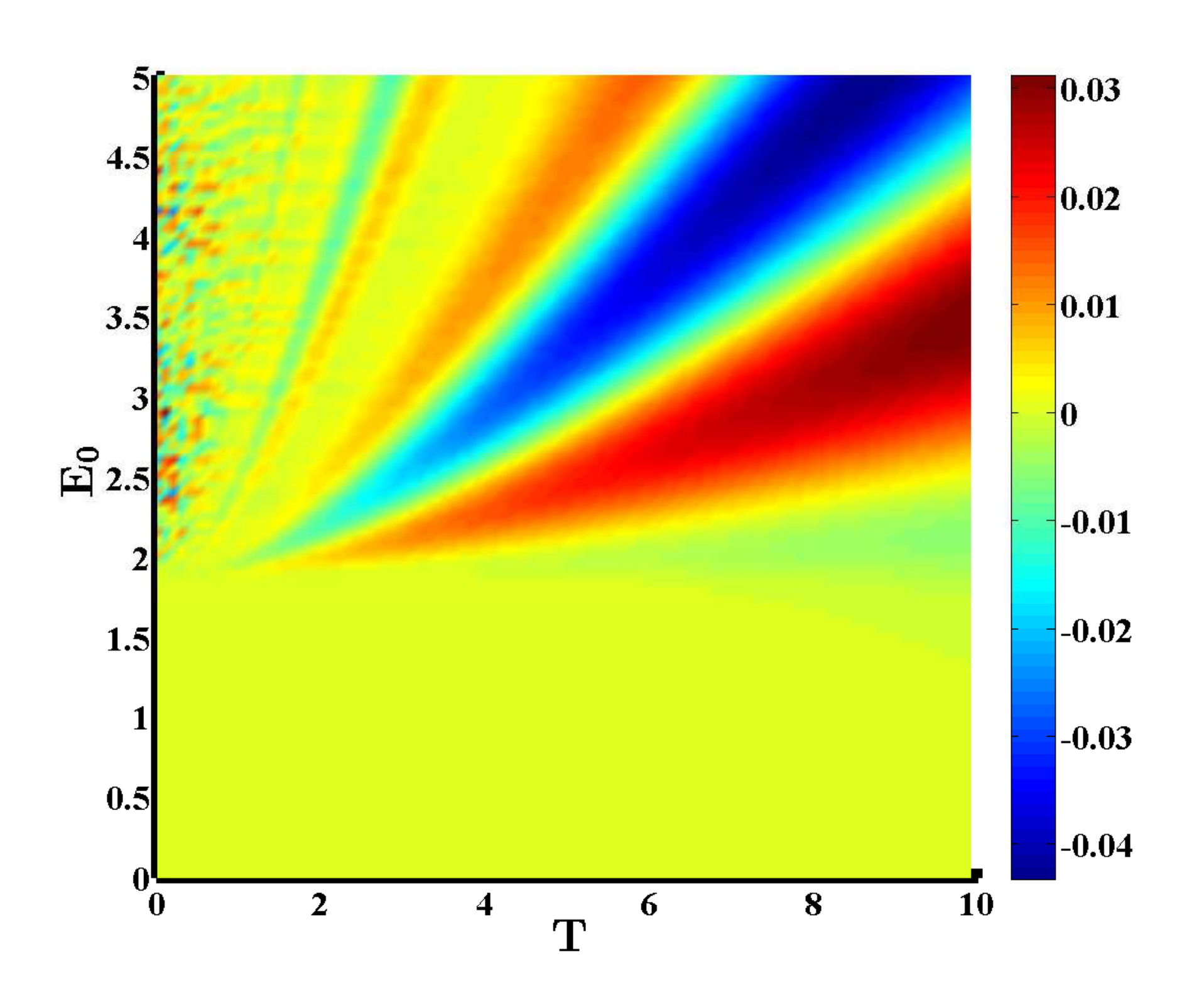}}
		\subfigure[]{\includegraphics[width=0.35\textwidth]{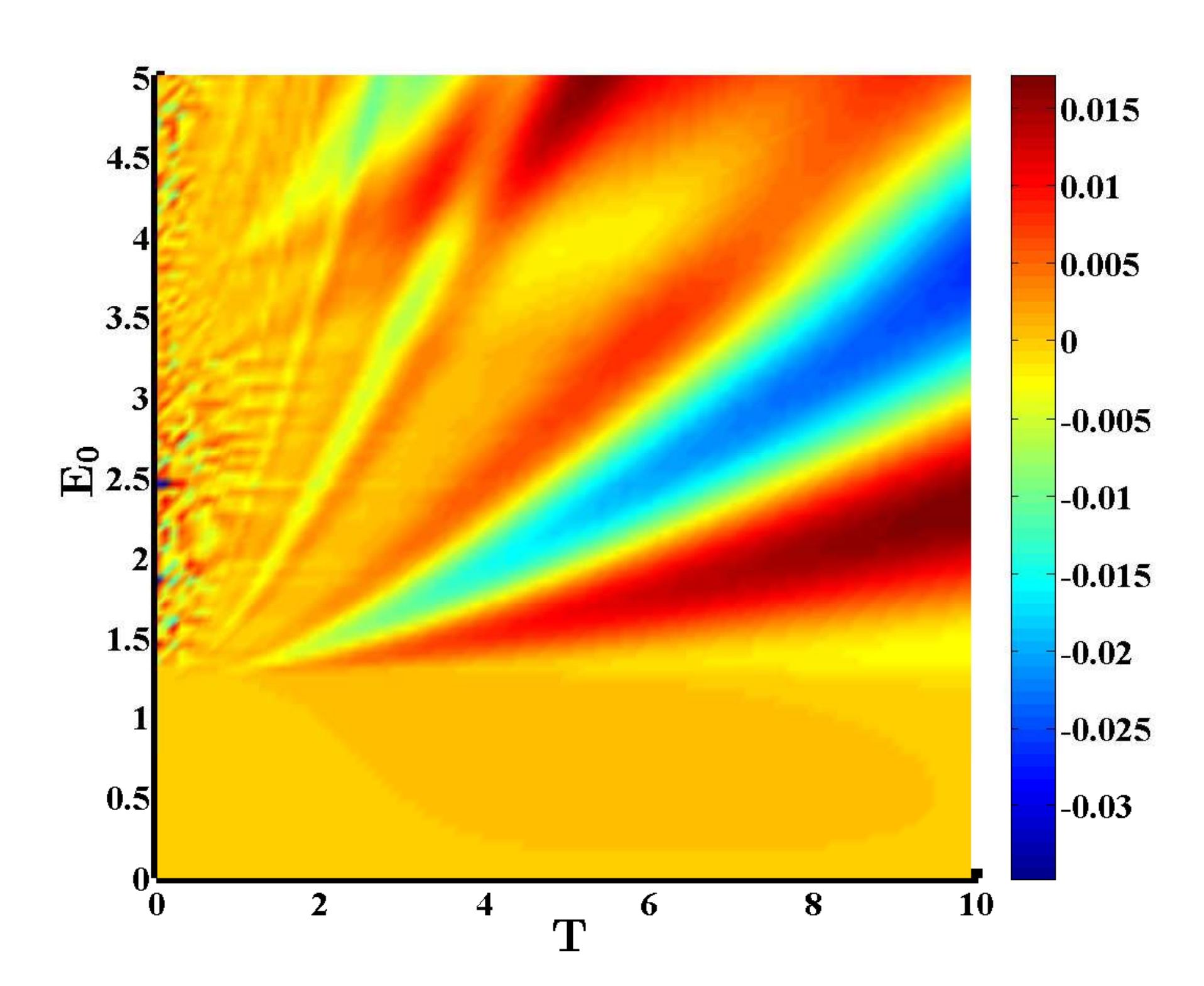}}
		\subfigure[]{\includegraphics[width=0.35\textwidth]{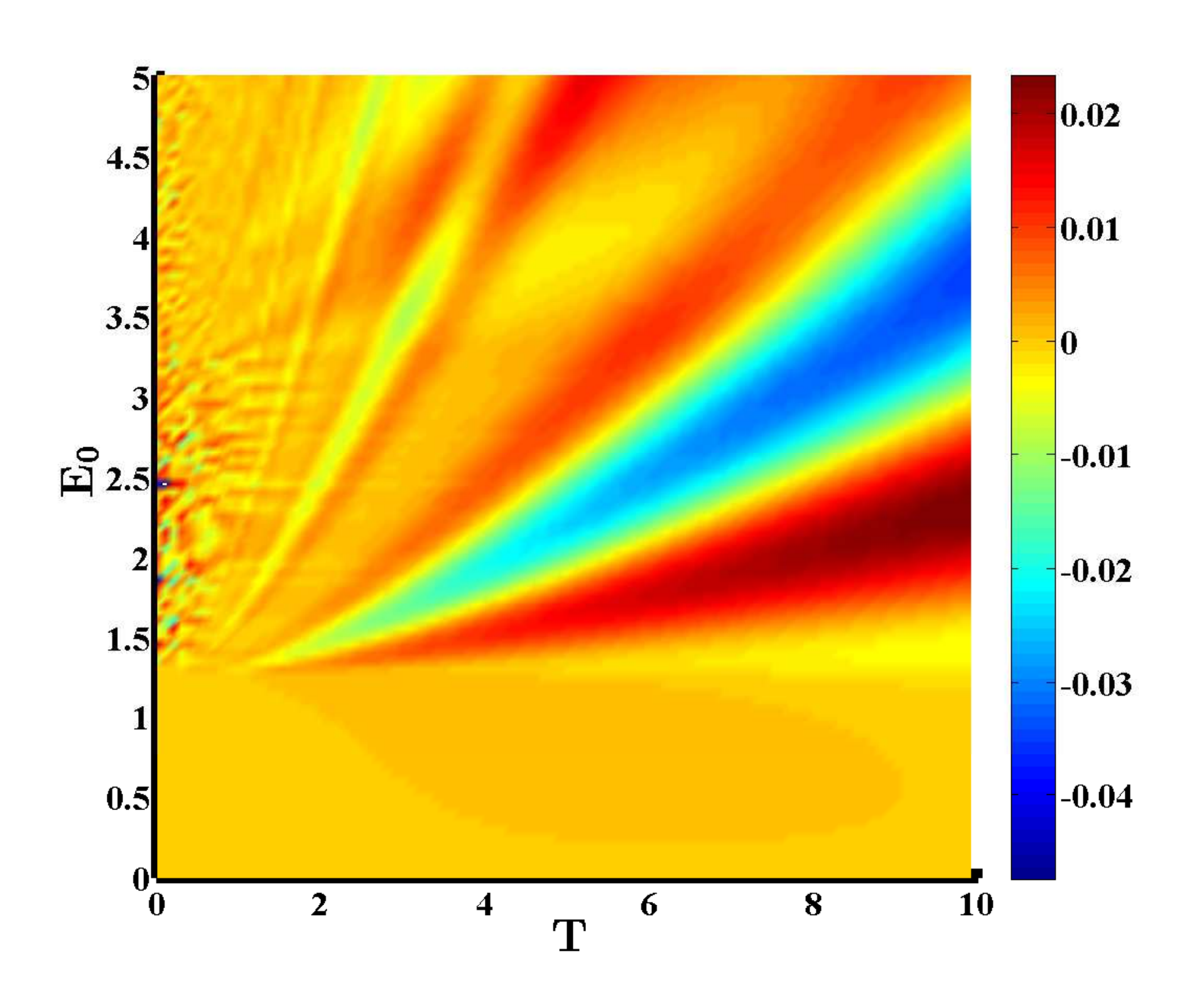}}
	}
	\caption{the surface plot of magnetoelectric polarizability against electric field and temperature with the following magnetic field, magnetic and electric site-dependent parameters :(a)$B_0 =0.15 mT, \lambda = \tau =  0$, (b) $ B_0 =0.15 mT, \lambda = 0.1, \tau =0.5$, (c) $B_0 =0.15 mT, \lambda = \tau =   0.5$. }
	\label{FO3}
\end{figure}

 
Furthermore, the phase diagram in term of the magnetoelectric polarizability obtained within the $E_0 - T$ plane, in  Fig.~\eqref{FO3}, shows three regions as function of approximated values of critical electric field and temperature: a region where the magnetoelectric polarizability varies slightly with the electric field ($E_0 \mathrm{<} 1.15 V/m$), a region where the magnetoelectric polarizability exhibits a peak-like points ($E_0 \mathrm{>} 1.15 V/m, T \mathrm{<} 1 K$) and a region where the magnetoelectric polarizability oscillates both with the electric field and the temperature ($E_0 \mathrm{>} 1.15 V/m , T \mathrm{>} 1 K$ ). Note that the critical electric field and the critical temperature are tunable by the magnetic and electric site-dependent parameters. 

\par
\par
\par
\section{Conclusion}\label{S5}
In this work, the influence of both the site-dependent magnetic and electric fields on the magneto-electrocaloric effect in a multiferroic antiferromagnetic quantum spin system has been investigated. The Hamiltonian is diagonalized by the help of the spin-wave theory and the free energy is derived via the statistical physic. As analytical results, the physical parameters such as the entropy, the specific heat capacity, the adiabatic magnetic, electric, and magnetoelectric cooling rate are calculated. The results displayed by the graphs obtained demonstrate that the site-dependent magnetic and electric fields strongly affect the caloric effect and quantum phase transitions operate in multiferroic antiferromagnets.  

\par 
Indeed, by interpreting the magnetic field dependence of entropy, It is showed that one peak-like point occurs in the case of the uniform magnetic field ($\lambda =0$) as observed in the previous work \cite{25}. However, two peaks-like points occur in the case of a site-dependent magnetic field ($\lambda \mathrm{\neq} 0$) indicating the formation of the intermediate phase between the order and the disordered phase. It is observed that the parameter $\tau $ considerable affects the amplitudes of the peaks whereas the parameter $\lambda $ affects the position of theses peaks (critical magnetic fields). In fact, when $\lambda $ increase the first critical point is shifted into the left meanwhile and the second critical point is shifted into the right. This implies that by increasing the parameter $\lambda $ one can effectively decrease and increase the values of the critical magnetic fields. That result is confirmed by the magnetic field dependence of the specific heat capacity. In addition, the electric field dependence of the entropy displays an oscillatory behaviour from certain value of the electric field (critical electric field).

\par 
On the other hand, the numerical results obtained for the adiabatic magnetic, electric, and magnetoelectric cooling rate show a characteristic behavior of the caloric or multi-caloric effect as observed by the experimentalist and in other theoretical works. Indeed, the result display by the graphs exhibits the alternating negative and positive peaks occur at corresponding change points (critical temperature, critical  magnetic or electric fields) tuneable by the electric or/and magnetic fields. Note that the site-dependent magnetic field enhance the magneto-electrocaloric effect.

\par
 Overall, it is demonstrated that the cumulative influence of the site-dependent magnetic and electric fields allows us not only to reveal quantum critical points hide in a multiferroic quantum spin system but also to control the caloric or multi-caloric effect essential in the construction of the solid-state refrigeration devices. 

\section*{Acknowledgement}
This research did not receive any specific grant from funding agencies in the public, commercial, or not-for-profit sectors.
%

\end{document}